\documentclass[nopacs,nofootinbib]{article}

\usepackage{feynmp}
\usepackage{graphicx}
\usepackage{amssymb}
\usepackage{amsmath}
\usepackage{bm}
\usepackage{sidecap}
\usepackage{multirow}
\usepackage[dvips]{color}
\usepackage{colordvi}
\usepackage{verbatim}
\usepackage{bbm}
\usepackage{bigstrut}
\usepackage{fix-cm}

\usepackage[labelfont={normalsize},subrefformat=parens]{subfig}
\usepackage{enumerate}
\usepackage[french,english,american]{babel}
\usepackage{array}
\usepackage{t1enc}

\newcommand{\fourmat}[4]{{\begin{pmatrix}
                         {#1} & {#2} \\ {#3} & {#4}
                         \end{pmatrix}}}
\newcommand{\twospinor}[2]{{\begin{pmatrix}
                           {#1} \\ {#2}
                           \end{pmatrix}}}
           
\newcommand{\bra}[1]{\langle #1 \vert}
\newcommand{\ket}[1]{\vert #1 \rangle}

\begin{document}

\title{The nuclear energy density functional formalism}

%%%%%%%%%%%%%%%%%%%%%%%%%%%%%%%%%%%%%%%%%%%%%%%%%%%%%%%%%%%%%%%%%%%%%%%%%%%%%%%
\author{T. Duguet\\
CEA-Saclay DSM/Irfu/SPhN, F-91191 Gif sur Yvette Cedex, France\\
NSCL and Department of Physics and Astronomy, \\ Michigan State University, East Lansing, MI 48824, USA\\
Email: thomas.duguet@cea.fr}
%\date{\today}
  
%\email{thomas.duguet@cea.fr}
%\affiliation{CEA-Saclay DSM/Irfu/SPhN, F-91191 Gif sur Yvette Cedex, France}
%\affiliation{National Superconducting Cyclotron Laboratory and Department of Physics and Astronomy, Michigan State University, East Lansing, MI 48824, USA}
  
\maketitle
 
\begin{abstract}
The present document focuses on the theoretical foundations of the nuclear energy density functional (EDF) method. As such, it does not aim at reviewing the status of the field, at covering all possible ramifications of the approach or at presenting recent achievements and applications. 
The objective is to provide a modern account of the nuclear EDF formalism that is at variance with traditional presentations that rely, at one point or another, on a {\it Hamiltonian-based} picture. The latter is not general enough to encompass what the nuclear EDF method represents as of today. Specifically, the traditional Hamiltonian-based picture does not allow one to grasp the difficulties associated with the fact that currently available parametrizations of the energy kernel $E[g',g]$ at play in the method do not derive from a genuine Hamilton operator, would the latter be effective. The method is formulated from the outset through the most general multi-reference, i.e. beyond mean-field, implementation such that the single-reference, i.e. "mean-field", derives as a particular case. As such, a key point of the presentation provided here is to demonstrate that the multi-reference EDF method can indeed be formulated in a {\it mathematically} meaningful fashion even if $E[g',g]$ does {\it not} derive from a genuine Hamilton operator. In particular, the restoration of symmetries can be entirely formulated without making {\it any} reference to a projected state, i.e. within a genuine EDF framework. However, and as is illustrated in the present document, a mathematically meaningful formulation does not guarantee that the formalism is sound from a {\it physical} standpoint. The price at which the latter can be enforced as well in the future is eventually alluded to.
\end{abstract}

\newpage

\tableofcontents

\newpage

\pagestyle{headings}

\section{Introduction}

\subsection{Generalities}

Low-energy nuclear physics aims at addressing several fundamental, yet only partially answered, questions. Among those are (i) {\it how do neutrons and protons bind inside a nucleus and what are the limits of existence of the latter regarding its mass, neutron-proton imbalance, angular momentum\ldots?} (ii) {\it How to explain the complex phenomenology of nuclei starting from elementary two-, three-\dots A-nucleon (AN) interactions?} (iii) {\it How do the latter interactions eventually emerge from quantum chromodynamics (QCD)?} Such questions have numerous ramifications and implications such that partial answers to them continuously impact other fields of physics (e.g. astrophysics, tests of the Standard Model). In spite of over eighty years of theoretical and experimental studies, low-energy nuclear physics remains an open and difficult problem. While extensive progress has been made, an accurate and universal description of low-energy nuclear systems from first principles is still beyond reach.

The first difficulty resides in the inter-particle interactions at play. Strong inter-nucleon interactions relevant to describing low-energy phenomena must be modelled within the non-perturbative regime of the gauge theory of interacting quarks and gluons, i.e. QCD. Within such a frame, nucleons are assigned to spin and isospin SU(2) doublets such that they are 4-component fermions interacting in various configurations stemming from invariances of the problem, e.g. they interact through central, spin-orbit, tensor, quadratic spin-orbit... couplings. In addition to its complex operator structure, the 2N force produces a weakly-bound neutron-proton state (i.e. the deuteron) in the coupled $^3$S$_1$-$^3$D$_1$ partial waves and a virtual di-neutron state in the $^1$S$_0$ partial wave. Associated large scattering lengths, together with the short-range repulsion between nucleons make the nuclear many-body problem highly non-perturbative. In addition to such difficulties, the treatment of 3N, 4N\ldots interactions in a theory of point-like nucleons is unavoidable. This has become clear over the last fifteen years as one was aiming at a consistent understanding of (i) differential nucleon-deuteron cross-sections~\cite{KalantarNayestanaki:2011wz}, (ii) the under-estimation of triton and light-nuclei binding energies~\cite{nogga00}, (iii) the Tjon line~\cite{nogga04b}, (iv) the violation of the Koltun sum rule~\cite{faessler75} and (v) the saturation of symmetric nuclear matter~\cite{fujita57,zuo02a} in connection with the Coester line problem~\cite{coester70,brockmann90a}.

The second difficulty stems from the nature of the system of interest. Most nuclei (i.e. those with masses typically between 10 and 350) are by essence intermediates between few- and many-body systems. As a result (i) most nuclei are beyond theoretical and computational limits of ab-initio techniques that describe the interacting system from basic AN forces, while (ii) finite-size effects play a significant role, which prevents statistical treatments. Furthermore, a unified view of low-energy nuclear physics implies a coherent description of small- and large-amplitude collective motions, as well as of closed and open systems, e.g. of the structure-reaction interface that is mandatory to understand spontaneous and induced fission, fusion, nucleon emission at the drip-line\ldots

The study of the atomic nucleus aims at accessing its ground-state (mass, radius, deformation and multipolar moments...) and excited-states (single-particle, vibrational, shape and spin isomers, high-spin and super-deformed rotational bands...) properties as well as the various decay modes between them (nuclear, electromagnetic and electroweak), together with reaction properties (elastic and inelastic scattering, transfer and pickup, fusion...). This is to be achieved for systems over the nuclear chart, i.e. not only for the nearly 3100 observed nuclei~\cite{sonzogni07} but also for the thousands that are still to be discovered. In that respect, a cross-fertilization between theoretical and experimental studies is topical, with the with the advent of (i) a new-generation of radioactive-ion-beam (RIB) facilities producing very short-lived systems with larger yields, and (ii) high-sensitivity and high-selectivity detectors allowing measurements with low statistics. Upcoming facilities based on in-flight fragmentation, stopped and reaccelerated beams or a combination of both are going to further explore the nuclear chart towards the limits of stability against nucleon emission, the so-called neutron and proton drip-lines. The study of highly neutron-rich nuclei will help understand the astrophysical nucleosynthesis of about half of the nuclei heavier than iron through the conjectured r-process. The access to nuclei with a large neutron-over-proton ratio has already started to modify certain cornerstones of nuclear structure, e.g. some of the "standard" magic numbers are significantly altered while others (may) appear~\cite{sorlin08}. When adding even more neutrons, the proximity of the Fermi energy to the particle continuum gives rise to exotic phenomena, such as the formation of light nuclear halos~\cite{tanihata85a,fukuda91} with anomalously large extensions~\cite{hansen87,jensen04} or the existence of di-proton emitters~\cite{blank07a,pfutzner12a}. In addition to reaching out to the most exotic nuclei, experiments closer to the valley of stability still provide critical information. For instance, precise mass measurements using Penning traps~\cite{blaum06a} or Schottky spectrometry~\cite{schlitt96} not only refine and extend mass difference formul\ae~\cite{wang11a} to  better understand nuclear structure properties, e.g. pairing correlations, but also contribute to testing the standard model of particle physics, e.g. recent mass measurements have helped refine the validation of the unitarity of the Cabibbo-Kobayashi-Maskawa (CKM) flavour-mixing matrix~\cite{Towner:2010zz}. Eventually, other limits of existence are of key importance, e.g. the quest for superheavy elements and the conjectured island of stability beyond the $Z=82$ magic number~\cite{Zagrebaev:2012hy}. In addition to the quoted references, we refer the interested reader to Vols. 1-3 of this series that contain many contributions relevant to the topics alluded to just above.

\subsection{Nuclear structure theory}

In such a context, the challenge of contemporary nuclear structure theory is to describe, in a controlled\footnote{The notion of "controlled" description refers to the capability of estimating uncertainties of various origins in the theoretical method employed.} and unified manner, the entire range of nuclei along with the equation of state of extended nuclear matter, from a fraction to few times nuclear saturation density and over a wide range of temperatures. All such properties find an interesting outcome in the physics of neutron stars and supernovae explosions as well as in the nucleosynthesis of heavy elements as already alluded to above.

\subsubsection{Ab initio methods}

While bulk properties of nuclei can be roughly explained using macroscopic approaches such as the liquid drop model (LDM)~\cite{moller02,royer06}, microscopic techniques are the tool of choice for a coherent description of static and dynamical nuclear properties. This leads to defining the class of so-called {\it ab-initio} methods that consists of solving, as exactly as possible, the nuclear many-body problem expressed in terms of elementary 2N, 3N, 4N\ldots interactions. For three- and four-nucleon systems, essentially exact solutions of the Faddeev or Yakubowski equations can be obtained using realistic vacuum forces~\cite{nogga00,friar88,nogga97}. Likewise, Green's function Monte-Carlo (GFMC) calculations~\cite{Pieper:2004qw,Pastore:2013ria} provide a numerically exact description of nuclei up to carbon starting from local 2N and 3N vacuum forces, although such a method already faces huge numerical challenges for $^{12}$C. Complementary ab-initio methods allow the treatment of nuclei up to $A\approx16$, e.g. (i)  the no-core shell model (NCSM)~\cite{Navratil:2009ut} that projects the interacting problem on a truncated harmonic oscillator model space or (ii) lattice effective field theory (LEFT)~\cite{Epelbaum:2012qn} that propagates nucleons as point-like particles on lattice sites interacting via pion exchanges and multi-nucleon operators. 

In the last ten years, a breakthrough has occurred that renders possible the ab-initio calculation of double closed-shell nuclei, along with those in their immediate vicinity, with masses up to $A\approx60$ on the basis of realistic 2N and 3N interactions. Three methods have been developed in order to move in this direction. First is Coupled-cluster (CC) theory~\cite{Hagen:2010gd,Binder:2012mk}, which constructs the correlated ground-state from a product state using an exponentiated cluster expansion, truncated to $B$-body operators (typ. $B\sim 2-3$). Second, self-consistent Green's function (SCGF)~\cite{Dickhoff:2004xx,Cipollone:2013zma} computes the approximate {\it dressed} one-body Green's function describing the propagation of a nucleon within the correlated medium. Last but not least, in-medium similarity renormalization group (IMSRG)~\cite{Tsukiyama:2010rj,Hergert:2012nb} proceeds to the decoupling of a finite-density reference state from excitations built on top of it via a sequence of infinitesimal renormalization group transformations. The frontier in the development of such ab-initio many-body methods is not only to push calculations to higher masses but also to extend them to truly open-shell systems. Decisive steps are taken in this direction for SCGF~\cite{Soma:2011aj,Soma:2012zd}, IMSRG~\cite{Hergert:2013uja} and CC~\cite{signoracci13a} theories. This is meant to extend the reach of ab-initio calculations from a few tens to several hundreds of mid-mass nuclei.

\subsubsection{The configuration interaction method}

Accessing even heavier systems requires more drastic approximations to the interacting many-body problem. Part of the physics that cannot be treated explicitly is accounted for through the formulation and use of so-called {\it in-medium interactions}. The configuration interaction (CI) model~\cite{caurier04}, i.e. shell model (SM), constructs a model space within which valence nucleons interact through an effective interaction that compensates for high-lying excitations outside that model space as well as for excitations of the core that are not treated explicitly. Even though such an effective interaction can be constructed starting explicitly from elementary interactions~\cite{Dean:2004ck}, certain combinations of two-body matrix elements\footnote{In the sd shell for example, it is necessary to (slightly) refit about 30 combinations of two-body matrix elements in order to reach about $140$ keV root mean square error on nearly 600 pieces of spectroscopic data~\cite{brown06a}.} need to be slightly refitted to experimental data within the chosen model space (sd, pf...) to correct for the so-called monopole part of the interaction. Based on the conjectures that wrong monopoles originate from the omission of the 3N force in the starting vacuum Hamiltonian~\cite{zuker03}, the non-empirical SM based on diagrammatic techniques\footnote{The adjective "diagrammatic" refers to many-body methods  relying on the use of Feynman or Goldstone diagrams.} from renormalized 2N and 3N interactions is currently being revived~\cite{Otsuka:2009cs} and shows promising results~\cite{Holt:2010yb,Holt:2011fj}. Eventually, spectroscopic properties can be described with high accuracy using refitted effective interactions~\cite{caurier04,brown06a}. Still, improved accuracy is needed in the SM to use nuclei as laboratories for fundamental symmetries, e.g. to provide the matrix elements needed for the search of neutrinoless double-beta decay~\cite{Holt:2013tda}. 

\subsubsection{The nuclear energy density functional method}

Last but not least, the theoretical tool of choice for the microscopic and systematic description of medium- and heavy-mass nuclei is the energy density functional (EDF) method~\cite{bender03b,Niksic:2011sg}, often referred to as "self-consistent mean-field and beyond-mean-field methods". Such method has been empirically adapted from well-defined wave-function- and Hamiltonian-based approaches. Based on a relativistic or a non-relativistic framework, the EDF method aims at providing, within one consistent frame, (i) the detailed and complete description of specific nuclei of interest, (ii) systematic trends over a large set of nuclei and (iii) trustful extrapolations in the region of the nuclear chart where experimental data are and will remain unavailable. Thanks to a favourable numerical scaling, the EDF method is indeed amenable to systematic studies of systems with large numbers of nucleons, independent of their expected shell structure. The idealized {\it infinite nuclear matter} system relevant to the description of compact astrophysical objects such as neutron stars is accessible to EDF calculations as well. 

A fundamental aspect of the method is that it relies heavily on the concept of spontaneous breaking and restoration of symmetries. As such, the nuclear EDF method is intrinsically a two-step approach,
\begin{enumerate}
\item The first step is constituted by the so-called single-reference EDF (SR-EDF) implementation, originally adapted from the symmetry-unrestricted Hartree Fock Bogoliubov (HFB) method by using a {\it density-dependent} effective Hamilton "operator"~\cite{negele72a}. Later, the approximate energy was formulated directly as a possibly richer functional of one-body density matrices computed from a symmetry-breaking HFB state of reference. The power of the approach relies on its ability to parametrize the bulk of many-body correlations under the form of a functional of one-body density (matrices) while authorizing the latter to break symmetries dictated by the underlying Hamiltonian in order to account for static collective correlations. It is however difficult, if not impossible, to capture in this way the subsequent dynamical correlations associated with good symmetries and quantum collective fluctuations.
\item It is thus the goal of the second step, carried out through the multi-reference (MR) extension of the SR-EDF method, to grasp such long-range correlations. The MR-EDF implementation has been adapted from the generator coordinate method (GCM) performed in terms of symmetry-projected HFB states~\cite{ring80a}. Within the EDF context, the MR step necessitates a prescription to extend the SR energy functional\footnote{i.e. the density-dependence of the effective Hamilton operator in the traditional formulation.} associated with a single auxiliary state of reference to the non-diagonal energy kernel associated with a pair of reference states. Although constraints based on physical requirements have been worked out that limit the number of possible prescriptions~\cite{Robledo07a}, no first-principle approach to the formulation of such an extension exists today. Although this could have simply remained an academic issue with no measurable consequence, it has been realized recently that the lack of rigorous roots of the EDF method, and in particular of its MR implementation, is responsible for problematic pathologies~\cite{dobaczewski07,Lacroix:2008rj,Bender:2008rn,Duguet:2008rr}.
\end{enumerate}

\begin{figure}[hptb]
\centering
\includegraphics[keepaspectratio, width = 0.75\textwidth]{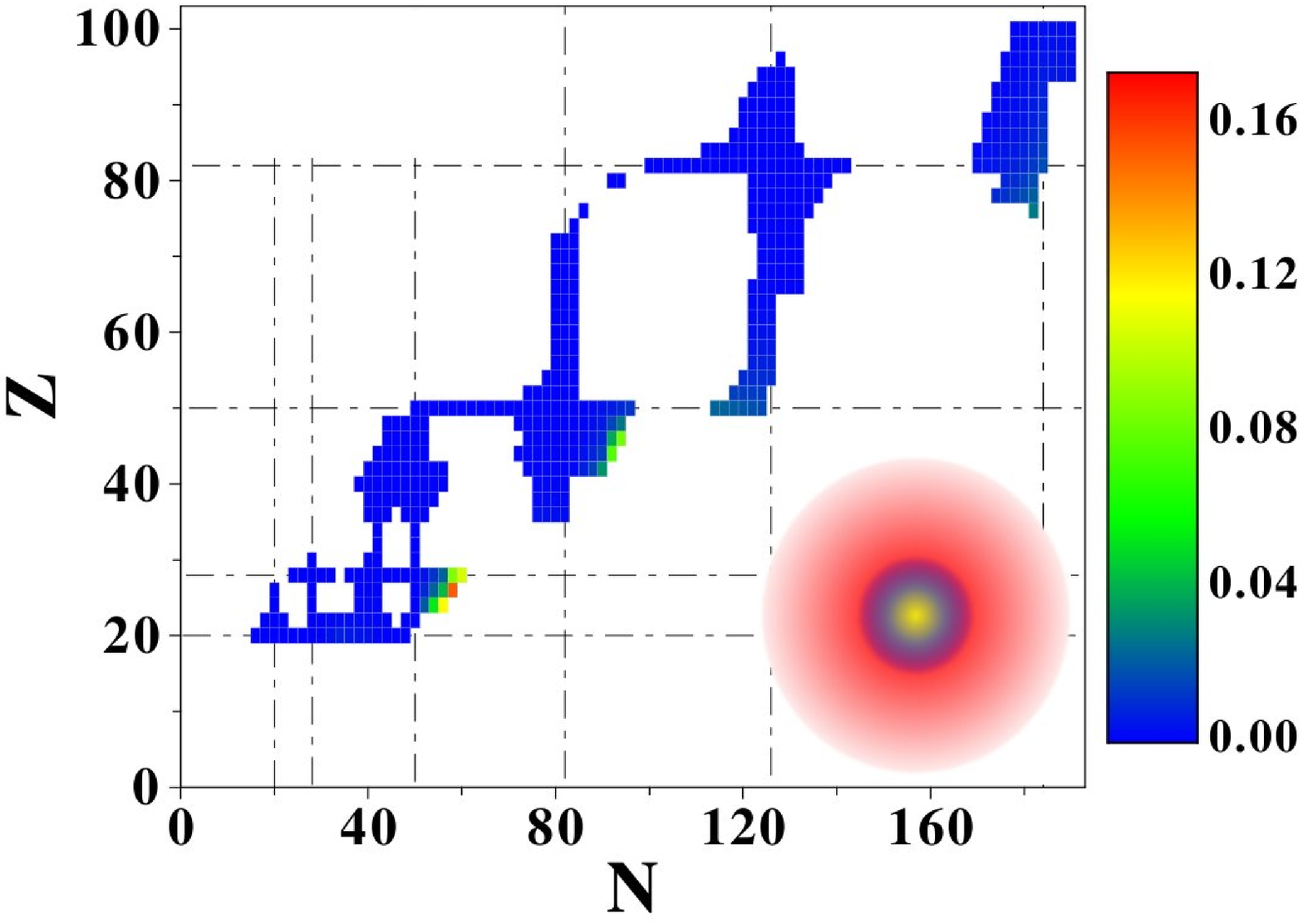}
\includegraphics[keepaspectratio, width = 0.75\textwidth]{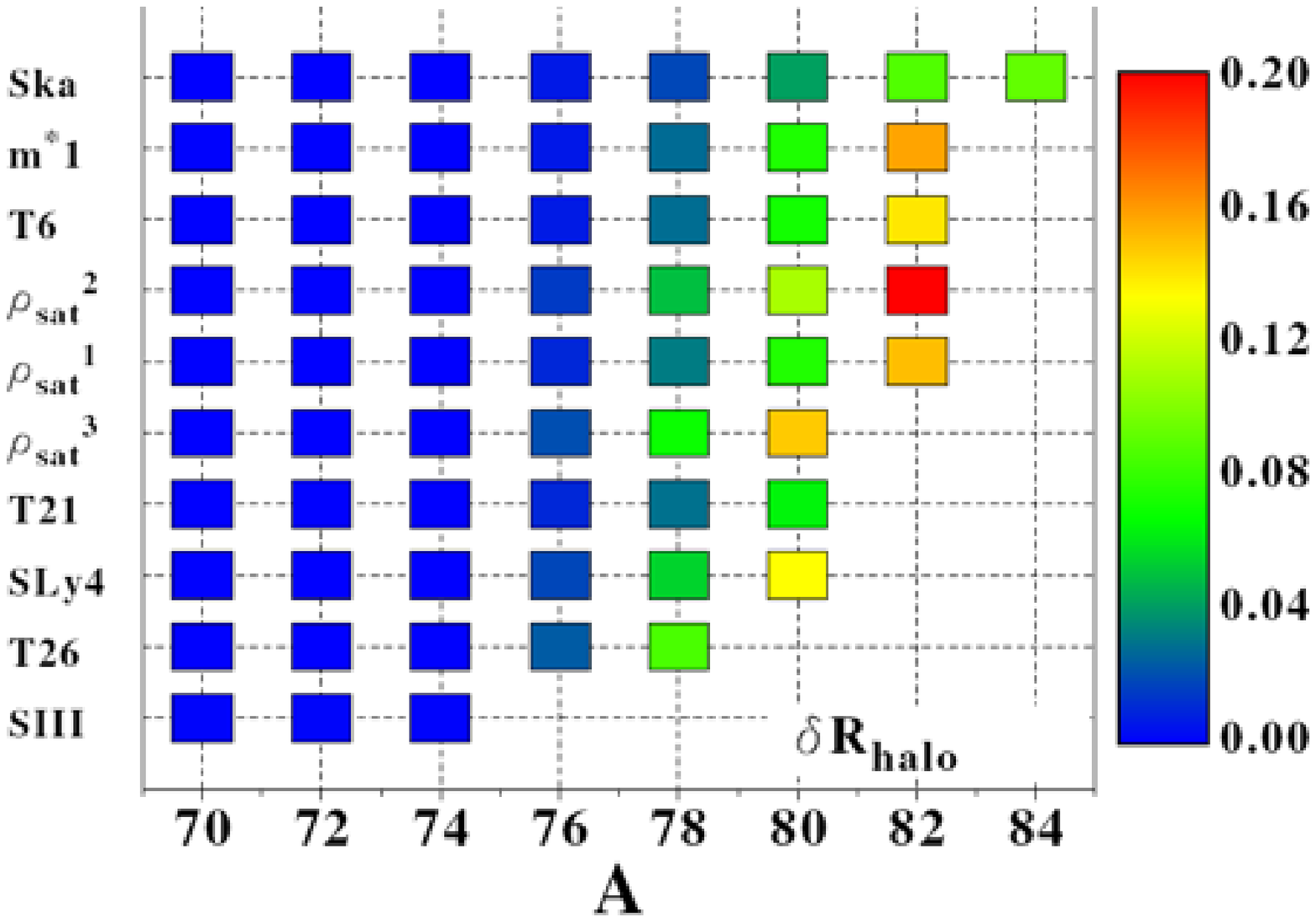}
\caption{\label{fig:halos_dvg} (Color online) Upper panel: halo parameter $\delta R_{\text{halo}}$~\cite{rotival07a} extracted for nearly five hundreds (predicted) spherical nuclei using the SLy4~\cite{chabanat98} Skyrme parametrization. Lower panel: halo parameter $\delta R_{\text{halo}}$ computed for drip-line chromium isotopes. The halo parameter $\delta R_{\text{halo}}$ quantifies in a model-independent fashion the contribution of the halo structure to the nuclear radius~\cite{rotival07a}. The colour scale refers to a length indicated in Fermi. Large discrepancies in the prediction of the drip-line position and in the extracted halo parameter are obtained from the selection of parametrizations used. Taken from Ref.~\cite{rotival07b}.}
\end{figure}

Modern parametrizations of the nuclear EDF, i.e. Skyrme, Gogny, or relativistic energy functionals, provide a good description of ground-state properties and, to a lesser extent, of spectroscopic features of known nuclei. Still, as of today, EDF  parametrizations are phenomenological as they rely on empirically-postulated functional forms whose free coupling constants are adjusted on a selected set of experimental data. This raises questions regarding (i) the connection between currently used EDF parametrizations and elementary AN forces, which is neither explicit nor qualitatively transparent, and regarding (ii) the predictive power of extrapolated EDF results into the experimentally unknown territory. Their lack of microscopic foundation often leads to parametrization-dependent predictions away from known data, i.e. to significant systematic errors, and makes difficult to design systematic improvements. Such a feature is illustrated in Fig.~\ref{fig:halos_dvg} for a particular observable of interest related to the prediction of halo structures and the location of the neutron drip-line in medium-mass nuclei~\cite{rotival07a,rotival07b}. Some systematic limitations of existing EDFs have been empirically identified~\cite{lesinski06a,Lesinski:2007zz,Kortelainen:2008rp,Bender:2009ty} over the last decade that relate to their (too) simple analytical representations and to the biases in their adjustment procedure, as well as to the lack of a solid microscopic foundation. Fuelled by interests in controlled extrapolations of nuclear properties in isospin, density, and temperature, efforts are currently being made to develop energy functionals with substantially reduced errors and improved predictive power. One possible path forward focuses on empirically improving the analytical form and the fitting procedure of existing phenomenological functionals~\cite{lesinski06a,Lesinski:2007zz,Margueron:2007uf,Niksic:2008vp,Carlsson:2008gm,Goriely:2009zz,Kortelainen:2010hv,Kortelainen:2011ft}. 

In order to improve on the limitations alluded to above and make EDF calculations truly reliable, several routes must be followed in the future. On the one hand, a better understanding of the foundations of the method and an explicit connection to elementary inter-nucleon interactions must be realized. On the other hand, empirically adjusted parametrizations must rely on advanced fitting and statistical analysis techniques.

\subsection{Goal of the present lecture notes}

The present lecture notes focus on the theoretical foundations of the nuclear energy density functional method. As such, they do not aim at reviewing the status of the field, at covering all possible ramifications of the approach or at presenting recent achievements and applications. For standard reviews that cover the connection to empirical data, we refer the reader to Refs.~\cite{bender03b,Niksic:2011sg}. In order to achieve our goal within the limits of the present document, the following choices are made in the following
\begin{enumerate}
\item the historical perspective is bypassed,
\item the presentation is limited to the non-relativistic framework,
\item time-dependent implementations of the method are not discussed,
\item the Skyrme family of parametrizations is used for illustration,
\item only the {\it full fledged} multi-reference formalism is discussed\footnote{Approximations such as the quasi-particle random phase approximation or the Schroedinger equation  based on a collective (e.g. Bohr) Hamiltonian are only mentioned in passing; see Sec.~\ref{approxMR}.},
\item applications are only shown to illustrate points of the formal discussion.
\end{enumerate}
The objective is to provide a {\it modern} account of the nuclear EDF formalism that is at variance with {\it traditional} presentations that rely, at one point or another, on a {\it Hamiltonian-based} picture. The latter is not general enough to encompass what the nuclear EDF method represents as of today. Specifically, the traditional Hamiltonian-based picture does not allow one to grasp the difficulties associated with the fact that currently available parametrizations of the energy kernel $E[g',g]$ at play in the method do not derive from a genuine Hamilton operator, would the latter be effective. As such, a key point of the presentation provided below is to demonstrate that the MR-EDF method can indeed be formulated in a {\it mathematically} meaningful fashion even if $E[g',g]$ does {\it not} derive from a genuine Hamilton operator. In particular, the restoration of symmetries can be entirely formulated without making {\it any} reference to a projected state, i.e. within a genuine EDF framework~\cite{Duguet:2010cv}. However, and as will be illustrated below, a mathematically meaningful formulation does not guarantee that the formalism is sound from a {\it physical} standpoint. We will eventually mention at which price the latter can be ensured as well.

\section{Prelude}
\label{MREDFsection}

\subsection{Reference states and Bogoliubov transformation}

The EDF method builds on the {\it effective} description of a nucleus made of an ensemble of quasi-particles moving independently in their self-created average field(s). As such, the approach relies on the use of product states of Bogoliubov type, which are nothing but a generalization of Slater determinants. To define such many-body states, let us introduce an arbitrary single-particle basis $\{| i \rangle\}$ of the one-body Hilbert space ${\cal H}_1$, where $\{i\}$ collects all spatial, spin and isospin quantum numbers necessary to define a given state. Basis states relate to particle creation operators through
\begin{equation}
a^{\dagger}_{i} | 0 \rangle = | i \rangle \, ,
\end{equation}
with $\big\{a_{i},a^{\dagger}_{j}\big\} = \delta_{ij}$. Associated single-particle wave-functions are given by
$\psi_{i} (\vec{r}\sigma \tau) \equiv \langle \vec{r}  \sigma  \tau  | i \rangle$, where $\sigma$ ($\tau$) denotes the $z$ component of the spin (isospin) 1/2 nucleon. From there, fully paired Bogoliubov vacua are defined as
\begin{equation}
\vert \Phi^{(g)} \rangle = \prod_{\mu} \beta^{(g)}_{\mu} \vert 0 \rangle \,\,\, ,
\label{Intro_met:product_state}
\end{equation}
and carry a collective label $g$ whose definition and meaning will be specified in Sec.~\ref{sec_orderparameters}. Quasi-particle creation and annihilation operators satisfy $\{\beta^{(g)}_{\mu},\beta^{(g)\dagger}_{\nu} \} =\delta_{\mu\nu}$ and relate to particle operators through the so-called Bogoliubov transformation
\begin{subequations}
\label{transfodetails}
\begin{align}
\beta^{(g)}_{\mu} &= \sum_i U^{(g)\ast}_{i \mu} a_i + V^{(g)\ast}_{i \mu} a^\dagger_i \,\,\,, \\
\beta^{(g)\dagger}_{\mu} &= \sum_i  V^{(g)}_{i \mu } a_i  +  U^{(g)}_{i \mu } a^\dagger_i \,\,\,.
\end{align}
\end{subequations}
Matrices $U^{(g)}$ and $V^{(g)}$, respectively made out of vectors $\mathbf{U}^{(g)}_{\mu}$ and $\mathbf{V}^{(g)}_{\mu}$ defined on ${\cal H}_1$, combine to make up the matrix representation of the Bogoliubov transformation~\cite{ring80a}
\begin{equation}
{\cal W}^{(g)}    \equiv     \fourmat{U}{V^{\ast}}
          {V}{U^{\ast}}^{(g)} \,  \label{Bogo}
\end{equation}
whose unitarity provides four identities
\begin{subequations}
\label{Bogounitarity}
\begin{eqnarray}
U^{(g)}  \, U^{(g)\dagger} + V^{(g)\ast} \, V^{(g)T}  & = & 1 \, , \label{Bogounitaritya} \\
U^{(g)\ast} \, V^{(g)T}  + V^{(g)} \, U^{(g)\dagger} & = & 0 \, , \label{Bogounitarityb}  \\
U^{(g)\dagger} \, U^{(g)}    + V^{(g)\dagger} \, V^{(g)}  & = & 1 \, ,\label{Bogounitarityc} \\
U^{(g)T} \, V^{(g)}   + V^{(g)T} \, U^{(g)}   & = & 0 \, . \label{Bogounitarityd}
\end{eqnarray}
\end{subequations}
Fully paired Bogoliubov states $\vert \Phi^{(g)} \rangle$ are denoted as "vacua" in the sense that they are annihilated by the set of quasi-particle annihilation operators, i.e.
\begin{eqnarray}
\beta^{(g)}_\mu \, | \Phi^{(g)} \rangle &=& 0 \,\, \,\, \forall \,\, \mu .
\end{eqnarray}
Such a notion generalizes the physical vacuum $| 0 \rangle$, which is annihilated by the set of particle annihilation operators $\{a_{i}\}$, and Slater determinants that are annihilated by the set of operators $\{a_{p}, a^{\dagger}_{h}\}$, where $p$ ($h$) denote unoccupied (occupied) single-particle states. Furthermore, Bogoliubov states $\vert \Phi^{(g)} \rangle$  break particle-number symmetry, i.e. as opposed to Slater determinants they are not eigenstates of the particle (neutron or proton) number operator $N$. Still, states defined through Eq.~\ref{Intro_met:product_state} carry an even number-parity quantum number, i.e. they are linear combinations of eigenstates of $N$ corresponding to even number of particles only. As such, they are appropriate to the description of even-even nuclei. In a more general setting, one may consider Bogoliubov states obtained by performing an even number of quasi-particle excitations on top of a fully paired vacuum or by performing an odd number of such excitations to access odd number-parity states appropriate to the description of odd nuclei~\cite{duguet02a,Bally:2011iz}. In such a situation, reference states carry an additional label, besides $g$, to denote the set of quasi-particle excitations that characterizes them. 

\subsection{Elements of group theory}

The nuclear EDF method relies heavily on breaking and restoring symmetries of the {\it underlying}, i.e. realistic, nuclear Hamiltonian. As of today, state-of-the-art calculations typically take advantage of breaking translational, rotational and particle-number symmetries, while only restoring the last two. There also exists few calculations treating (solely) the restoration of linear momentum~\cite{rodriguezguzman04a}. In order to tackle such a key aspect of the method, let us introduce basic elements of group theory.

We consider the symmetry group ${\cal G}$ of the nuclear Hamiltonian $H$. Because it is the case for the most relevant symmetries, we consider ${\cal G}$ to be a continuous, possibly non-abelian, compact Lie group ${\cal G}=\{R(\alpha)\}$ parametrized by a set of $r$ real parameters $\alpha \equiv\{\alpha_i \in D_i \, ; \,i=1,\ldots, r\}$ defined over a domain of definition $D_{{\cal G}}\equiv\{D_i \, ; \,i=1,\ldots, r\}$. We thus have $[R(\alpha),H]=0$ for any $R(\alpha) \in {\cal G}$. The invariant measure on ${\cal G}$ is defined as $dm(\alpha)$ and  its volume is given by
\begin{eqnarray}
v_{{\cal G}} &\equiv& \int_{D_{{\cal G}}} dm(\alpha) \, .
\end{eqnarray}
Next, we introduce the set of infinitesimal generators $\vec{C}=\{C_i ; i=1,\ldots, r\}$ that make up the Lie algebra and in terms of which any transformation $R(\alpha)$ of the group can be expressed via an exponential map $R_{\vec{C}}(\alpha)$. 

We further consider irreducible representations (Irreps) $S^{\lambda}_{ab}(\alpha)$ of the group labelled by eigenvalues $\lambda$ of the Casimir operator $\Lambda$. Irreducible representations of dimension $d_{\lambda}$ are spanned by states that are also eigenstates of one of the generators, e.g. $C_1$. Indices $"a"$ and $"b"$ in $S^{\lambda}_{ab}(\alpha)$ refer to the $d_{\lambda}$ corresponding eigenvalues. The unitarity of the Irreps, together with the combination law of two successive transformations, can be read off
\begin{eqnarray}
\sum_{c} S^{\lambda \, \ast}_{ca}(\alpha') \, S^{\lambda}_{cb}(\alpha)  &=& \sum_{c} S^{\lambda}_{ac}(-\alpha') \, S^{\lambda}_{cb}(\alpha) = S^{\lambda}_{ab}(\alpha\!-\!\alpha') \,\,\, ,
\label{Symmetries:unitarity}
\end{eqnarray}
where arguments $-\alpha$ and $\alpha-\alpha'$ symbolically denote parameters of transformations $R^{-1}(\alpha)$ and $R^{-1}(\alpha')R(\alpha)$, respectively. Additionally, the orthogonality of the Irreps reads
\begin{equation}
\int_{{\cal G}} dm(\alpha) \, S^{\lambda \, \ast}_{ab}(\alpha)  \, S^{\lambda'}_{a'b'}(\alpha) = \frac{v_{{\cal G}}}{d_\lambda} \, \delta_{\lambda\lambda'} \, \delta_{aa'} \, \delta_{bb'} \,  \,\,\, . \label{orthogonality}
\end{equation}
Any function $f(\alpha)$ defined on $D_{{\cal G}}$ can be decomposed over the Irreps of the group according to
\begin{equation}
f(\alpha) \equiv \sum_{\lambda ab} \, f^{\lambda}_{ab} \, \, S^{\lambda}_{ab}(\alpha) \, , \label{decomposition_general}
\end{equation}
which defines the set of expansion coefficients $\{f^{\lambda}_{ab}\}$.
\begin{table}[htbp]
\begin{tabular}{|c|ccccccccc|}
\hline
${\cal G}$ &  $\alpha$ &  $dm(\alpha)$ & $v_{{\cal G}}$ & $\vec{C}$ & $\Lambda$ & $C_1$ & $R_{\vec{C}}(\alpha)$ & $S^{\lambda}_{ab}(\alpha)$ & $d_{\lambda}$ \\
\hline
  &    &   &  &  &  &  &  &   &  \\
$U(1)$ &  $\varphi$  & $d\varphi$  & $2\pi$ & $N$ & $N^2$ & - & $e^{iN\varphi}$& $e^{im\varphi}$ & $1$ \\
  &    &   &  &  &  &  &  &   &  \\
$SO(3)$ &  $\alpha,\beta,\gamma$  &  $\sin \beta d\alpha d\beta d\gamma$ & $16\pi^2$ & $\vec{J}$ & $J^{2}$ & $J_z$ & $e^{-i\alpha J_{z}} \, e^{-i\beta
J_{y}} \, e^{-i\gamma J_{z}}$ &  ${\cal D}^{J}_{MK}(\Omega)$ & $2J\!+\!1$ \\
  &    &   &  &  &  &  &  &   &  \\
\hline
\end{tabular}
\caption{Characteristics of $SO(3)$ and $U(1)$ relevant to the present study. The gauge angle parametrizing $U(1)$ is $\varphi \in [0,2\pi]$ whereas Euler angles parameterizing $SO(3)$ are $\Omega \equiv (\alpha,\beta,\gamma) \in [0,4\pi] \times [0,\pi] \times [0,2\pi]$. One-dimensional Irreps of $U(1)$ are labeled by $m \in \mathbb{Z}$ whereas $(2J\!+\!1)$-dimensional Irreps of $SO(3)$ are labeled by $2J \in \mathbb{N}$ and are given by the so-called Wigner functions ${\cal D}^{J}_{MK}(\Omega)$~\cite{varshalovich88a}, where $(2M,2K)\in
\mathbb{Z}^2$ with $-2J\leq 2M,2K \leq+2J$.}
\label{Symmetries:detailsgroup} 
\end{table}

Later on, we wish to apply above considerations to two groups of particular interest, i.e. the abelian group $U(1)$ associated with particle-number symmetry and the non-abelian group $SO(3)$ associated with rotational symmetry. The relevant elements and equations at play for each of these two cases can be deduced from above using correspondence Tab.~\ref{Symmetries:detailsgroup}. In the case of $U(1)$, decomposition~(\ref{decomposition_general}) of a function $f(\varphi)$ defined on $D_{U(1)}=[0,2\pi]$, i.e. its Fourier expansion, reads
\begin{equation}
f(\varphi) \equiv \sum_{m} \, f^{m} \, \, e^{im\varphi } \, . \label{decomposition_U1}
\end{equation}
Similarly, the decomposition of a function $f(\Omega)$ defined on $D_{SO(3)}=[0,4\pi] \times [0,\pi] \times [0,2\pi]$ over Irreps of $SO(3)$ reads
\begin{equation}
f(\Omega) \equiv \sum_{JMK} \, f^{J}_{MK} \, \, {\cal D}^{J}_{MK}(\Omega) \label{decomposition_SO3} \, ,
\end{equation}
where ${\cal D}^{J}_{MK}(\Omega)$ denotes the so-called Wigner function~\cite{varshalovich88a}.

\subsection{Collective variable and symmetry breaking}
\label{sectionsymbreaking}

\subsubsection{Order parameters}
\label{sec_orderparameters}

Whenever $| \Phi^{(g)} \rangle$ breaks a symmetry of the nuclear Hamiltonian, it does not carry the associated symmetry quantum number(s). The three main symmetries considered here lead to loosing good total linear momentum $\vec{P}$, total angular momentum $(J^2, J_z)$ and neutron/proton $N/Z$ quantum numbers. Doing so, $| \Phi^{(g)} \rangle$ acquires non-zero order parameters, i.e. one per broken symmetry, which we group under the generic notation $g\equiv |g|\,e^{i \alpha}\equiv \langle \Phi^{(g)} | G | \Phi^{(g)} \rangle$, where $G$ is an appropriate operator whose average value in a symmetry conserving state is zero. The norm $|g|$ of the order parameter tracks the extent to which $| \Phi^{(g)} \rangle$ breaks the symmetry, i.e. its "deformation", whereas the phase $\alpha = {\rm Arg}(g)$ characterizes the orientation of the deformed body with respect to the chosen reference frame\footnote{For certain symmetries, e.g. SO(3), the phase $\alpha$ collects in fact several angles. See Tab.~\ref{Symmetries:detailsgroup} for two relevant examples.}. 
\begin{table}[htbp]
\begin{tabular}{|c|ccc|}
\hline
${\cal G}$ &  $|g|$ & \hspace{0.2cm} &  $\alpha = {\rm Arg}(g)$ \\
\hline
&&& \\
$U(1)$ &  $||\kappa||$  &&  $\varphi$  \\
&&& \\
$SO(3)$ & $\,\, \rho_{\lambda\mu} \,\, (\lambda > 2J)$    &&  $\alpha,\beta,\gamma$  \\
&&& \\
\hline
\end{tabular}
\caption{Norm and phase of the order parameters associated with broken $U(1)$ and $SO(3)$ symmetries.}
\label{Symmetries:orderparameters} 
\end{table}
In the present study, order parameters associated with the breaking of translational, rotational and particle-number symmetries should be specified. As only the latter two are effectively restored in state-of-the-art calculations, Tab.~\ref{Symmetries:orderparameters} provides the order parameters used to track the breaking of $U(1)$ and $SO(3)$ symmetries. As $|g|$ must be zero/non-zero for good/broken symmetry states, the anomalous density\footnote{Although it can be done rigorously, we do not state explicitly here the definition of the norm of $\kappa$.} $\kappa^{gg}$ (see Eq.~\ref{Intro_met:OBDM}) is a good candidate for $U(1)$. For $SO(3)$, one uses  multipole moments $\rho_{\lambda\mu}$ of the matter density distribution $\rho^{gg}_{0}(\vec{r})$ (see Eq.~\ref{eq:Skyrme_int:locdensities}) with $\lambda > 2J$~\cite{sadoudi11thesis}. As for $U(1)$ the phase $\alpha = {\rm Arg}(g)$ provides the orientation $\varphi$ of $\mathbf{\kappa}^{gg}$ in gauge space, while for $SO(3)$ it gives the orientation $\Omega \equiv (\alpha,\beta,\gamma)$ of the deformed density distribution in real space. 

\subsubsection{Symmetry-breaking reference state}

Eventually, states $| \Phi^{(g)} \rangle$ that are typically dealt with in state-of-the-art calculations can be written in full glory as
\begin{eqnarray}
| \Phi^{(\rho_{\lambda\mu} \Omega  ; ||\kappa_p|| \varphi_p ; ||\kappa_n|| \varphi_n)} \rangle \!\!
& \equiv &  \!\!  R_{\vec{J}}(\Omega)   R_N(\varphi_{n})  R_Z(\varphi_{p}) 
      | \Phi^{(\rho_{\lambda\mu}  0  ; ||\kappa_p|| 0 ; ||\kappa_n|| 0)} \rangle \, , \label{refstates}
\end{eqnarray}
where the breaking of $U(1)$ appears once for protons ($\varphi_{p}$) and once for neutrons ($\varphi_{n}$). Equation~\ref{refstates} indicates that the state corresponding to a finite value of the phase $\alpha$, i.e. to a given orientation of the "deformed" body, can be obtained from the one at $\alpha=0$ through the application of the rotation operator
\begin{eqnarray}
| \Phi^{(g)} \rangle  & \equiv &  R(\alpha) | \Phi^{(|g|0)} \rangle  \, .
\end{eqnarray}

\section{Energy and norm kernels}
\label{constraints}

The basic inputs to the nuclear EDF method take the form of the so-called off-diagonal energy and norm kernels
\begin{subequations}
\label{Intro_met:EDF_kernel}
\begin{eqnarray}
E[g',g] &\equiv& E[\langle \Phi^{(g')} |  , | \Phi^{(g)} \rangle]  \,\,\,, \label{Intro_met:EDF_kernela} \\
N[g',g] &\equiv& \langle \Phi^{(g')} |  \Phi^{(g)} \rangle  \,\,\,, \label{Intro_met:EDF_kernelb}
\end{eqnarray}
\end{subequations}
that define quantities associated with two product states $|  \Phi^{(g)} \rangle$ and $|  \Phi^{(g')} \rangle$ possibly carrying different values of the order parameters.

\subsection{Norm kernel}
\label{norm_kernel}

The definition of the norm kernel in Eq.~\ref{Intro_met:EDF_kernelb} is fully explicit and does not pose any problem. However, the actual computation of both its phase and its norm has posed a great challenge to nuclear theorists over the years. It is only recently that a method to compute $N[g',g]$ unambiguously in terms of Pfaffian was proposed~\cite{Robledo:2009yd}. This constitutes a rather involved technical discussion that goes beyond the scope of the present lecture notes. We  refer the interested readers to Refs.~\cite{Robledo:2009yd,Robledo:2011ce,Avez:2011wr,Oi:2011qp,Gao:2013vaa}.

\subsection{Energy kernel}
\label{energy_kernel}

The energy kernel $E[g',g]$ is postulated under the form of a general, possibly complicated, functional of $| \Phi^{(g')} \rangle$ and $| \Phi^{(g)} \rangle$. Such a feature lies at the heart of the EDF approach as a way to effectively sum up the bulk of many-body correlations. Having no a priori knowledge of the most appropriate functional, one must at least constrain it to fulfil a minimal set~\cite{Robledo07a,robledo10a} of basic properties.  

The first requirement states that transforming both $| \Phi^{(g')} \rangle$ and $| \Phi^{(g)} \rangle$ via any element $R(\alpha'') \in {\cal G}$ must leave the kernel invariant, i.e.  
\begin{eqnarray}
E[\langle \Phi^{(g')} |R^{\dagger}(\alpha'')  , R(\alpha'')| \Phi^{(g)} \rangle] &=& E[\langle \Phi^{(g')} |  , | \Phi^{(g)} \rangle] \, , \label{invariancekernel}
\end{eqnarray}
which is equivalent to demanding that the kernel only depends on the {\it difference} of phases of the order parameters labelling the two states, i.e.
\begin{eqnarray}
E[|g'|\, \alpha',|g|\, \alpha] &=& E[|g'| 0,|g|\, \alpha\!-\!\alpha'] \, . \label{invariancekernelsuite}
\end{eqnarray}
Such a property is necessary and sufficient to ensure later on that the energy is real and independent of the reference frame.

Other requirements relate to the behaviour of the kernel in the limit where $| \Phi^{(g')} \rangle$ and $| \Phi^{(g)} \rangle$ are "close" to each other. In case diagonal and off-diagonal kernels were to be defined through separate means, one must first ensure that they are consistent, i.e. one must ensure that the former is obtained from the latter when taking $| \Phi^{(g')} \rangle = | \Phi^{(g)} \rangle$. Probing the kernel in the vicinity of the diagonal, one further requires that (i) the chemical potentials $\lambda_N$ and $\lambda_Z$ obtained through SR calculations are consistent with their extraction from the Kamlah expansion~\cite{kamlah68a} of the particle number restored MR energy and that (ii) the quasi-particle random-phase approximation is recovered from the most general MR scheme whenever $| \Phi^{(g')} \rangle$ and $| \Phi^{(g)} \rangle$ differ harmonically from a common reference state~\cite{jancovici64,brink68}. The latter two requirements are fulfilled~\cite{Robledo07a,robledo10a} if, and only if, $E[\langle \Phi^{(g')} |  , | \Phi^{(g)} \rangle]$
does indeed only depend on the bra $\langle \Phi^{(g')} |$ and on the ket $| \Phi^{(g)} \rangle$, as was so far implied by the notation used. 

It happens that a sufficient condition for all above properties to be fulfilled is to postulate that the off-diagonal energy kernel is a functional
\begin{equation}
E[g',g]\equiv E[\mathbf{\rho}^{g'\!g}, \mathbf{\kappa}^{g'\!g}, \mathbf{\kappa}^{gg'  \ast}] \,\,\, , \label{Intro_met:EDF_kernelc}
\end{equation}
in the mathematical sense, of normal and anomalous one-body transition (i.e. off-diagonal) density matrices computed from  $\langle \Phi^{(g')} |$ and $| \Phi^{(g)} \rangle$, respectively defined through
\begin{subequations}
\label{Intro_met:OBDM}
\begin{align}
\rho_{ij}^{g'\! g} &\equiv \frac{\langle \Phi^{(g')} | a^{\dagger}_{j} a_{i}| \Phi^{(g)} \rangle}
                {\langle \Phi^{(g')}  | \Phi^{(g)}  \rangle}   \,\,\,, \label{Intro_met:OBDM1} \\
\kappa_{ij}^{g'\! g} &\equiv \frac{\langle \Phi^{(g')} | a_{j} a_{i}| \Phi^{(g)} \rangle}
                {\langle \Phi^{(g')}  | \Phi^{(g)}  \rangle}     \,\,\,, \label{Intro_met:OBDM2} \\
\kappa^{gg' \ast}_{ij} &\equiv \frac{\langle \Phi^{(g')} | a^{\dagger}_{i} a^{\dagger}_{j}| \Phi^{(g)} \rangle}
                {\langle \Phi^{(g')}  | \Phi^{(g)}  \rangle}     \,\,\,. \label{Intro_met:OBDM3}
\end{align}
\end{subequations}
One observes that $\rho_{ij}^{g'\! g\ast} = \rho_{ji}^{g g'}$, $\kappa_{ij}^{g'\! g}=-\kappa_{ji}^{g'\! g}$ and $\kappa^{gg' \ast}_{ij}=-\kappa^{gg' \ast}_{ji}$, i.e. the two anomalous densities are antisymmetric whereas the normal density matrix is hermitian whenever $g=g'$.

\subsection{Pseudo-potential-based energy kernel}
\label{pseudopotential}

A particular implementation of the EDF method consists of deriving the EDF kernel from a pseudo Hamiltonian
\begin{eqnarray}
H_{\text{pseudo}} &\equiv& \,\, \sum_{ij} t^{1N\,\text{pseudo}}_{ij} a^\dagger_i a_j^{\,} \nonumber \\
&&\!\!\!\!+
\left(\frac{1}{2!}\right)^{\! 2} \sum_{ijkl} \bar{v}^{2N\,\text{pseudo}}_{ijkl} a^\dagger_i a^\dagger_j a_l^{\,} a_k^{\,} \nonumber \\
&&\!\!\!\!+
\left(\frac{1}{3!}\right)^{\! 2} \sum_{ijklmn} \bar{v}^{3N\,\text{pseudo}}_{ijklmn} a^\dagger_i a^\dagger_j a^\dagger_k a_n^{\,} a_m^{\,} a_l^{\,} 
+
\, \cdots
\,\,\,, \label{Intro_met:hamiltonian}
\end{eqnarray}
where $t^{1N\,\text{pseudo}}$ embodies an effective one-body kinetic energy operator while $\bar{v}^{AN\,\text{pseudo}}_{ijkl}$ denotes antisymmetrized matrix-elements of a A-body pseudo-potential, i.e. of a A-body effective interaction. The word "pseudo" refers to the fact that operators entering Eq.~\ref{Intro_met:hamiltonian} are not the same as the elementary operators entering ab-initio theories; e.g. $\bar{v}^{AN\,\text{pseudo}}$ should not be confused with realistic AN interactions. Eventually, $H_{\text{pseudo}}$ is only to be seen as a mere intermediary used to generate the fundamental ingredient of the theory, i.e. the off-diagonal energy kernel. In such a context, the latter is computed through
\begin{subequations}
\label{Intro_met:HFB_kernel}
\begin{eqnarray}
 E_{\text{pseudo}}[g',g] &\equiv& \frac{\langle \Phi^{(g')} | H_{\text{pseudo}} | \Phi^{(g)} \rangle}{\langle \Phi^{(g')} | \Phi^{(g)} \rangle} \label{defpseudoEDF} \\
 &=&  \,\, \sum_{ij} t_{ij} \, \rho^{g'\!g}_{ij}  \label{energyspelledout} \\
&& \!\!\!\! +
\frac{1}{2} \sum_{ijkl} \bar{v}^{2N\,\text{pseudo}}_{ijkl}  \, \rho^{g'\!g}_{ki} \, \rho^{g'\!g}_{lj}
+
\frac{1}{6} \sum_{ijklmn} \bar{v}^{3N\,\text{pseudo}}_{ijklmn}   \, \rho^{g'\!g}_{li} \, \rho^{g'\!g}_{mj} \, \rho^{g'\!g}_{nk}
+
\, \cdots \nonumber  \\
&&\!\!\!\! + \frac{1}{4} \sum_{ijkl} \bar{v}^{2N\,\text{pseudo}}_{ijkl}  \, \kappa^{gg' \! \ast}_{ij} \, \kappa^{g'\!g}_{kl}
+
\frac{1}{4} \sum_{ijklmn} \bar{v}^{3N\,\text{pseudo}}_{ijklmn}  \, \kappa^{gg' \! \ast}_{ij} \, \kappa^{g'\!g}_{lm}   \, \rho^{g'\!g}_{nk} 
+
\, \cdots \nonumber \\
&\equiv& E_{\text{pseudo}}[\mathbf{\rho}^{g'\!g}, \mathbf{\kappa}^{g'\!g}, \mathbf{\kappa}^{gg'  \, \ast}] \,\,\,, 
\end{eqnarray}
\end{subequations}
and is indeed a functional of one-body transition density matrices in virtue of the generalized (i.e. off-diagonal) Wick theorem~\cite{balian69a}. As long as $H_{\text{pseudo}}$ possesses the same symmetries as the underlying nuclear Hamiltonian, Eq.~\ref{invariancekernel} is automatically fulfilled for any $R(\alpha'') \in {\cal G}$.

\subsection{Skyrme parametrization}
\label{skyrme}

We now introduce a particular family of EDF parametrizations in view of illustrating some of the points alluded to in the previous section. The Skyrme parametrization\footnote{Coulomb and center-of-mass correction contributions are omitted here for simplicity.} is a local energy functional, i.e. it is expressed as a single integral in coordinate space
of a local energy density involving a set of local densities derived from the density matrices introduced in Eq.~\ref{Intro_met:OBDM}.

\subsubsection{Local densities}

Introducing the creation $a^{\dagger} (\vec{r} \sigma \tau)$ and annihilation $a(\vec{r} \sigma \tau)$ operators in the coordinate representation
\begin{subequations}
\label{eq:Skyrme_int:crtoci}
\begin{align}
a(\vec{r} \sigma \tau) \equiv & \,  \sum_i \varphi^{\,}_i (\vec{r} \sigma \tau) \; a^{\,}_{i}  \,\,\,, \\
a^{\dagger}(\vec{r} \sigma \tau) \equiv & \, \sum_i \varphi_i ^\ast(\vec{r} \sigma \tau) \; a^{\dagger}_{i}  \,\,\, ,
\end{align}
\end{subequations}
one obtains the transition density matrices in that representation
\begin{subequations}
\label{eq:Skyrme_int:intro:nonlocdensity}
\begin{eqnarray}
\rho^{g'\! g} (\vec{r} \sigma \tau,\vec{r}\,' \sigma ' \tau') \, &\equiv& \, \frac
{\langle \Phi^{(g')} | a^{\dagger}(\vec{r}\,' \sigma ' \tau') a(\vec{r} \sigma \tau) | \Phi^{(g)} \rangle}
{\langle \Phi^{(g')}  | \Phi^{(g)}  \rangle}  = \sum_{ij} \varphi^{\dagger}_{j} (\vec{r}\,' \sigma ' \tau') \, \varphi^{\,}_{i} (\vec{r} \sigma \tau)  \, \rho^{g'\! g}_{ij} \,\,\, , \nonumber
\\
\kappa^{g'\! g} (\vec{r} \sigma \tau,\vec{r}\,' \sigma ' \tau') \, &\equiv& \, \frac
{\langle \Phi^{(g')} | a(\vec{r}\,' \sigma ' \tau') a(\vec{r} \sigma \tau) | \Phi^{(g)} \rangle}
{\langle \Phi^{(g')}  | \Phi^{(g)}  \rangle}  = \sum_{ij} \varphi_{j} (\vec{r}\,' \sigma ' \tau') \, \varphi^{\,}_{i} (\vec{r} \sigma \tau) \,  \kappa^{g'\! g}_{ij} \,\,\, . \nonumber
\end{eqnarray}
\end{subequations}
Further considering spin Pauli matrices\footnote{Proton/neutron mixing is presently ignored such that $\rho^{g'\! g} (\vec{r} \sigma \tau,\vec{r}\,' \sigma ' \tau')=\kappa^{g'\! g} (\vec{r} \sigma \tau,\vec{r}\,' \sigma ' \tau')=0$ for $\tau \neq \tau'$.  This does not correspond to the most general situation~\cite{perlinska04a}.}
\begin{equation}
\label{eq:Skyrme_int:pauli_mat}
\sigma_x \equiv
\begin{pmatrix}
    0  \; \; \; \; 1 \\
    1  \; \; \; \; 0
\end{pmatrix}
\; , \;
\sigma_y \equiv
\begin{pmatrix}
    0  \,  -{\mathrm i} \\
    {\mathrm i}  \; \; \; \; 0
\end{pmatrix}
\; , \;
\sigma_z \equiv
\begin{pmatrix}
    1  \; \; \; \; 0 \\
    0  \,   -1
\end{pmatrix}
\,\,\, ,
\end{equation}
a set of non-local densities containing up to two gradients can be defined
\begin{subequations}
\label{eq:Skyrme_int:nonlocdensities}
\begin{align} 
\rho^{g'\! g}_\tau (\vec{r}  ,\vec{r}\,' ) \, 
\equiv&\,
 \, \sum_{\sigma} \rho^{g'\! g} (\vec{r} \sigma \tau ,\vec{r}\,' \sigma \tau)
  \,\,\, , \label{eq:Skyrme_int:nonlocdensities:rho}\\
 s^{g'\! g}_{\tau, \nu} (\vec{r}  ,\vec{r}\,' ) \, 
\equiv&\,  \sum_{\sigma' \sigma} \rho^{g'\! g} (\vec{r} \sigma \tau,\vec{r}\,' \sigma ' \tau) \langle \sigma ' \vert \sigma_{\nu} \vert \sigma \rangle
 \,\,\, , \label{eq:Skyrme_int:nonlocdensities:s}\\
 \tilde{\rho}^{g'\! g}_\tau (\vec{r}  ,\vec{r}\,' ) \, 
\equiv&\,
 \, \sum_{\sigma} 2\bar{\sigma} \kappa^{g'\! g} (\vec{r} \sigma \tau,\vec{r}\,' \bar{\sigma} \tau)
  \,\,\, , \label{eq:Skyrme_int:nonlocdensities:trho}\\
 \tilde{s}^{g'\! g}_{\tau, \nu} (\vec{r}  ,\vec{r}\,' ) \, 
\equiv&\,  \sum_{\sigma' \sigma}  2 \bar{\sigma}' \kappa^{g'\! g} (\vec{r} \sigma \tau,\vec{r}\,' \bar{\sigma}' \tau) \langle \sigma ' \vert \sigma_{\nu} \vert \sigma \rangle
 \,\,\, , \label{eq:Skyrme_int:nonlocdensities:ts}\\
\tau^{g'\! g}_\tau (\vec{r}  ,\vec{r}\,' ) \, 
\equiv&\,
 \, \sum_\mu \nabla_{\vec{r}, \mu} \, \nabla_{\vec{r}\,', \mu} \, \rho^{g'\! g}_\tau (\vec{r}  ,\vec{r}\,' ) \, 
 \,\,\, , \\
T^{g'\! g}_{\tau, \nu} (\vec{r}  ,\vec{r}\,' ) \, 
\equiv&\,
 \, \sum_\mu \nabla_{\vec{r}, \mu} \, \nabla_{\vec{r}\,', \mu} \, s^{g'\! g}_{\tau, \nu} (\vec{r}  ,\vec{r}\,' ) \, 
 \,\,\, , \\
 \tilde{\tau}^{g'\! g}_\tau (\vec{r}  ,\vec{r}\,' ) \, 
\equiv&\,
 \, \sum_\mu \nabla_{\vec{r}, \mu} \, \nabla_{\vec{r}\,', \mu} \, \tilde{\rho}^{g'\! g}_\tau (\vec{r}  ,\vec{r}\,' ) \, 
 \,\,\, , \\
 \tilde{T}^{g'\! g}_{\tau, \nu} (\vec{r}  ,\vec{r}\,' ) \, 
\equiv&\,
 \, \sum_\mu \nabla_{\vec{r}, \mu} \, \nabla_{\vec{r}\,', \mu} \, \tilde{s}^{g'\! g}_{\tau, \nu} (\vec{r}  ,\vec{r}\,' ) \, 
 \,\,\, , \\
j^{g'\! g}_{\tau,\mu} (\vec{r}  ,\vec{r}\,' ) \, 
\equiv&\,
 \, - \frac{{\mathrm i}}{2} \left( \nabla_{\vec{r}, \mu} \, -  \, \nabla_{\vec{r}\,', \mu} \right)  \, \rho^{g'\! g}_\tau (\vec{r}  ,\vec{r}\,' ) \, 
 \,\,\, , \\
J^{g'\! g}_{\tau, \mu \nu} (\vec{r}  ,\vec{r}\,' ) \, 
\equiv&\,
 \,  - \frac{{\mathrm i}}{2} \left( \nabla_{\vec{r}, \mu} \, -  \, \nabla_{\vec{r}\,', \mu} \right) \, s^{g'\! g}_{\tau, \nu} (\vec{r}  ,\vec{r}\,' ) \, 
 \,\,\, , \\
 \tilde{j}^{g'\! g}_{\tau,\mu} (\vec{r}  ,\vec{r}\,' ) \, 
\equiv&\,
 \, - \frac{{\mathrm i}}{2} \left( \nabla_{\vec{r}, \mu} \, -  \, \nabla_{\vec{r}\,', \mu} \right)  \, \tilde{\rho}^{g'\! g}_\tau (\vec{r}  ,\vec{r}\,' ) \, 
 \,\,\, , \\
 \tilde{J}^{g'\! g}_{\tau, \mu \nu} (\vec{r}  ,\vec{r}\,' ) \, 
\equiv&\,
 \,  - \frac{{\mathrm i}}{2} \left( \nabla_{\vec{r}, \mu} \, -  \, \nabla_{\vec{r}\,', \mu} \right) \, \tilde{s}^{g'\! g}_{\tau, \nu} (\vec{r}  ,\vec{r}\,' ) \, 
 \,\,\, ,
\end{align}
\end{subequations}
where $\vec{\nabla}^{\,}_{\vec{r}}$ denotes the gradient acting on coordinate $\vec{r}$ while $\bar{\sigma} \equiv - \sigma$. Greek indexes refer to cartesian components of a vector ($\mu$) or a tensor ($\mu,\nu$). Densities without Greek index such as $\rho^{g'\! g}_{\tau}$, $\tilde{\rho}^{g'\! g}_{\tau}$ are scalar densities. Equation~\ref{eq:Skyrme_int:nonlocdensities} provides non-local matter, spin, pair, pair-spin, kinetic, spin-kinetic, pair-kinetic, pair-spin-kinetic, current, spin-current, pair-current and pair-spin-current densities for a given isospin projection, respectively.

Eventually, corresponding local densities are trivially obtained through
\begin{subequations}
\label{eq:Skyrme_int:locdensities}
\begin{alignat}{4}
 \rho^{g'\! g}_\tau (\vec{r}) 
 \equiv & \; \rho^{g'\! g}_\tau (\vec{r}  ,\vec{r} ) 
   \,\,\, &, \;\;
 s^{g'\! g}_{\tau,\mu} (\vec{r}) 
 \equiv & \; s^{g'\! g}_{\tau, \mu} (\vec{r}  ,\vec{r} ) 
    \,\,\, ,   \\
 \tilde{\rho}^{g'\! g}_\tau (\vec{r}) 
 \equiv & \; \tilde{\rho}^{g'\! g}_\tau (\vec{r}  ,\vec{r} ) 
   \,\,\, &,  \;\;
 \tilde{s}^{g'\! g}_{\tau,\mu} (\vec{r}) 
 \equiv & \; \tilde{s}^{g'\! g}_{\tau, \mu} (\vec{r}  ,\vec{r} ) 
    \,\,\, ,   \\
 \tau^{g'\! g}_\tau (\vec{r}) 
 \equiv & \; \tau^{g'\! g}_\tau (\vec{r}  ,\vec{r}) 
   \,\,\, &,  \;\;
 T^{g'\! g}_{\tau, \mu}(\vec{r}) 
 \equiv & \; T^{g'\! g}_{\tau, \mu} (\vec{r}  ,\vec{r} )  
     \,\,\, ,  \\
 \tilde{\tau}^{g'\! g}_\tau (\vec{r}) 
 \equiv & \; \tilde{\tau}^{g'\! g}_\tau (\vec{r}  ,\vec{r}) 
   \,\,\, &,  \;\;
 \tilde{T}^{g'\! g}_{\tau, \mu}(\vec{r}) 
 \equiv & \; \tilde{T}^{g'\! g}_{\tau, \mu} (\vec{r}  ,\vec{r} )  
     \,\,\, ,  \\
 j^{g'\! g}_{\tau,\mu} (\vec{r}) 
\equiv & \; j^{g'\! g}_{\tau,\mu} (\vec{r}  ,\vec{r} )
   \,\,\, &,  \;\;
 J^{g'\! g}_{\tau,\mu \nu}(\vec{r}) 
\equiv & \; J^{g'\! g}_{\tau, \mu \nu} (\vec{r}  ,\vec{r} ) 
      \,\,\, , \\
 \tilde{j}^{g'\! g}_{\tau,\mu} (\vec{r}) 
\equiv & \; \tilde{j}^{g'\! g}_{\tau,\mu} (\vec{r}  ,\vec{r} )
   \,\,\, &,  \;\;
 \tilde{J}^{g'\! g}_{\tau,\mu \nu}(\vec{r}) 
\equiv & \; \tilde{J}^{g'\! g}_{\tau, \mu \nu} (\vec{r}  ,\vec{r} ) 
      \,\,\, .
\end{alignat}
\end{subequations}
Considering neutron-neutron and proton-proton pairing only, densities $ \tilde{s}^{g'\! g}_{\tau, \nu},  \tilde{T}^{g'\! g}_{\tau, \nu}$ and $\tilde{j}^{g'\! g}_{\tau, \mu}$ are null~\cite{perlinska04a}. We finally introduce the spin-orbit current as the pseudo-vector part of the spin-orbit tensor
\begin{equation}
J^{g'\! g}_{\tau,\lambda}(\vec{r}) \equiv \sum_{\mu\nu} \epsilon_{\lambda \mu \nu} J^{g'\! g}_{\tau,\mu \nu}(\vec{r}) \,.
\end{equation}

\subsubsection{Energy kernel}

The basic parametrization of the Skyrme energy kernel is a bilinear local functional built out of the above local densities such that each term may contain up to two gradients and two spin Pauli matrices. It is written as 
\begin{eqnarray}
E[\mathbf{\rho}^{g'\!g}, \mathbf{\kappa}^{g'\!g}, \mathbf{\kappa}^{gg'  \ast}]  &\equiv&  \int \! d\vec{r} \, \left\{ {\cal E}^{g'\!g}_{\rho}(\vec{r}) + {\cal E}^{g'\!g}_{\rho\rho} (\vec{r}) + {\cal E}^{g'\!g}_{\kappa\kappa} (\vec{r})\right\} \, \, ,
\end{eqnarray}
where the term linear in the normal density denotes the effective kinetic energy while the terms bilinear in the normal and anomalous density matrices model the effective nuclear interaction energy. Suppressing the spatial argument $\vec{r}$ for simplicity, the three contributions to the local energy density read
\begin{subequations}
\label{BilinearSkyrmeEDF}
\begin{eqnarray}
{\cal E}^{g'\!g}_{\rho} &=&  \frac{\hbar^2}{2m} \sum_{\tau} \tau^{g'\!g}_{\tau} \, , \label{BilinearSkyrmeEDF1}\\
{\cal E}^{g'\!g}_{\rho\rho}  &=&
\sum_{\tau\tau'} \bigg[
              C_{\tau\tau'}^{\rho\rho} \; \rho^{g'\! g}_{\tau} \, \rho^{g'\! g}_{\tau'}  + C_{\tau\tau'}^{\rho\Delta\rho} \; \rho^{g'\! g}_{\tau} \Delta \rho^{g'\! g}_{\tau'}  + C_{\tau\tau'}^{\rho\tau} \; \Big( \rho^{g'\! g}_{\tau} \, \tau^{g'\! g}_{\tau'} - \vec{j}^{g'\! g}_{\tau} \cdot \vec{j}^{g'\! g}_{\tau'} \Big) \nonumber \\
 && 
  + C_{\tau\tau'}^{ss} \; \vec{s}^{\,g'\! g}_{\tau} \cdot \vec{s}^{\,g'\! g}_{\tau'} + C_{\tau\tau'}^{s\Delta s} \; \vec{s}^{\,g'\! g}_{\tau} \cdot \Delta \vec{s}^{\,g'\! g}_{\tau'} + C_{\tau\tau'}^{\rho\nabla J} \; \Big( \rho^{g'\! g}_{\tau}  \vec{\nabla} \cdot \vec{J}^{g'\! g}_{\tau'} + \vec{j}^{g'\! g}_{\tau} \cdot \vec{\nabla} \times \vec{s}^{\,g'\! g}_{\tau'} \Big)
               \nonumber \\
  &&  
   + C_{\tau\tau'}^{J\bar{J}} \Big( \sum_{\mu\nu} J^{g'\! g}_{\tau, \mu\mu} \, J^{g'\! g}_{\tau',\nu \nu} +
                             J^{g'\! g}_{\tau, \mu \nu} \, J^{g'\! g}_{\tau', \nu \mu} -  2 \; \vec{s}^{\,g'\! g}_{\tau} \cdot \vec{F}^{g'\! g}_{\tau'}  \Big)     \nonumber \\
 && 
   + C_{\tau\tau'}^{JJ} \Big( \sum_{\mu\nu} J^{g'\! g}_{\tau, \mu \nu} \, J^{g'\! g}_{\tau', \mu \nu} - \vec{s}^{\,g'\! g}_{\tau} \cdot \vec{T}^{g'\! g}_{\tau'} \Big)   + C_{\tau\tau'}^{\nabla s\nabla s} \; \vec{\nabla} \cdot \vec{s}^{\,g'\! g}_{\tau} \, \, \vec{\nabla} \cdot \vec{s}^{\,g'\! g}_{\tau'}  \bigg]\,,  \label{BilinearSkyrmeEDF2} \\  
{\cal E}^{g'\!g}_{\kappa\kappa}  &=& \sum_{\tau} \Big\{ \nonumber
 C^{\tilde{\rho} \tilde{\rho}}_{\tau\tau}  \tilde{\rho}^{gg' \ast}_{\tau}  \tilde{\rho}^{g'\! g}_{\tau}
+ C^{\tilde{\tau} \tilde{\rho}}_{\tau\tau} \left(\tilde{\rho}^{gg' \ast}_{\tau}  \tilde{\tau}^{g'\! g}_{\tau}  
+      \tilde{\tau}^{gg' \ast}_{\tau} \tilde{\rho}^{g'\! g}_{\tau}
+ \frac{1}{2}   \vec{\nabla} \tilde{\rho}^{gg' \ast}_{\tau}  \cdot \vec{\nabla} \tilde{\rho}^{g'\! g}_{\tau}\right)
 \\ 
&& 
+ \sum_{\mu\nu} \Big(
   C^{\tilde{J} \tilde{J}1}_{\tau\tau} \tilde{J}^{gg' \ast}_{\tau, \mu \nu}   \tilde{J}^{g'\! g}_{\tau, \mu \nu} 
 + C^{\tilde{J} \tilde{J}2}_{\tau\tau}  \tilde{J}^{gg' \ast}_{\tau, \nu \nu}   \tilde{J}^{g'\! g}_{\tau, \mu \mu}
+ C^{\tilde{J} \tilde{J}3}_{\tau\tau} \tilde{J}^{gg' \ast}_{\tau, \nu \mu} \tilde{J}^{g'\! g}_{\tau, \mu \nu}   
\Big)
\Big\}  \, , \label{BilinearSkyrmeEDF3}
\end{eqnarray}
\end{subequations}
A key feature of expressions~\ref{BilinearSkyrmeEDF2} and~\ref{BilinearSkyrmeEDF3} relates to the fact that local densities are not combined arbitrarily to build the various bilinear terms at play. Given $R(\alpha'') \in {\cal G}$, one must characterize the transformation law of each local density induced by the transformation of $\langle \Phi^{(g')} |$ and $| \Phi^{(g)} \rangle$ in order to identify which bilinear combinations can be formed to fulfil Eq.~\ref{invariancekernel}. Such a procedure must be typically conducted for Galilean transformations, rotations in coordinate, gauge and isospin spaces, as well as for a time-reversal transformation. We refer the reader to Refs.~\cite{doba95a,perlinska04a} for a detailed discussion regarding the constraints generated by Eq.~\ref{invariancekernel} on the diagonal energy kernel $E[\mathbf{\rho}^{gg}, \mathbf{\kappa}^{gg}, \mathbf{\kappa}^{gg  \ast}]$. To give a taste of the constraints at play, let us however exemplify the situation by briefly discussing four transformations of interest. 

Fulfilling Eq.~\ref{invariancekernel} under Galilean transformations leads to the necessity to {\it group} several bilinear terms together, i.e. only the sum of terms grouped in between parenthesis in Eq.~\ref{BilinearSkyrmeEDF} are invariant. This per se reduces the number of free coupling constants entering the EDF kernel. Turning to space rotations, the set of local densities transform according to
\begin{subequations}
\label{transfolocaldensities}
\begin{eqnarray}
 \rho^{\Omega'\!-\!\Omega'' \Omega\!-\!\Omega''}_{\tau} (\vec{r}) & = &  \rho^{\Omega' \Omega}_{\tau} ({\cal R}^{-1}(\Omega'')\vec{r})
      \\
 \tau^{\Omega'\!-\!\Omega'' \Omega\!-\!\Omega''}_{\tau} (\vec{r}) & = & \tau^{\Omega' \Omega}_{\tau} ({\cal R}^{-1}(\Omega'')\vec{r}) 
      \\
 \vec{s}^{\,\Omega'\!-\!\Omega'' \Omega\!-\!\Omega''}_{\tau} (\vec{r}) & = & {\cal R}^{-1}(\Omega'') \, \vec{s}^{\,\Omega' \Omega}_{\tau} ({\cal R}^{-1}(\Omega'')\vec{r})
      \\
& \vdots & \nonumber
\end{eqnarray}
\end{subequations}
where ${\cal R}(\Omega)$ is the 3-dimensional matrix representation of the rotation, i.e. local densities transform according to their scalar, vector or tensor field character. In order to fulfil Eq.~\ref{invariancekernel}, densities are combined in Eq.~\ref{BilinearSkyrmeEDF} such that each bilinear term eventually transforms as a scalar field. As result, integrating over $\vec{r}$ provides a scalar independent of ${\cal R}^{-1}(\Omega'')$. Although the realistic nuclear Hamiltonian contains a slight breaking of the isospin invariance and of the isospin symmetry, only the latter can anyway be characterized in a functional that does not mix protons and neutrons. Enforcing it requires that $C^{ff'}_{nn}=C^{ff'}_{pp}$ and $C^{ff'}_{np}=C^{ff'}_{pn}$. Last but not least, fulfilling Eq.~\ref{invariancekernel} under a rotation in gauge space does not impose any constraint on the part of the EDF kernel that depends on the normal density matrix $\rho^{g'\!g}$ but imposes that anomalous densities enter under the form of bilinear products of the form $\kappa^{gg'  \ast}  \kappa^{g'\! g}$, which is indeed the case of each term appearing in Eq.~\ref{BilinearSkyrmeEDF3}.

\subsubsection{Pseudo-potential-based kernel}
\label{pseudopotSkyrme}

Let us now illustrate the pseudo-potential based approach within the Skyrme family of parametrizations. To make the discussion transparent, we simplify it by considering a toy two-body Skyrme pseudo-potential, i.e. the operators considered in Eq.~\ref{Intro_met:hamiltonian} are  
\begin{subequations}
\begin{eqnarray}
t^{1N\,\text{pseudo}} &\equiv& - \frac{\hbar^2}{2m} \, \delta(\vec{r}_1 - \vec{r}_2) \, \triangle \,\,\, , \\
v^{2N\,\text{pseudo/toy}} &\equiv& t_0 \, (1- \,P_{\sigma}) \, \delta(\vec{r}_1 - \vec{r}_2)  \label{Skyrme_int:2body_int:t2} 
 \,\,\,,
\end{eqnarray}
\end{subequations}
where $P_{\sigma} \equiv (1+\sigma_1 \cdot \sigma_{2})/2$ is the two-body spin-exchange operator. Further neglecting isospin for simplicity, the EDF kernel computed through Eq.~\ref{Intro_met:HFB_kernel} can be put under the form
\begin{eqnarray}
\label{Intro_met:example:HFB_kernel}
E^{\text{toy}}_{\text{pseudo}}[\mathbf{\rho}^{g'\!g}, \mathbf{\kappa}^{g'\!g}, \mathbf{\kappa}^{gg'  \ast}] &\equiv&  \int d\vec{r} \, \left[
\frac{\hbar^2}{2m} \, \tau^{g'\!g }(\vec{r})  
+ A^{\rho\rho} \; \rho^{g'\! g}(\vec{r}) \, \rho^{g'\! g}(\vec{r}) \right. \nonumber \\
&& \,\,\, \left. +A^{ss} \, \vec{s}^{\,g'\! g}(\vec{r}) \cdot \vec{s}^{\,g'\! g}(\vec{r}) + A^{\tilde{\rho}\tilde{\rho}} \; \tilde{\rho}^{g g'\ast}(\vec{r}) \, \tilde{\rho}^{g'\! g}(\vec{r})\right]
\,\,\, .
\end{eqnarray}
In Eq.~\ref{Intro_met:example:HFB_kernel}, functional coefficients $A^{\rho\rho}$, $A^{ss}$ and $A^{\tilde{\rho}\tilde{\rho}}$ are related to the free parameter $t_0$ entering the pseudo potential through 
\begin{subequations}
\label{correlationcoeffs}
\begin{eqnarray}
A^{\rho\rho} = - A^{ss}  = \frac{t_0}{2} \,\,\, , \label{correlationcoeffsA} \\
A^{\rho\rho} = + A^{\tilde{\rho}\tilde{\rho}}  = \frac{t_0}{2} \, \, \, , \label{correlationcoeffsB}
\end{eqnarray}
\end{subequations}
and are thus interrelated. 

If we now come back to the generic Skyrme parametrization~\ref{BilinearSkyrmeEDF}, it is possible to identify the reduced form that formally matches the above pseudo-potential-based toy functional. It obviously reads
\begin{eqnarray}
 \label{Intro_met:example:EDF_kernel2}
E^{\text{toy}}[\mathbf{\rho}^{g'\!g}, \mathbf{\kappa}^{g'\!g}, \mathbf{\kappa}^{gg'  \ast}]  &\equiv&  \int d\vec{r} \,
\left[
\frac{\hbar^2}{2m} \, \tau^{g'\!g }(\vec{r})  
+ C^{\rho\rho} \; \rho^{g'\! g}(\vec{r}) \, \rho^{g'\! g}(\vec{r}) \right. \nonumber \\
&& \,\,\, \left. + C^{ss} \, \vec{s}^{\,g'\! g}(\vec{r}) \cdot \vec{s}^{\,g'\! g}(\vec{r})+ C^{\tilde{\rho}\tilde{\rho}} \; \tilde{\rho}^{g g'\ast}(\vec{r}) \, \tilde{\rho}^{g'\! g}(\vec{r}) \right]
\,\,\,,
\end{eqnarray}
and looks indeed formally identical to Eq.~\ref{Intro_met:example:HFB_kernel}. Still, crucial differences exist between the two. Contrarily to the pseudo-potential-based approach, parameters $C^{\rho\rho}$, $C^{ss}$ and $C^{\tilde{\rho}\tilde{\rho}}$ are not a priori interrelated in the general EDF approach\footnote{In the case of the present toy functional, the fulfilment of Eq.~\ref{invariancekernel} under Galilean transformations does not correlate any of the couplings.}. Such a feature comes from the fact that the functional is postulated rather than computed as the matrix element of an operator. Interrelations between the functional couplings entering a pseudo-potential based EDF kernel are a manifestation of Pauli's principle that is automatically enforced by definition~\ref{defpseudoEDF}. On the contrary, Pauli's principle is violated in the more general approach to the EDF kernel. Let us now try to illustrate such a key point more transparently.

The energy kernel can always be expressed under the generic form~\ref{Intro_met:HFB_kernel}, {\it as long as its dependence on transition densities is polynomial}, which is the case of the above toy functionals. For the local Skyrme parametrization, this is achieved by expanding local densities according to
\begin{subequations}
\label{genericlocal}
\begin{eqnarray}
f^{g'\! g}_{\tau} (\vec{r})
&\equiv&  \sum_{ij} W^{f}_{ji} (\vec{r} \tau)  \,  \rho^{g'\! g}_{ij} \, , \label{genericlocalnormal} \\
\tilde{f}^{g'\! g}_{\tau} (\vec{r}\tau)
&\equiv&  \sum_{ij} W^{\tilde{f}}_{ji} (\vec{r} \tau)  \,  \kappa^{g'\! g}_{ij} \, , \label{genericlocalanomalous}
\end{eqnarray}
\end{subequations}
where $W^{f}_{ji} (\vec{r}\tau)$ and $W^{\tilde{f}}_{ji} (\vec{r}\tau)$ can be deduced from the definition of the various local densities at play. In the case of toy bilinear functionals~\ref{Intro_met:example:HFB_kernel} and~\ref{Intro_met:example:EDF_kernel2}, one finds
\begin{subequations}
\label{2bodyWF}
\begin{eqnarray}
W^{\rho}_{ji} (\vec{r}) &=& \varphi^{\dagger}_{j} (\vec{r}) \,  \varphi_{i} (\vec{r}) \, , \label{2bodyWFa} \\
\vec{W}^{\vec{s}}_{ji} (\vec{r}) &=& \varphi^{\dagger}_{j} (\vec{r}) \, \vec{\sigma} \, \varphi_{i} (\vec{r}) \, , \label{2bodyWFb} \\
W^{\tilde{\rho}}_{ji} (\vec{r}) &=& \sum_\sigma \sigma \,
		\varphi_j\big(\vec{r} \sigma\big) \,
		\varphi_i\big(\vec{r} \bar{\sigma}\big) \, , \label{2bodyWFc}
\end{eqnarray}
\end{subequations}
where $\varphi_{i} (\vec{r})$ $[\varphi^{\dagger}_{i} (\vec{r})]$ denotes a spinor with components $\varphi_{i} (\vec{r}\sigma)$ $[\varphi^{\ast}_{i} (\vec{r}\sigma)]$. With such definitions at hand, the effective two-body matrix elements $\bar{v}^{2N\,\text{toy}}_{ijkl}$ entering Eq.~\ref{normalenergy}-\ref{anomalousenergy} can be extracted in two different ways, i.e. either focusing on the term proportional to $\rho^{g'\!g}_{ki} \, \rho^{g'\!g}_{lj}$  or focusing on the term proportional to $\kappa^{gg' \! \ast}_{ij} \, \kappa^{g'\!g}_{kl}$, i.e.
\begin{subequations}
\label{matrixelements}
\begin{eqnarray}
\bar{v}^{2N\,\text{toy}\rho\rho}_{ijkl} &\equiv&  2 \int d\vec{r} \,  \Big[B^{\rho\rho} \, W^{\rho}_{ik} (\vec{r}) \, W^{\rho}_{jl} (\vec{r}) + B^{ss} \, \vec{W}^{\vec{s}}_{ik} (\vec{r}) \cdot \vec{W}^{\vec{s}}_{jl} (\vec{r})  \Big] \label{matrixelementsb}
\\
&=& 2 \int d\vec{r} \, \sum_{\sigma\sigma'} \varphi^{\ast}_{i} (\vec{r}\sigma)  \varphi^{\ast}_{j} (\vec{r}\sigma') \Big[ B^{\rho\rho} \varphi_{k} (\vec{r}\sigma)  \varphi_{l} (\vec{r}\sigma') \nonumber \\
&& + B^{ss} \Big(\varphi_{k} (\vec{r}\bar{\sigma})  \varphi_{l} (\vec{r}\bar{\sigma}') \! - \! \bar{\sigma} \! \bar{\sigma}' \! \varphi_{k} (\vec{r}\bar{\sigma}) \varphi_{l} (\vec{r}\bar{\sigma}') \! + \! \sigma \sigma' \! \varphi_{k} (\vec{r}\sigma)  \varphi_{l} (\vec{r}\sigma') \Big)\Big]   \, , \nonumber \\
\bar{v}^{2N\,\text{toy}\kappa\kappa}_{ijkl} &\equiv&  4 \int d\vec{r} \,  B^{\tilde{\rho}\tilde{\rho}}  \, W^{\tilde{\rho}\ast}_{ij} (\vec{r}) \, W^{\tilde{\rho}}_{kl} (\vec{r}) \label{matrixelementsb2}
\\
&=& 4 \int d\vec{r} \,  B^{\tilde{\rho}\tilde{\rho}}   \sum_{\sigma\sigma'} \bar{\sigma}\bar{\sigma}' \,
		\varphi^{\ast}_i\big(\vec{r} \sigma\big) \,
		\varphi^{\ast}_j\big(\vec{r} \bar{\sigma}\big) \,
		\varphi_k\big(\vec{r} \sigma'\big) \,
		\varphi_l\big(\vec{r} \bar{\sigma}'\big)   \, , \nonumber
\end{eqnarray}
\end{subequations}
with $B^{ff'}\equiv A^{ff'}$ for Eq.~\ref{Intro_met:example:HFB_kernel} and $B^{ff'}\equiv C^{ff'}$ for Eq.~\ref{Intro_met:example:EDF_kernel2}. 
Such an extraction of effective two-body matrix elements\footnote{The present analysis can be easily extended to trilinear functional terms and effective three-body matrix elements.} is instrumental to pin down the potential violation of Pauli's principle in the EDF kernel. 

\subsubsection{Spurious self-interaction and self-pairing contributions}
\label{pauli}

In the nuclear EDF framework, Pauli's principle is always satisfied at the level of the individual densities given that one-body density matrices are computed from antisymmetric many-body states (Eq.~\ref{Intro_met:OBDM}). The violation we now wish to briefly discuss may arise when multiplying several such densities together to build the interaction part of the energy kernel.

The first issue relates to the behaviour of $\bar{v}^{2N\,\rho\rho}_{ijkl}$ in the particular case where $k=l$ (or $i=j$). Pauli's principle requires such effective matrix elements to be zero given that two nucleons occupy the same single-particle state. It is easy to check that $\bar{v}^{2\,\text{toy}\rho\rho}_{ijkk}=0$ in Eq.~\ref{matrixelementsb} if, and only if, $B^{\rho\rho} = - B^{ss}$, i.e. if the pseudo-potential-based relationship~\ref{correlationcoeffsA} is satisfied. In the general EDF framework, such interrelations between functional parameters are not enforced and Pauli's principle is violated\footnote{This encompasses the  intermediate case where the EDF kernel is computed as the matrix elements of a {\it density-dependent} effective "Hamiltonian". Indeed, in such a case no exchange or pairing term corresponding to the density dependence of the effective vertex appears in the EDF kernel.}, e.g. $\bar{v}^{2N\, \rho\rho}_{ijkk}\neq0$. Such a violation eventually leads to a contamination of the EDF kernel by spurious self-interaction contributions, i.e. part of the interaction energy originates from individual nucleons interacting with themselves~\cite{perdew81a,Chamel:2010ac}. The self-interaction problem has been extensively studied within DFT for electronic systems and has been shown to contaminate significantly many observables, e.g. ionization energies and, thus, the asymptotic of the electronic density distribution~\cite{Ruz07aDFT}.
 
The self-interaction issue does not concern $\bar{v}^{2N\,\kappa\kappa}_{ijkl}$. Indeed, such a matrix element is multiplied by $\kappa^{gg' \! \ast}_{ij}$ and $\kappa^{g'\!g}_{kl}$  whose antisymmetry ensures that the corresponding contribution to the energy kernel is anyway zero for $i=k$ and/or $k=l$. However, a second issue relates to the link between $\bar{v}^{2N\,\rho\rho}_{ijkl}$ and $\bar{v}^{2N\,\kappa\kappa}_{ijkl}$. Equation~\ref{Intro_met:HFB_kernel} suggests that those two sets of matrix elements should be identical. As a matter of fact, it is straightforward to check that $\bar{v}^{2N\,\text{toy}\rho\rho}_{ijkl}=\bar{v}^{2N\,\text{toy}\kappa\kappa}_{ijkl}$ if, and only if, $B^{\rho\rho} = - B^{ss}=B^{\tilde{\rho}\tilde{\rho}}$, i.e. if pseudo-potential-based relationships~\ref{correlationcoeffsA} and~\ref{correlationcoeffsB} are satisfied. In the general EDF framework, such interrelations between functional parameters are not a priori enforced and Pauli's principle is violated, e.g. $\bar{v}^{2N\, \rho\rho}\neq\bar{v}^{2N\,\kappa\kappa}$. Such a violation eventually leads to a contamination of the EDF kernel by spurious {\it self-pairing} contributions. The notion of self-pairing was introduced for the first time in Refs.~\cite{Lacroix:2008rj,Bender:2008rn} and generalizes the well-known notion of self-interaction.

Within the nuclear context, the contamination of SR results by self-interaction and self-pairing processes has never been characterized. It thus  deserves attention in the future. In Sec.~\ref{difficulties}, we will however see that such spurious contributions to the energy kernel have already been understood to be responsible for critical pathologies in MR-EDF calculations.

\subsubsection{Modern parametrizations}
\label{modernparam}

On the one hand, the bilinear form of the Skyrme parametrization given in Eq.~\ref{BilinearSkyrmeEDF} constitutes the basis of any modern Skyrme parametrization. On the other hand, none of the modern Skyrme parametrizations strictly corresponds to such a form~\cite{bender03b,Lacroix:2008rj,Bender:2008rn}. The most common departures from it relate to the fact that~\cite{bender03b}
\begin{enumerate}
\item Couplings $C_{\tau\tau'}^{ff'}$ may further depend on a set of local densities in order to enrich the parametrization and provide more flexibility. Of course, such additional density dependences must not jeopardize Eq.~\ref{invariancekernel}. Common parametrizations are such that $C_{\tau\tau'}^{\rho\rho}$, $C_{\tau\tau'}^{ss}$ and $C^{\tilde{\rho} \tilde{\rho}}_{\tau\tau}$ depend on the {\it isoscalar} matter density $\rho^{g'\! g}_{0}(\vec{r})\equiv \rho^{g'\! g}_n(\vec{r})+\rho^{g'\! g}_p(\vec{r})$.
\item Specific couplings might be put to zero for (numerical) convenience, simplicity or because of the difficulty to identify empirical data that can help fix their value unambiguously. Typical examples concern $C_{\tau\tau'}^{JJ}$, $C_{\tau\tau'}^{J\bar{J}}$, $C_{\tau\tau'}^{\nabla s\nabla s}$, $C^{\tilde{\tau} \tilde{\rho}}_{\tau\tau}$ and $C^{\tilde{J} \tilde{J}1/2/3}_{\tau\tau}$.
\item The form of certain terms might be approximated. This is the case of the so-called exchange term originating from the Coulomb interaction (not shown here) that is usually treated in the Slater approximation.
\end{enumerate}
In the very large majority of cases, such deviations from the strict and complete bilinear form constitute a departure from the pseudo-potential based method, independent of whether or not the bilinear baseline was originally derived from a pseudo potential. Consequently, ad hoc modifications of the EDF parametrizations cause or reinforce a breaking of Pauli's principle and induce pathologies associated with it (see Secs.~\ref{pauli} and~\ref{difficulties}). Note that the latter statements apply equally to Gogny or relativistic parametrization of the EDF kernel. Still, most of the enrichments of the analytical form of the Skyrme family of parametrizations have been performed along this line in recent years. With no ambition of being exhaustive, let us mention some of the recent attempts at empirically enriching the Skyrme parametrization in order to improve its global performance and/or overcome a specific limitation. Such developments relate to
\begin{enumerate}
\item A dependence of $C^{\tilde{\rho} \tilde{\rho}}_{\tau\tau}$ on the scalar-isovector density to better reproduce pairing gaps in neutron-rich nuclei and asymmetric nuclear matter~\cite{Margueron:2007uf,Duguet:2003yi,Yamagami:2008ks,Chamel:2010rw,Yamagami:2012ga}.
\item A dependence of $C_{\tau\tau'}^{\rho\rho}$ and $C_{\tau\tau'}^{ss}$ on vector-isoscalar and vector-isovector densities to control infinite wavelength spin and isospin instabilities of nuclear matter beyond saturation density~\cite{Margueron:2009rn}.
\item An enriched dependence of $C_{\tau\tau'}^{\rho\rho}$ on the scalar-isoscalar density to fully decouple the isoscalar effective mass from the compressibility~\cite{Cochet:2003ex,Cochet:2003sy}. 
\item The pairing part of the EDF derived from a {\it regularized} zero-range two-body pseudo potential with {\it separable} Gaussian regulators~\cite{Duguet:2003yi,Lesinski:2008cd,Lesinski:2011rn} with the goal to have (i) a way to handle a finite-range pairing vertex that is numerically cost efficient and (ii) the possibility to connect to realistic nuclear forces.
\item A density dependence of $C_{\tau\tau'}^{\rho\Delta\rho}$ to produce a surface-peaked effective mass~\cite{Zalewski:2010ni,Fantina:2010iq}.
\item Use of $C_{nn}^{\rho\nabla J}\neq C_{pp}^{\rho\nabla J}$ to offer more flexibility in the reproduction of spin-orbit splittings~\cite{MoyadeGuerra:2011zz}.
\end{enumerate}

Even more recently, an effort towards the construction of new families of EDF parametrizations that derive strictly from a pseudo potential has emerged. This new trend is motivated by the identification of pathologies in MR-EDF calculations that originate from the breaking of Pauli's principle in any of the existing EDF parametrizations (see Sec.~\ref{pauli} and~\ref{difficulties}). Associated on-going developments relate to the construction of
\begin{enumerate}
\item A bilinear EDF derived from a zero-range Skyrme-like two-body pseudo potential containing up to six gradient operators~\cite{Carlsson:2008gm,Davesne:2013aja}.
\item The complete bilinear and trilinear EDF derived from zero-range Skyrme-like two- and three-body pseudo potentials containing up to two gradient operators~\cite{sadoudi11thesis,Sadoudi:2012jg}.
\item A bilinear EDF derived from a {\it regularized} zero-range Skyrme-like two-body pseudo potential with up to two gradient operators and Gaussian regulators~\cite{Dobaczewski:2012cv,Bennaceur:2013fua}.
\end{enumerate}

\section{Single-reference implementation}
\label{SREDFsubsection}

\begin{figure}[htbp]
\begin{center}
\includegraphics[width = 0.43\textwidth, keepaspectratio]{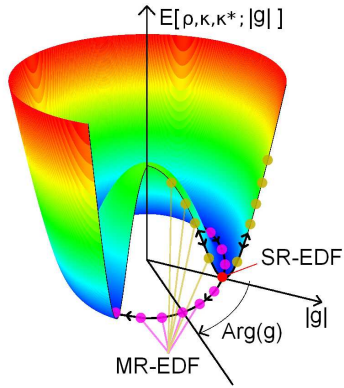}
\caption{Schematic view of the diagonal energy kernel $E[g,g]$ as a function of both the phase and the magnitude of the order parameter associated with a spontaneously broken symmetry.}
\label{Intro_met:hat}
\end{center}
\end{figure}

The single-reference implementation of the nuclear EDF method exclusively invokes the {\it diagonal} kernel $E[g,g]$. State $| \Phi^{(g)} \rangle$ is entitled to break as many symmetries of the nuclear Hamiltonian as it finds energetically favourable. That a certain symmetry does break spontaneously usually depends on the number of elementary constituents of the system under consideration (see Sec.~\ref{breaksymfinitesystems}). As state $| \Phi^{(g)} \rangle$ acquires a finite order parameter $g$, the diagonal kernel remains independent of its phase $\alpha$, as schematically pictured in Fig.~\ref{Intro_met:hat}. Such a degeneracy derives trivially from Eq.~\ref{invariancekernel}. Whenever the system does break the symmetry spontaneously, i.e. whenever the minimal energy is obtained for a non zero value of $g$, the two-dimensional profile of $E[g,g]$ takes the typical form of a "mexican hat". The degeneracy of $E[g,g]$ with respect to $\alpha$ relates to the fact that a spontaneous symmetry breaking at the SR level gives rise to a zero-energy Goldstone mode. One practical consequence is that SR calculations can be performed at any fixed value of $\alpha$, e.g. $\alpha=0$.

\subsection{Equation of motion}
\label{SRequation}

The SR energy is obtained, for a targeted value of $|g|$, through the minimization
\begin{eqnarray}
\label{Intro_met:SR-Min}
E^{SR}_{|g|} &\equiv& \text{Min}_{\{| \Phi^{(|g|0)} \rangle\}}  \Big\{ {\cal E}_{|g|} \Big\} \,\,\,,
\end{eqnarray}
within the manifold of (symmetry-breaking) Bogoliubov states. The diagonal energy kernel to be actually minimized reads\footnote{One way to ensure that the minimization is indeed performed within the manyfold of product states consists of adding an additional Lagrange constraint requiring that the {\it generalized} density matrix~\cite{ring80a} ${\cal R}$ remains idempotent.}
\begin{eqnarray}
\label{modifiedkernel}
{\cal E}_{|g|} &\equiv& E[g,g] - \lambda \, \Big[N - \langle \Phi^{(g)} | N | \Phi^{(g)} \rangle \Big] - \lambda_{|g|} \, \Big[|g| - |\langle \Phi^{(g)} | G | \Phi^{(g)} \rangle |\Big] \,\,\,.
\end{eqnarray}
The last two terms in Eq.~\ref{modifiedkernel} introduce Lagrange parameters\footnote{Expressions are given here for linear constraints although practical calculations often rely on quadratic constraints~\cite{staszczak10a}.} that are to be adjusted such that the average number of nucleons in $\vert \Phi^{(|g|0)} \rangle$ is equal to its actual number in the nucleus under study and such that the norm of the order parameter is equal to the desired value $|g|$. 

Equations~\ref{Intro_met:SR-Min}-\ref{modifiedkernel} lead to solving an equation of motion that takes the form of a constrained Bogoliubov-De Gennes eigenvalue problem\footnote{Depending on the isospin projection $\tau$ considered, $\lambda = \lambda_{n}$ or $\lambda_{p}$.}
\begin{equation}
         \fourmat{\mathbf{h} - \lambda \, \mathbf{1} }{\mathbf{\Delta}        }
          {-\mathbf{\Delta}^{\ast}}{-\mathbf{h}^{\ast} + \lambda \, \mathbf{1} }^{(g)} \, \twospinor{\mathbf{U}}{\mathbf{V}}^{(g)}_{\mu}  = E^{|g|}_{\mu}  \,
\twospinor{\mathbf{U}}{\mathbf{V}}^{(g)}_{\mu} \, , \label{Bogoequation}
\end{equation}
which is to be realized iteratively and where the (constrained) one-body fields are defined through functional derivatives of the (modified) diagonal energy kernel
\begin{eqnarray}
\mathbf{h}^{(g)} - \lambda \, \mathbf{1}   &\equiv & \frac{\delta {\cal E}_{|g|}}{\delta \mathbf{\rho}^{gg \ast}}  \hspace{0.6cm} ;  \hspace{0.6cm}  \mathbf{\Delta}^{(g)} \equiv \frac{\delta {\cal E}_{|g|}}{\delta \mathbf{\kappa}^{gg \, \ast}} \, .
\end{eqnarray}
The field $\mathbf{h}^{(g)}$ governs the {\it effective} single-particle motion while the anomalous field $\mathbf{\Delta}^{(g)}$ drives pairing correlations. Explicit expressions of the fields are easily obtained given a specific (e.g. Skyrme) parametrization of the EDF kernel. Equation~\ref{Bogoequation} provides the set of quasi-particle energies $E^{|g|}_{\mu}$ at "deformation" $g$ and the corresponding wave-functions $(\mathbf{U},\mathbf{V})^{(g)}_{\mu}$ from which density matrices $\rho^{gg}=V^{(g)\ast}V^{(g)T}$ and $\kappa^{gg}=V^{(g)\ast}U^{(g)T}$, as well as the total energy, can be computed.

\begin{figure}[htbp]
\begin{center}
\includegraphics[width = 0.45\textwidth, keepaspectratio]{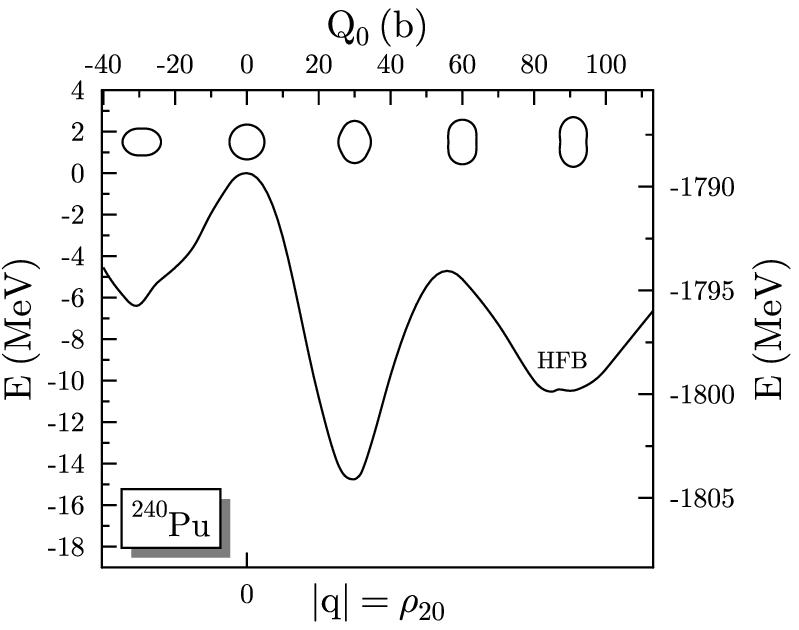} \hspace{0.7cm} \includegraphics[width = 0.45\textwidth, keepaspectratio]{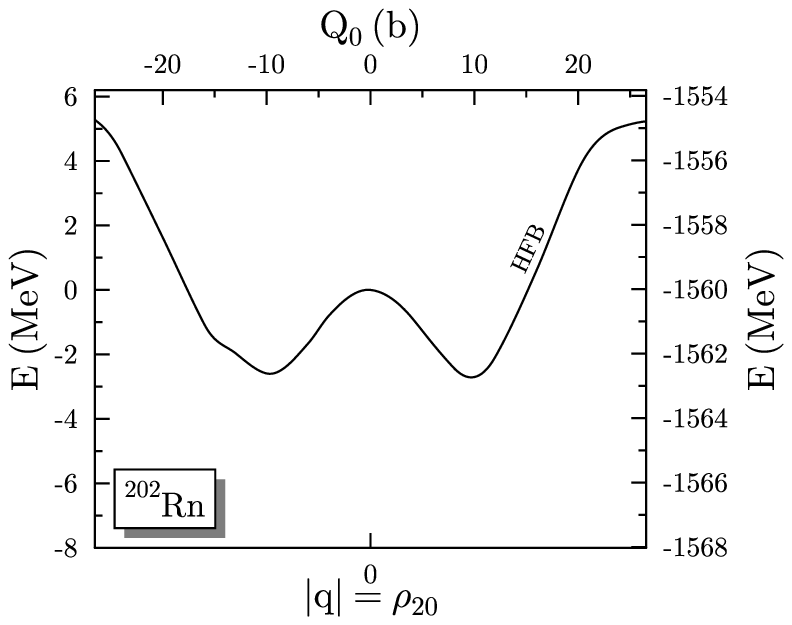} \includegraphics[width = 0.45\textwidth, keepaspectratio]{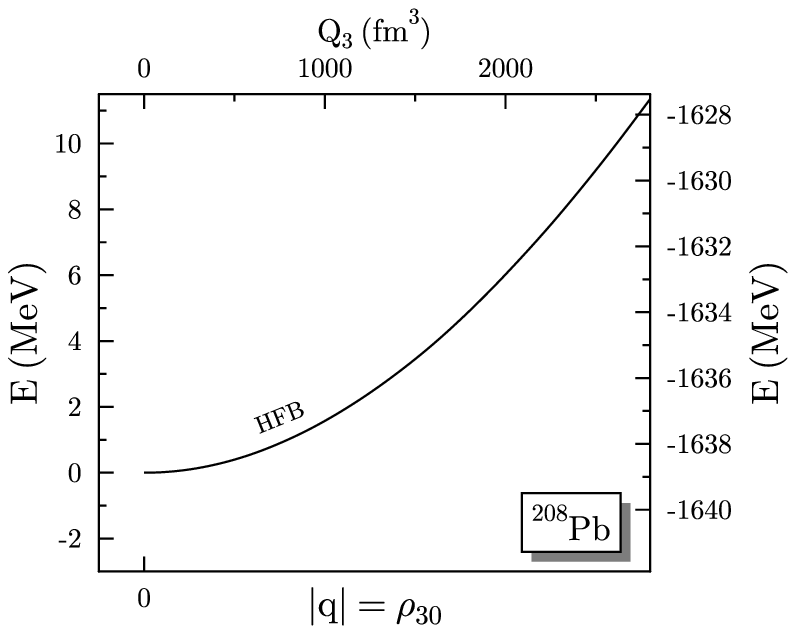}  \hspace{0.7cm} \includegraphics[width = 0.45\textwidth, keepaspectratio]{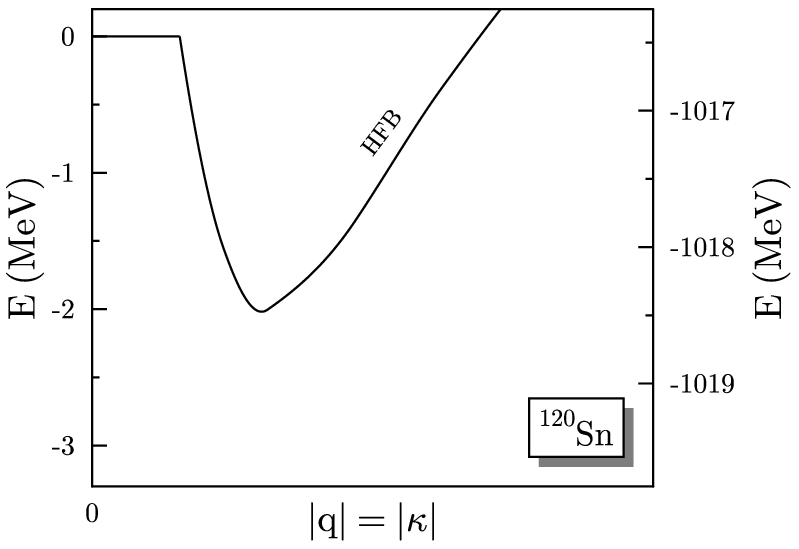}
\caption{Energy landscapes as a function of the norm of various order parameters~\cite{bender13a}. Note that $|q|$ stands for $|g|$ in the figure. Upper panels: SR-EDF energy of $^{240}$Pu and $^{202}$Rn as a function of axial quadrupole deformation ($|g|\equiv \rho_{20}$). Lower left panel: SR-EDF energy of $^{208}$Pb as a function of axial octupole deformation ($|g|\equiv \rho_{30}$). Lower right panel: SR-EDF energy of $^{120}$Sn as a function of pairing deformation ($|g|\equiv ||\kappa||$). Left vertical axes are rescaled with respect to the symmetry conserving, i.e. non-deformed, reference point.}
\label{PES}
\end{center}
\end{figure}

The full SR energy landscape, associated with the complete set of reference states $\{| \Phi^{|g|\alpha} \rangle = R(\alpha) | \Phi^{(|g|0)} \rangle \, ; \, |g| \in [0,+\infty[ \, ; \, \alpha \in D_{{\cal G}}\}$, is generated through repeated calculations performed for various targeted values of $|g|$. The degeneracy of $E[g,g]$ with respect to $\alpha$ makes it unnecessary to solve the equation of motion for $\alpha\neq0$. As an illustration, Fig.~\ref{PES} displays the energy landscapes associated with various order parameters, i.e. various operators $G$. First, the energy landscape of $^{240}$Pu and $^{202}$Rn as a function of axial quadrupole deformation ($|g|\equiv \rho_{20}$) demonstrates that rotational symmetry is spontaneously broken in those nuclei. Second, the energy landscape of $^{208}$Pb as a function of axial octupole deformation ($|g|\equiv \rho_{30}$) illustrates that this nucleus is found to remain spherical at the SR-EDF level. Last but not least, the energy landscape of $^{120}$Sn as a function of pairing deformation  ($|g|\equiv ||\kappa||$) shows that such a nucleus is superfluid.

The absolute minimum of the SR landscape $E^{SR}_{\text{GS}} \equiv \text{Min}_{|g|} \big\{E^{SR}_{|g|}\big\}$ provides a first approximation to the ground-state binding energy that incorporates static collective correlations via the breaking of symmetries. Such a solution provides a first approximation to other quantities of interest, e.g. ground-state's charge and matter radii as well as nucleonic density distributions, one-nucleon separation energies and effective single-particle energies (see Sec.~\ref{SR-ESPE}), along with individual excitations through an even number of quasi-particle excitations. Using one projection of the angular-momentum vector, e.g. $J_{x}$, as the constrain operator gives access to rotational excitations of the nucleus when solving  Eq.~\ref{Bogoequation} for appropriate values of $\langle \Phi^{(g)} | J_{x} | \Phi^{(g)} \rangle$. This actually corresponds to using the {\it velocity} along the phase of the order parameter as a collective degree of freedom.

The full SR landscape provides a richer information. Along the radial direction $|g|$, in particular, the curvature around the minimum characterizes the sensitivity of the system to a change of collective "deformation", whereas the existence of a secondary minimum can be tentatively associated with a shape isomer. Such an analysis is the starting point of the more advanced MR implementation detailed in Sec.~\ref{MRformalismsection} below.

\subsection{One-nucleon addition and removal processes}
\label{SR-sepEnergies}

In the context of SR-EDF calculations, the description of states in the N$\pm$1 systems rely on Bogoliubov states having the form of one quasi-particle excitations on top of an even number-parity vacuum
\begin{equation}
\vert \Phi^{(g)}_k \rangle \equiv  \beta^{(g)\dagger}_k \, \vert \Phi^{(g)} \rangle \,\,\, .
\label{qp_state}
\end{equation}
The even-number parity vacuum being associated with an even-even system, one-nucleon addition and removal energies to final states of the A$\pm$1 systems are obtained through
\begin{eqnarray}
E_k^{|g| \pm} &=&  \pm \left\{E[\mathbf{\rho}^{gg}_{k}, \mathbf{\kappa}^{gg}_{k}, \mathbf{\kappa}^{gg  \, \ast}_{k}] - E[\mathbf{\rho}^{gg}, \mathbf{\kappa}^{gg}, \mathbf{\kappa}^{gg  \, \ast}]\right\}  \mp \lambda \left\{\langle  \Phi^{(g)}_k | N |  \Phi^{(g)}_k \rangle - (N\pm 1)\right\} \nonumber \\
&=&  \lambda  \pm E^{|g|}_{k} \,\,\,, \label{eq:epmk2}
\end{eqnarray}
where $\mathbf{\rho}^{gg}_{k}$ and $\mathbf{\kappa}^{gg}_{k}$ denote the density matrices computed from $|  \Phi^{(g)}_k \rangle$~\cite{ring80a}. The error associated with the difference between the average number of particles in state $|  \Phi^{(g)}_k \rangle$ and the targeted particle number N$\pm1$ is compensated for by the last term in the definition of $E_k^{|g| \pm}$. In the perturbative approach (Eq.~\ref{eq:epmk2}), the chemical potential $\lambda$  and quasi-particle energies $E^{|g|}_{k}$ are outputs of Eq.~\ref{Bogoequation} solved for the even number-parity vacuum.

Spectroscopic amplitudes associated with the (perturbative) addition and removal of a nucleon are obtained as
\begin{subequations}
\label{spectrobogo}
\begin{eqnarray}
\bra {\Phi^{(g)}_k} a^\dagger_p \ket {\Phi^{(g)}} &=& U^{(g)\ast}_{pk}  \, , \\
\bra {\Phi^{(g)}_k} a_p \ket {\Phi^{(g)}}  &=& V^{(g)\ast}_{pk}  .
\end{eqnarray}
\end{subequations}
From these amplitudes, spectroscopic probability matrices are introduced through  $\mathbf{S}_{k}^{(g)+} \equiv \mathbf{U}^{(g)}_{k} \mathbf{U}^{(g)\dagger}_{k}$ and $\mathbf{S}_{k}^{(g)-} \equiv \mathbf{V}^{(g)\ast}_{k}\mathbf{V}^{(g)T}_{k}$ and satisfy, according to Eq.~\ref{Bogounitaritya}, the sum rule
\begin{eqnarray}
\sum_{k} \mathbf{S}_{k}^{(g)+} + \sum_{k} \mathbf{S}_{k}^{(g)-} &=& \mathbf{1} \, .
\end{eqnarray}
Corresponding spectroscopic factors are nothing but the norm of spectroscopic probability matrices and are thus given~\cite{rotival07a} by
\begin{eqnarray}
SF_{k}^{(g)\pm} &\equiv& \text{Tr}_{{\cal H}_{1}}\!\left[ \mathbf{S}_{k}^{(g)\pm}\right] \, \, .  \label{SF_SREDF}
\end{eqnarray}

Any inclusion of many-body correlations leads to a fragmentation of the spectroscopic strength associated with one-nucleon addition and removal processes\footnote{It is specific to the EDF method to {\it implicitly} account for correlations via the functional character of $E[g,g]$. As such, one-nucleon separation energies $E_k^{|g|\pm}$ obtained through SR-EDF calculations can be seen as effective centroids of a more fragmented underlying spectrum generated via a theory that explicitly accounts for those correlations.}. Within the SR-EDF method, this is the case of static collective correlations that are incorporated via the breaking of symmetries. For example, pairing correlations fragment the strength near the Fermi energy into two peaks belonging, respectively, to addition and removal channels. Similarly, the lifting of the $2j\!+\!1$ degeneracy seen at sphericity in the additional/removal spectrum $E_k^{|g|\pm}$ is nothing but the fragmentation of the strength induced by the correlations grasped via the breaking of rotational invariance. Still, this happens at the price of losing good symmetry quantum numbers, which makes difficult to interpret the additional/removal spectrum $E_k^{|g|\pm}$. One must thus await for the MR-EDF description to restore symmetries and achieve a meaningful comparison with experimental data. This will bring further correlations to the description and additional fragmentation of the strength. The latter reveals that separation energies  $E_k^{|g|\pm}$ do not target experimental values yet; i.e. absolute values of experimental one-nucleon addition (removal) energies are typically underestimated (overestimated) on purpose by SR-EDF calculations\footnote{Inaccuracies associated with the quality of empirical EDF parametrizations are responsible for quantitative discrepancies while the present discussion relates to qualitative differences that are built in on purpose.} in magic nuclei~\cite{bender03b}.

\subsection{Effective single-particle energies}
\label{SR-ESPE}

In an ab-initio context, meaningful effective single-particle energies (ESPEs) providing the underlying shell structure relate to the Baranger centroid Hamiltonian. The latter is computed from outputs of the A-body Schroedinger equation through~\cite{baranger70a,Duguet:2011sq}
\begin{eqnarray}
\mathbf{h}^{\text{cent}} &\equiv& \sum_{\mu\in {\cal H}_{A\!+\!1}} \mathbf{S}_{\mu}^{+} E_{\mu}^{+} + \sum_{\nu\in {\cal H}_{A\!-\!1}}  \mathbf{S}_{\nu}^{-} E_{\nu}^{-} \label{defsumrule} \,\,\, ,
\end{eqnarray}
where ${\cal H}_{A\!\pm\!1}$ denotes the A$\pm$1 Hilbert space. Specifically, ESPEs are the {\it eigenvalues} $\{e^{\text{cent}}_{p}\}$ of the centroid field~\cite{baranger70a}
\begin{eqnarray}
\mathbf{h}^{\text{cent}} \, \psi^{\text{cent}}_p &=& e^{\text{cent}}_{p} \, \psi^{\text{cent}}_p \,\,\, , \label{HFfield3}
\end{eqnarray}
and are nothing but barycentre of one-nucleon separation energies weighted by the probability to reach the corresponding A+1 (A-1) eigenstates through the addition (removal) a nucleon to (from) single-particle state $\psi^{\text{cent}}_p$. As such, they recollect the strength fragmented by many-body correlations. 

Let us now transpose the discussion to the context of SR-EDF calculations. Following Baranger, the objective is to build meaningful centroids of the fragmented strength. As discussed above, the only fragmentation of the strength {\it explicitly} accounted for within the SR-EDF method relates to the breaking of symmetries. Let us illustrate the situation by taking the breaking of particle number and angular momentum as examples. Below, the breaking of the former is explicitly embodied by the Bogoliubov algebra whereas the breaking of the latter is materialized by the labels $|g| \equiv \rho_{\lambda\mu}$ and $\text{Arg}(g)\equiv \Omega$.

As far as gathering the strength fragmented by pairing correlations, one can indeed reach an interesting result~\cite{sadoudi12b}. Multiplying the first (second) line of Eq.~\ref{Bogoequation} by $\mathbf{U}^{(g)\dagger}_{k}$ ($\mathbf{V}^{(g)\dagger}_{k}$) and summing over $k$, one obtains
\begin{subequations}
\label{identity}
\begin{eqnarray}
\sum_{k} \mathbf{h}^{(g)} \, \mathbf{U}^{(g)}_{k} \, \mathbf{U}^{(g)\dagger}_{k} + \sum_{k} \mathbf{\Delta}^{(g)} \, \mathbf{V}^{(g)}_{k} \, \mathbf{U}^{(g)\dagger}_{k} &=& \sum_{k}  (\lambda + E^{|g|}_{k}) \, \mathbf{U}^{(g)}_{k} \, \mathbf{U}^{(g)\dagger}_{k} \, ,  \\
\sum_{k} \mathbf{\Delta}^{(g)} \, \mathbf{U}^{(g)\ast}_{k} \, \mathbf{V}^{(g)T}_{k} + \sum_{k} \mathbf{h}^{(g)} \, \mathbf{V}^{(g)\ast}_{k} \, \mathbf{V}^{(g)T}_{k} &=& \sum_{k}  (\lambda - E^{|g|}_{k}) \, \mathbf{V}^{(g)\ast}_{k} \, \mathbf{V}^{(g)T}_{k} \,\, . 
\end{eqnarray}
\end{subequations}
Adding up both lines, using Eqs.~\ref{Bogounitaritya} and~\ref{Bogounitarityb} eventually provides
\begin{eqnarray}
\mathbf{h}^{(g)} &=&  \sum_{k}  \mathbf{S}_{k}^{(g)+} \, E_k^{|g| +} +  \sum_{k} \mathbf{S}_{k}^{(g)-} \, E_k^{|g|-} \,, \label{sumruleEDF}
\end{eqnarray}
which is analogous to Eq.~\ref{defsumrule} and provides $\mathbf{h}^{(g)}$ with the meaning of a centroid field. The coupling of addition and removal spectroscopic amplitudes via the anomalous field $\mathbf{\Delta}^{(g)}$ in Eq.~\ref{Bogoequation} is screened out from the Baranger sum rule. This is an a priori non-trivial result, though straightforward to obtain. Of course, the explicit tackling of pairing correlations does impact the centroid field indirectly via the feedback of such correlations on the normal  density matrix and the dependence of $\mathbf{h}^{(g)}$ on the latter. Interestingly, Eq.~\ref{sumruleEDF} justifies the traditional use by practitioners of the eigenvalues of $\mathbf{h}^{(g)}$ as effective single-particle energies\footnote{In view of Eq.~\ref{sumruleEDF}, it thus appears more justified to use eigenvalues of $\mathbf{h}^{(g)}$ as ESPEs rather than its diagonal matrix elements in the basis diagonalizing $\rho^{gg}$, i.e. the so-called {\it canonical} basis, as it is often done by practitioners, e.g. see Ref.~\cite{rotival07a}.}, i.e.
\begin{eqnarray}
\mathbf{h}^{(g)} \, \psi^{(g)}_p &=& e^{|g|}_{p} \, \psi^{(g)}_p \,\,\, . \label{HFfieldEDF}
\end{eqnarray}
It is remarkable that Eq.~\ref{sumruleEDF} could be obtained without making any explicit reference to a Hamilton operator, i.e. within the strict spirit of the EDF method. This is at variance with the standard proof that allows one to connect the centroid field with the static part of the one-nucleon self energy~\cite{baranger70a,Duguet:2011sq}.
\begin{figure}[htbp]
\begin{center}
\includegraphics[width = 0.7\textwidth, keepaspectratio]{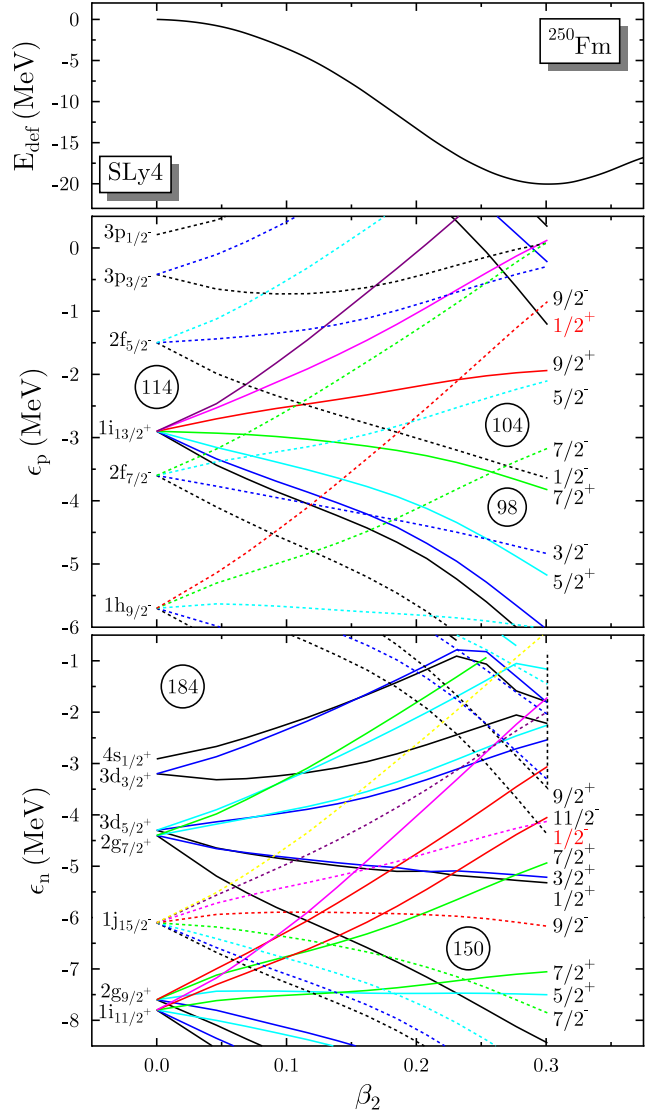}
\caption{Energy landscape and effective single-particle energies of $^{250}$Fm as a function of axial quadrupole deformation ($|g|=\rho_{20}=\beta_2$)~\cite{bender13a}.}
\label{ESPE}
\end{center}
\end{figure}

Sum rule~\ref{sumruleEDF} only gathers the strength fragmented by correlations associated with the breaking of particle number, not yet the strength fragmented by the breaking of angular momentum. As a matter of fact, $\mathbf{h}^{(g)}$ does break rotational symmetry such that the ESPE spectrum $e^{|g|}_{p}$ displays the same lifting of the $2j\!+\!1$ degeneracy as $ E_k^{|g| \pm}$. Plotted against $|g|=\rho_{20}$, the spectrum $e^{|g|}_{p}$ takes the form of a so-called Nilsson diagram as is illustrated in Fig.~\ref{ESPE} for $^{250}$Fm. One observes that the minimum of the energy landscape is obtained for a deformation that reflects a compromise between $N = 150$ and $Z \sim 100$ deformed shell gaps in the ESPE spectrum. 

One can now go one step further and recollect the strength associated with the breaking of rotational symmetry\footnote{Such a procedure can be extended to any subgroup of ${\cal G}$.}. To do so, one notices that $\mathbf{h}^{(g)}$, as any one-body operator, transforms under rotation according to\footnote{Equation~\ref{rotationHFfield} can be recovered by expressing matrices $\mathbf{S}_{k}^{(g) \pm}$ in a spherical basis $p=n\pi jm\tau$ and by working out how such matrices transform under the rotation of $\ket {\Phi^{(g)}}$ and $\ket {\Phi^{(g)}_k}$.}
\begin{eqnarray}
\mathbf{h}^{(\rho_{\lambda\mu} \Omega)}  &=&  R(\Omega) \, \mathbf{h}^{(\rho_{\lambda\mu} 0)} \, R^{\dagger}(\Omega)  \,\,\, . \label{rotationHFfield}
\end{eqnarray}
The fragmented strength is recollected by extracting the monopole, i.e. angular-averaged, part of $\mathbf{h}^{(\rho_{\lambda\mu} \Omega)}$. Expressing Eq.~\ref{rotationHFfield} in a spherical basis, omitting isospin projection and parity\footnote{If  $\mathbf{h}^{(\rho_{\lambda\mu} 0)}$ breaks parity, one further needs to extract the component belonging to the trivial Irreps of $C_i$, i.e. the inversion center group. Indeed, restoring spherical symmetry does not ensure that parity is a good quantum number, e.g. a $j=3/2$ single-particle state can be a linear combination of $d_{3/2}$ and $p_{3/2}$ states. Proceeding to such an extraction would deliver a one-body field that is block-diagonal with respect to parity $\pi$ as well.} quantum numbers for simplicity, as well as using orthogonality relationship~\ref{orthogonality}, the monopole operator satisfies~\cite{duguet12a}
\begin{subequations}
\label{monopolefield}
\begin{eqnarray}
h^{\text{mon}[\rho_{\lambda\mu}]}_{n jm n' j'm'}  &\equiv&  \frac{2J\!+\!1}{16\pi^2} \int_{D_{\Omega}}\! d\Omega \, {\cal D}^{0 \, \ast}_{00}(\Omega)  \,\, h^{(\rho_{\lambda\mu} \Omega)}_{n jm n' j'm'} \label{monopolefieldA} \\
&=&  \delta_{jj'} \, \delta_{mm'}  \sum_{m''} h^{(\rho_{\lambda\mu} 0)}_{n jm'' n' jm''} \,\,\,. \label{monopolefieldB}
\end{eqnarray}
\end{subequations}
Equation~\ref{monopolefieldB} demonstrates that $\mathbf{h}^{\text{mon}[\rho_{\lambda\mu}]}$ displays spherical symmetry and is built out of the $(j,m)$ blocks of the deformed operator $\mathbf{h}^{(\rho_{\lambda\mu} 0)}$, including an averaging over the magnetic quantum number $m$. The monopole field thus extracted carries the deformation label $\rho_{\lambda\mu}$ as a memory of the symmetry breaking field it has been extracted from. Spherical ESPEs gathering the strength of the fragmented spectrum $e^{|g|}_{p}$ are then obtained through
\begin{eqnarray}
\mathbf{h}^{\text{mon}[\rho_{\lambda\mu}]} \, \psi^{\text{mon}[\rho_{\lambda\mu}]}_{n jm} &=& e^{\text{mon}[\rho_{\lambda\mu}]}_{n jm} \, \psi^{\text{mon}[\rho_{\lambda\mu}]}_{n jm} \,\,\, . \label{sphericalESPEEDF}
\end{eqnarray}
Equation~\ref{sphericalESPEEDF} defines the way to extract a spherical effective single-particle energy spectrum out of any SR-EDF calculation. Such a procedure has neither been defined nor used so far\footnote{Practically speaking, Eqs.~\ref{monopolefield}-\ref{sphericalESPEEDF} are particularly trivial to implement in numerical codes that expend deformed solutions out of a spherical, e.g. harmonic oscillator, basis.}. As already mentioned, the above procedure is not limited to $SO(3)$ and can be extended to any broken symmetry.

\subsection{Equation of state of infinite nuclear matter}

Infinite nuclear matter (INM) is an idealized nuclear system that has relevance to the study of several real systems, e.g. the physics of neutron stars or the dynamic of supernovae explosions. The system is made of protons and neutrons and is considered to be homogeneous. The Coulomb interaction between protons is switched off. One is first and foremost interested in computing the equation of state (EOS) of such a system, i.e. its energy per nucleon as a function of its density. This can easily be done at the SR level. Below, we illustrate the procedure at zero temperature on the basis of the strict bilinear Skyrme parametrization introduced in Eq.~\ref{BilinearSkyrmeEDF}. Furthermore, pairing correlations are omitted as they little impact bulk properties such as the EOS. However, one should note that pairing properties, e.g. pairing gaps, of INM are of importance to the physics of neutron stars~\cite{Chamel:2012br}.

\subsubsection{Definitions}

The four basic degrees of freedom characterizing INM are the scalar-isoscalar $\rho_0$, scalar-isovector $\rho_1$, vector-isoscalar $s_0$ and vector-isovector $s_1$ densities. They can be expressed through neutron and proton as well as spin-up and spin-down densities in the following way
\begin{subequations}
\label{eq:INM:rho01s01}
\begin{align}
\rho_0 =& \rho_{n\uparrow} + \rho_{n\downarrow} + \rho_{p\uparrow} + \rho_{p\downarrow} %= \rho_n + \rho_p
\,\,\,, \\
\rho_1 =& \rho_{n\uparrow} + \rho_{n\downarrow} - \rho_{p\uparrow} - \rho_{p\downarrow} %= \rho_n - \rho_p
\,\,\,, \\
s_0 =& \rho_{n\uparrow} - \rho_{n\downarrow} + \rho_{p\uparrow} - \rho_{p\downarrow} %= s_n + s_p
\,\,\,, \\
s_1 =& \rho_{n\uparrow} - \rho_{n\downarrow} - \rho_{p\uparrow} + \rho_{p\downarrow} %= s_n - s_p
\,\,\,,
\end{align}
\end{subequations}
such that the inverse relationships read
\begin{subequations}
\label{eq:INM:rhonpsnp}
\begin{align}
\rho_{n\uparrow} =& \frac{1}{4}\Big( 1 + I_\tau + I_\sigma + I_{\sigma\tau} \Big) \rho_0
\,\,\,, \\
\rho_{n\downarrow} =& \frac{1}{4}\Big( 1 + I_\tau - I_\sigma - I_{\sigma\tau} \Big) \rho_0 
\,\,\,, \\
\rho_{p\uparrow} =& \frac{1}{4}\Big( 1 - I_\tau + I_\sigma - I_{\sigma\tau} \Big) \rho_0
\,\,\,, \\
\rho_{p\downarrow} =& \frac{1}{4}\Big( 1 - I_\tau - I_\sigma + I_{\sigma\tau} \Big) \rho_0
\,\,\,,
\end{align}
\end{subequations} 
where isospin $I_\tau \equiv \rho_1/\rho_0$, spin $I_\sigma \equiv s_0/\rho_0$ and spin-isospin $I_{\sigma\tau} \equiv s_1/\rho_0$ excesses ($-1 \leq I_i \leq 1$) have been introduced. The typical cases of interest are (i) symmetric nuclear matter ($I_\tau=I_\sigma=I_{\sigma\tau}=0$), (ii) isospin-asymmetric nuclear matter ($I_\tau \neq 0$), (iii) spin-polarized nuclear matter ($I_\sigma \neq 0$) and (iv) isospin-asymmetric spin-polarized nuclear matter ($I_\tau \neq 0$, $I_\sigma \neq 0$ and $I_{\sigma\tau} \neq 0$). 

Infinite nuclear matter being translationally invariant, it is convenient to use a plane wave basis
\begin{equation}
\label{eq:INM:Intro:WF}
\langle \vec{r} \sigma \tau | \vec{k} \sigma' \tau' \rangle = \varphi_{\vec{k} \sigma' \tau'}(\vec{r} \sigma \tau)  = \left( 2\pi \right)^{-\frac{3}{2}} \, \exp({\mathrm i} \vec{k} \cdot \vec{r})  \; \delta_{\sigma\sigma'} \, \delta_{\tau\tau'} \,\,\, ,
\end{equation}
where $\tau\sigma=\{n\uparrow,n\downarrow,p\uparrow,p\downarrow\}$. Neglecting pairing, the SR state reduces to a Slater determinant obtained by filling individual orbitals $\varphi_{\vec{k}\sigma' \tau'}(\vec{r} \sigma \tau)$ up to the Fermi momentum, i.e. the normal density matrix is diagonal in the plane-wave basis and equal to 1 for states characterized by $|\vec{k}| \leq k_{F,\tau\sigma}$ and 0 otherwise, where $k_{F,\tau\sigma}$ denotes the spin- and isospin-dependent Fermi momentum. The SR state does not carry any non-zero order parameter such that the label $g$ can be dropped in the present section.

Starting from Eq.~\ref{eq:INM:Intro:WF}, local densities can be computed explicitly.  The sum over basis states in Eq.~\ref{eq:Skyrme_int:crtoci} becomes an integral over the sphere of radius $k_{F,\tau\sigma}$. Eventually, local densities of interest are constant in space and read as
\begin{subequations}
\begin{eqnarray}
\rho_{\tau\sigma} &=& \int_{|\vec{k}|\leq k_{F,\tau\sigma}} \hspace{-0.3cm} d\vec{k}  \; 
      \varphi^{\ast}_{\vec{k}} (\vec{r} \sigma \tau) \, \varphi_{\vec{k}} (\vec{r} \sigma \tau) 
     = \frac{1}{6\pi^2} k_{F,\tau\sigma}^3 \,\,\,,
\label{eq:INM:Intro:rho:npud} \\
\tau_{\tau\sigma} &=& \int_{|\vec{k}|\leq k_{F,\tau\sigma}} \hspace{-0.3cm}  d\vec{k}  \; 
      \left[ \vec{\nabla} \varphi^{\ast}_{\vec{k}} (\vec{r} \sigma \tau) \right] \,\cdot \,  \left[ \vec{\nabla} \varphi_{\vec{k}} (\vec{r} \sigma \tau)  \right]
   = \frac{3}{20} \; \frac{2}{3\pi^2} \; k_{F,\tau\sigma}^5 \,\,\, .
\label{eq:INM:Intro:kin:npud} 
\end{eqnarray}
\end{subequations}
With the choice of a Fermi surface centred at \mbox{$\vec{k} = 0$},  current densities vanish \mbox{$\vec{j}_{q\sigma} = 0$}. 
Also, all gradients of local densities are zero 
\mbox{$\nabla_\nu \rho_{q\sigma} = 0$} by construction, as are the pair densities. Using Eqs.~\ref{eq:INM:rhonpsnp}, \ref{eq:INM:Intro:rho:npud} and~\ref{eq:INM:Intro:kin:npud}, one relates spin-isospin kinetic densities to spin, isospin and spin-isospin excesses
\begin{subequations}
\label{eq:INM:Intro:kin}
\begin{align}
\tau_0 =& \, \tau_{n\uparrow} + \tau_{n\downarrow} + \tau_{p\uparrow} + \tau_{p\downarrow} 
= \,            \frac{3}{5} c_s \rho_0^{5/3} F^{(0)}_{5/3}(I_\tau,I_\sigma,I_{\sigma\tau})
\,\,\,, \\
\tau_1 =& \, \tau_{n\uparrow} + \tau_{n\downarrow} - \tau_{p\uparrow} - \tau_{p\downarrow}
= \,            \frac{3}{5} c_s \rho_0^{5/3} F^{(\tau)}_{5/3}(I_\tau,I_\sigma,I_{\sigma\tau})
\,\,\,, \\
T_0 =& \, \tau_{n\uparrow} - \tau_{n\downarrow} + \tau_{p\uparrow} - \tau_{p\downarrow}
= \,        \frac{3}{5} c_s \rho_0^{5/3} F^{(\sigma)}_{5/3}(I_\tau,I_\sigma,I_{\sigma\tau})
\,\,\,, \\
T_1 =& \, \tau_{n\uparrow} - \tau_{n\downarrow} - \tau_{p\uparrow} + \tau_{p\downarrow}
%\nonumber \\ &
= \,        \frac{3}{5} c_s \rho_0^{5/3} F^{(\sigma\tau)}_{5/3}(I_\tau,I_\sigma,I_{\sigma\tau})
\,\,\,,
\end{align}
\end{subequations}
where $F$-functions~\cite{bender02a} are explicated in App.~\ref{sec:F-functions}. We further introduce $\displaystyle c_s\equiv(3\pi^2/2)^{2/3}$ and $\displaystyle c_n\equiv(3\pi^2)^{2/3}$.

Last but not least, the results are expressed below in terms of isoscalar $C^{ff'}_{0}$ and isovector $C^{ff'}_{1}$ couplings. The latter are related to the couplings in the neutron/proton representation (under the assumption of isospin symmetry) used in Eq.~\ref{BilinearSkyrmeEDF} through  
\begin{subequations}
\begin{eqnarray}
C^{ff'}_{0}&=& \frac{1}{2} (C^{ff'}_{\tau\tau}+C^{ff'}_{\tau\bar{\tau}}) \, , \\
C^{ff'}_{1}&=& \frac{1}{2} (C^{ff'}_{\tau\tau}-C^{ff'}_{\tau\bar{\tau}}) \, .
\end{eqnarray}
\end{subequations}
The fact that most of the local densities are zero in INM implies that properties will be expressed in terms of a limited number of couplings. 

\subsubsection{Symmetric nuclear matter}

Symmetric nuclear matter (SNM) is characterized by an equal number of protons and neutrons as well as of spin up and spin down nucleons. Consequently, $\rho_1=I_\tau=0$ and $I_\sigma = I_{\sigma\tau} = 0$. Only two local densities $\rho_0$ and $\tau_0$ subsist, i.e. $\rho_n=\rho_p=\frac{1}{2}\rho_0$ and $\tau_n=\tau_p=\frac{1}{2}\tau_0$, with
\begin{equation}
\rho_{0} =  \frac{2}{3\pi^2} \, k_{F}^3 \,\, ; \,\,
\tau_{0} = \frac{3}{5} \, c_s \,\rho_0^{5/3} \,\, .
\end{equation}
The EOS is obtained from Eq.~\ref{BilinearSkyrmeEDF} as
\begin{eqnarray}
\frac{E}{A} &\equiv& \frac{{\cal E}_\rho+{\cal E}_{\rho\rho}}{\rho_0} =
\frac{3}{5}\,\frac{\hbar^2}{2m}\,c_s\,\rho_0^{2/3}
+ C^{\rho\rho}_{0}\; \rho_0^2  + \frac{3}{5} \,c_s   \, C^{\rho\tau}_{0}  \, \rho_0^{5/3}   
\label{eq:INM:SNM:E}
\,\,\,.
\end{eqnarray}
Symmetric nuclear matter presents a stable state such that a minimum energy is obtained for a finite density $\rho_{\text{sat}}$. The pressure of the fluid relates to the first derivative of the EOS with respect to the isoscalar density, which in SNM reads
\begin{eqnarray}
P &\equiv & \rho_0^2 \frac{\partial E/A }{\partial \rho_0} \Big\vert_A =
\frac{2}{5}\,\frac{\hbar^2}{2m}\,c_s\,\rho_0^{5/3}
+ C^{\rho\rho}_{0} \, \rho_0^{2}  
 + \,c_s  \, C^{\rho\tau}_{0}  \, \rho_0^{8/3}    
 \,\,\,. \label{eq:INM:SNM:P}
\end{eqnarray}
The equilibrium density $\rho_{\text{sat}}$ is obtained as the solution of $P(\rho_{\text{sat}}) = 0$.

The incompressibility of the nuclear fluid relates to the second derivative of the EOS with respect to the isoscalar density and expresses the energy cost to compress the nuclear fluid. It is defined as
\begin{align}
K\equiv &\frac{18P}{\rho_0}+9\rho_0^2 \frac{\partial^2 E/A}{\partial \rho_0^2}
\,\,\,,
\label{eq:INM:SNM:K}
\end{align}
such that at equilibrium
\begin{eqnarray}
K_\infty &\equiv & 9\rho_0^2 \frac{\partial^2 E/A}{\partial \rho_0^2} \Big\vert_{\rho_0=\rho_{\text{sat}}}
= -\frac{6}{5}\,\frac{\hbar^2}{2m}\,c_s\,\rho_{\text{sat}}^{2/3}
   + 6 \,c_s  \, C^{\rho\tau}_{0}  \, \rho_{\text{sat}}^{5/3}    
\,\,\,, \label{eq:INM:SNM:Kinf}
\end{eqnarray}
which needs to be positive for the system to be stable against density fluctuations.

\subsubsection{Asymmetric nuclear matter}

In general, INM is characterized by (i)  unequal proton and neutron matter densities, i.e. $I_\tau \neq 0$, (ii) a global spin polarization, i.e. $I_\sigma \neq 0$ and (iii) a spin polarization that differs for neutron and proton species, i.e. $I_{\sigma\tau} \neq 0$. The EOS of such a nuclear fluid is given by
\begin{eqnarray}
\frac{E}{A}&=&
\frac{3}{5}\frac{\hbar^2}{2m}\,c_s\,F^{(0)}_{5/3}(I_{\tau},I_{\sigma},I_{\sigma \tau}) \, \rho_0^{2/3}
+ C^{\rho\rho}_{0}  \, \rho_0  
+ C^{\rho\rho}_{1}   \, \rho_0 \, I_{\tau}^{2}
+ C^{ss}_{0}  \, \rho_0 \, I_{\sigma}^{2}  + C^{ss}_{1}  \, \rho_0 \, I_{\sigma\tau}^{2} 
 \nonumber \\ 
&& +\frac{3}{5}  \Big[ C^{\rho\tau}_{0}   \, F^{(0)}_{5/3}(I_{\tau},I_{\sigma},I_{\sigma \tau}) 
+ C^{\rho\tau}_{1}\; I_{\tau} \, F^{(\tau)}_{5/3}(I_{\tau},I_{\sigma},I_{\sigma \tau}) 
\nonumber \\
&& \quad \quad - C^{JJ}_{0}\; I_{\sigma}  \, F^{(\sigma)}_{5/3}(I_{\tau},I_{\sigma},I_{\sigma \tau}) 
- C^{JJ}_{1}\; I_{\sigma \tau}  \, F^{(\sigma \tau)}_{5/3}(I_{\tau},I_{\sigma},I_{\sigma \tau}) \Big]  c_s \rho_0^{5/3} 
\nonumber
\,\,\,.
\end{eqnarray}
Spin, isospin and spin-isospin symmetry energies are analogues of $K_\infty$ with respect to spin, isospin and spin-isospin excesses, respectively. As such, they characterize the stiffness of the EOS with respect to generating such non-zero excesses. At saturation of SNM, i.e. when $I_\sigma=I_\tau=I_{\sigma\tau}=0$ and $\rho_0=\rho_{\text{sat}}$, the three symmetry energies are given by 
\begin{subequations}
\begin{eqnarray}
a_\tau 
&\equiv& \frac{1}{2}\frac{\partial^2 E_H/A}{\partial I_\tau^2} \Big\vert_{I_\sigma = I_\tau = I_{\sigma\tau} = 0} 
\label{eq:INM:ANM:a_i}  \\ 
&=&
\frac{1}{3}\,\frac{\hbar^2}{2m}\,c_s\,\rho_0^{2/3}
+ C^{\rho\rho}_{1}  \; \rho_0  
+ \bigg[ \frac{1}{3}   C^{\rho\tau}_{0}
+  C^{\rho\tau}_{1}  \bigg] \,c_s \, \rho_0^{5/3}   
\nonumber 
\,, \\
a_\sigma 
&\equiv& \frac{1}{2}\frac{\partial^2 E_H/A}{\partial I_\sigma^2} \Big\vert_{I_\sigma = I_\tau = I_{\sigma\tau} = 0} 
\label{eq:INM:ANM:a_s}  \\ 
&=&
\frac{1}{3}\,\frac{\hbar^2}{2m}\,c_s\,\rho_0^{2/3}
+ C^{ss}_{0} \; \rho_0  
+ \bigg[ \frac{1}{3}   C^{\rho\tau}_{0}
- C^{JJ}_{0}   \bigg] \,c_s \, \rho_0^{5/3}   
\nonumber
\,, \\
a_{\sigma\tau}
&\equiv& \frac{1}{2}\frac{\partial^2 E_H/A}{\partial I_{\sigma\tau}^2} \Big\vert_{I_\sigma = I_\tau = I_{\sigma\tau} = 0} 
\label{eq:INM:ANM:a_si}  \\ 
&=&
\frac{1}{3}\,\frac{\hbar^2}{2m}\,c_s\,\rho_0^{2/3}
+ C^{ss}_{1}  \; \rho_0  
+ \bigg[ \frac{1}{3}   C^{\rho\tau}_{0}
- C^{JJ}_{1}   \bigg] \,c_s \, \rho_0^{5/3}   
\nonumber \,,
\end{eqnarray}
\end{subequations}
and must be positive for the minimum of the EOS to be stable. 

Two quantities of interest are intimately connected to the skin thickness of heavy isospin-asymmetric nuclei, i.e. to the difference between their neutron and proton radii. These quantities are the density-symmetry coefficient $L$
\begin{eqnarray}
L & \equiv &   
3 \rho 
\frac{\partial}{\partial \rho} \left(
\frac{1}{2}\frac{\partial^2 E/A}{\partial I_\tau^2} \right)
\Big\vert_{I_\sigma = I_\tau = I_{\sigma\tau} = 0} 
 \\
& =&
 \frac{2}{3}\,\frac{\hbar^2}{2m}\,c_s\rho_0^{2/3}
+3  C^{\rho\rho}_{1}  \rho_0  
 + \bigg[  \frac{5}{3} C^{\rho\tau}_{0}  + 5 C^{\rho\tau}_{1}
\bigg] \,c_s \, \rho_0^{5/3}   
\nonumber \,,
\end{eqnarray}
and the symmetry compressibility
\begin{eqnarray}
K_{sym} &\equiv&
9 \rho^2 
\frac{\partial^2}{\partial \rho^2} \left(
\frac{1}{2}\frac{\partial^2 E/A}{\partial I_\tau^2} \right)
\Big\vert_{I_\sigma = I_\tau = I_{\sigma\tau} = 0} 
\\
&=&
-\frac{2}{3}\,\frac{\hbar^2}{2m}\,c_s\rho_0^{2/3}
+\frac{10}{3}  \, c_s \,  C^{\rho\tau}_{0} \rho_0^{5/3}  
+  
10 \, c_s \, C^{\rho\tau}_{1} 
 \rho_0^{5/3}  
\,. \nonumber
\end{eqnarray}

\subsubsection{Pure neutron matter}

A particular case of isospin-asymmetric and spin-symmetric nuclear matter is pure neutron matter (PNM) obtained for $I_\tau=1$ and $I_\sigma=I_{\sigma\tau} = 0$. The EOS of PNM reads
\begin{eqnarray}
\frac{E}{A} &=&
\frac{3}{5}\frac{\hbar^2}{2m}\,c_n \rho_0^{2/3}
+ C^{\rho\rho}_{0} \; \rho_0  
+ C^{\rho\rho}_{1} \; \rho_0  
+\frac{3}{5}  \,c_n  \, C^{\rho\tau}_{0} \;  \rho_0^{5/3}   
 +\frac{3}{5}  \,c_n  \, C^{\rho\tau}_{1} \;  \rho_0^{5/3}   \label{eq:INM:PNM:E} 
\, .
\end{eqnarray}

\subsection{Symmetry breaking and "deformation"}

\label{breaksymfinitesystems}
\begin{table}[htbp]
\begin{tabular}{|l|c|l|}
\cline{2-3} \multicolumn{1}{c|}{}
 & Nuclei &  Excitation pattern \\
\hline
Space translation $\vec{a}$ & All & Surface vibrations  \\
Gauge rotation $\varphi$ & All but double magic ones  &  Energy gap \\
Space rotation $\alpha,\beta,\gamma$ & All but singly-magic ones & Ground-state rotational bands  \\
\hline
\end{tabular}
\caption{Categories of nuclei that tend to break translational, rotational and particle number symmetries as well as associated patterns in their excitation spectrum.}
\label{excitation_patterns} 
\end{table}

There are important points to underline regarding the notions of symmetry breaking and "deformation" in finite systems. To do so, let us take rotational symmetry and the deformation of the density distribution as an example. Of course, the discussion conducted below applies to any of the symmetries of interest.
\begin{enumerate}
\item The breaking of a symmetry is never quite real in a finite system. Eventually, any quantum state of the nucleus does carry good angular momentum $(J,M)$ such that it is improper to describe it as a wave packet mixing states belonging to different irreducible representations of $SO(3)$, i.e. carrying different values of $J$. Only in infinite systems characterized by infinite inertia would the sequence of states belonging to a rotational band be truly degenerate. This makes the symmetry breaking real in infinite systems as it offers the possibility to describe the true ground state as a linear combination of states with different $J$ values. In a finite system, quantum fluctuations associated with finite inertia eventually lift the degeneracy such that good symmetry quantum numbers must eventually be restored.
\item In a finite system, the notion of "deformation" that characterizes the breaking of a symmetry is thus necessarily an artefact associated with an {\it incomplete} theoretical description. As such, the $J^{\pi}=0^{+}$ ground state of an even-even nucleus is {\it never} "deformed", given that the density distribution of {\it any} $J=0$ quantum state is spherically symmetric. It is only within an incomplete theoretical description such as the SR-EDF method that one may speak improperly of a "deformed" $J^{\pi}=0^{+}$ ground state\footnote{It is important to underline at this point that the notion of "deformation" differs depending on the angular momentum of the targeted many-body state. This is due to the fact that a symmetry-{\it conserving} state with angular momentum $J$ does display non-zero multipole moments of the density for $\lambda \leq 2J$~\cite{sadoudi11thesis}. For example, having a reference state with non-zero quadrupole and hexadecapole moments does {\it not} characterize a breaking of rotational symmetry if one means to describe a $J = 2$ state. In such a case, one must check multipoles with $\lambda >4$ (or any odd multipole) to state whether rotational symmetry is broken or not. It happens that product states of the Bogoliubov type usually generate non-zero multipole moments of all (e.g. even) multipolarities as soon as they display a non-zero collective quadrupole moment. As such, they break rotational symmetry independent of the angular momentum of the good-symmetry state one is eventually after.}. Once rotational symmetry is restored, the corresponding density distribution is indeed spherically symmetric.
\item Within, e.g., the SR-EDF method, one notices that the breaking of the rotational symmetry depends on the number of elementary constituents of the even-even nucleus under consideration; i.e. the symmetry does not break in double and single magic nuclei while it breaks in essentially all double open-shell nuclei\footnote{Of course, the fact that the neutron or proton number is magic is not known a priori but is based on a posteriori observations and experimental facts. In particular, the fact that traditional magic numbers, i.e. $N,Z=2, 8, 20, 28, 50, 82, 126$, remain as one goes to very isospin-asymmetric nuclei is the subject of intense on-going experimental and theoretical investigations~\cite{sorlin08}.}. This raises an important question. If all $J^{\pi}=0^{+}$ states are eventually equally spherical in front of god, are "spherical" $J^{\pi}=0^{+}$ states more spherical than "deformed" ones!? To rephrase it, one may ask in what way the intermediate artefact of "deformation" tells us anything real about the nucleus under consideration? As a matter of fact, the artefact of ground-state "deformation" does not tell us anything about the ground state but rather about the way the nucleus primarily {\it excites}. In the case of rotational symmetry, the fact that the ground state comes out to be deformed at the SR-EDF level tells us, at a low theoretical cost, that a rotational band built on top of it should exist. To reverse engineer the statement, any experimental spectrum containing a set of states that can be convincingly ordered as a $J(J+1)$ sequence above the ground state will see the latter being deformed within the (incomplete) SR-EDF description.
\end{enumerate}
To conclude, even though the symmetry breaking is fictitious in a finite system it leaves its fingerprint on excitation spectra.  Such a connection between the two notions is schematically illustrated in Tab.~\ref{excitation_patterns} for the three symmetries of present interest.

\subsection{Connection to density functional theory?}

It has become customary in nuclear physics to assimilate the SR-EDF method, eventually including corrections {\it a la} Lipkin or Kamlah, with density functional theory (DFT) at play in electronic systems, i.e. to state that the Hohenberg-Kohn (HK) theorem~\cite{hohenberg64} underlays nuclear SR-EDF calculations. This is a misconception as distinct strategies actually support both methods. Whereas the SR-EDF method minimizes the energy with respect to a symmetry-breaking trial density, DFT relies on an energy functional whose minimum must be reached for a local one-body density\footnote{The scheme can be extended to a set of local densities or to the full density matrix.} that possesses {\it all} symmetries of the actual ground-state density, i.e. that displays fingerprints of the symmetry quantum-numbers carried by the exact ground-state~\cite{fertig00a}. As a matter of fact, generating a symmetry-breaking solution is known to be problematic in DFT, as it lies outside the frame of the HK theorem, and is usually referred to as the {\it symmetry dilemma}. To bypass that dilemma and grasp kinematical correlations associated with good symmetries, several reformulations of DFT have been proposed over the years, e.g. see Refs.~\cite{gross88a,gorling93a}.

Recent efforts within the nuclear community have been devoted to formulating a HK-like theorem in terms of the internal density, i.e. the matter distribution relative to the center of mass of the self-bound system~\cite{Eng07a,Messud:2009jh}. Together with an appropriate Kohn-Sham scheme~\cite{Messud:2009jh}, it allows one to reinterpret the SR-EDF method as a functional of the internal density rather than as a functional of a laboratory density that breaks translational invariance. This constitutes an interesting route whose ultimate consequence would be to remove entirely the notion of breaking and restoration of symmetries from the EDF approach and make the SR formulation a complete many-body method, at least in principle. To reach such a point though, the work of Refs.~\cite{Eng07a,Messud:2009jh} must be extended, at least, to rotational and particle-number symmetries, knowing that translational symmetry was somewhat the easy case to deal with given the explicit decoupling of internal and center of mass motions. Going in such a direction, an interesting formulation was recently proposed that provides the Schroedinger equation based on collective Hamiltonian with a firm ground~\cite{lesinski13a}. This problem deserves significant attention in the future.

\section{Multi-reference implementation}
\label{MRformalismsection}

In a finite system, quantum fluctuations eventually make the symmetry breaking fictitious such that good symmetries must eventually be restored. From a group theory perspective, the diagonal energy kernel $E[g,g]$ associated with a symmetry breaking state $\vert \Phi^{(g)} \rangle$ mixes irreducible representations of the symmetry group of interest, and so does $E^{SR}_{\text{GS}}$. The symmetry restoration consists of extracting energies that can be put in one-to-one correspondence with Irreps of the group. In terms of the schematic "mexican-hat" of Fig.~\ref{Intro_met:hat}, doing so corresponds to incorporating zero-energy fluctuations along the phase of the order parameter. 

Furthermore, fluctuations of $|g|$, i.e. configuration mixing along the radial coordinate of the "mexican-hat", must be considered at the same time. This is well illustrated by Fig.~\ref{PES}. On the one hand, the SR energy landscape of $^{240}$Pu is stiff in the vicinity of its minimum and well separated from the secondary minimum tentatively associated with a fission isomer. On the other hand, $^{202}$Rn is "soft" with respect to axial quadrupole deformation and displays two equally pertinent oblate and prolate minima that are separated by a small barrier of about $2$\,MeV height. While the SR minimum provides a reasonable picture of what the intrinsic state of $^{240}$Pu might be, no single reference state characterized by a fixed value of $|g|=\rho_{20}$ is entitled to do so for $^{202}$Rn, i.e. fluctuations in $|g|=\rho_{20}$ are expected to be large a priori.

Within the EDF method, the large amplitude collective motions associated with the fluctuations of both the phase $\alpha$ and the magnitude $|g|$ of the order parameters are accounted for by the multi-reference framework. In doing so, a MR-EDF calculation accesses collective, i.e. "rotational" and "vibrational", excitations while incorporating associated correlations in the ground state. Technically speaking, the MR step invokes the complete set of product states $\{| \Phi^{(|g|\alpha)} \rangle = R(\alpha) | \Phi^{(|g|0)} \rangle \, ; \, |g| \in [0,+\infty[ \, ; \, \alpha \in D_{{\cal G}}\}$ such that the MR energy mixes off-diagonal energy $E[g',g]$ and norm  $N[g',g]$ kernels associated with all pairs of states belonging to that set (see below). The restoration of symmetries performed after variation is presently considered, i.e. the states $\{| \Phi^{(|g|0)} \rangle\}$ are determined {\it prior to} the MR step through repeated SR calculations. A more involved and performing approach consists of determining $| \Phi^{(|g|0)} \rangle$ through the minimization of the symmetry-restored energy ${\cal E}^{\lambda}_{|g|}$ defined below, i.e. while including the effect of the fluctuations associated with the restoration of the good symmetry~\cite{ring80a}.  

As mentioned in the introduction, a key aspect of the MR formulation provided below is that it is conducted rigorously from a {\it mathematical} viewpoint on the basis of a generic EDF kernel $E[g',g]$ that does not necessarily refer to a pseudo Hamilton operator. In particular, the restoration of symmetries is shown to be properly formulated without making any reference to a projected state~\cite{Duguet:2010cv}, which is a necessity in the general EDF context. This however does not guarantee that the MR formalism is sound from a {\it physical} standpoint as will be illustrated in Sec.~\ref{difficulties}.

\subsection{Symmetry-restored kernels}
\label{section_symmetry_restoration}

One starts by considering energy and norm kernels as two functions defined over the domain\footnote{We take advantage of property~\ref{invariancekernel} to fix one of the two phases involved to zero.} $D_{{\cal G}}$ and by decomposing them over the Irreps of ${\cal G}$ according to Eq.~\ref{decomposition_general}, i.e.
\begin{subequations}
\label{decompo_kernel}
\begin{eqnarray}
N[|g'| \, 0,|g|\, \alpha] &\equiv& \sum_{\lambda ab} \, {\cal N}^{\lambda}_{ab}[|g'|,|g|] \, \, S^{\lambda}_{ab}(\alpha) \, , \label{decompo_kernel2} \\
E[|g'| \, 0,|g|\, \alpha]  \,\, N[|g'| \, 0,|g|\, \alpha] &\equiv& \sum_{\lambda ab} \, {\cal E}^{\lambda}_{ab}[|g'|,|g|] \,\, {\cal N}^{\lambda}_{ab}[|g'|,|g|] \, \, S^{\lambda}_{ab}(\alpha) \, , \label{decompo_kernel1}
\end{eqnarray}
\end{subequations}
where the sum runs over all Irreps. Multiplying Eq.~\ref{decompo_kernel} by $S^{\lambda \, \ast}_{ab}(\alpha)$, integrating it over the domain of the group and using orthogonality relationship~\ref{orthogonality} allows one to extract the expansion coefficients associated with a specific Irrep, i.e.
\begin{subequations}
\label{projected_kernels2}
\begin{eqnarray}
\label{projected_norm_kernel2a}  {\cal N}^{\lambda}_{ab}[|g'|,|g|] &=& \frac{d_\lambda}{v_{{\cal G}}} \int_{D_{{\cal G}}}\! dm(\alpha) \, S^{\lambda \, \ast}_{ab}(\alpha) \, \, N[|g'| \, 0,|g|\, \alpha]  \,\,\, ,\\
\label{projected_energy_kernel2b} {\cal E}^{\lambda}_{ab}[|g'|,|g|]\,\, {\cal N}^{\lambda}_{ab}[|g'|,|g|] &=& \frac{d_\lambda}{v_{{\cal G}}} \int_{D_{{\cal G}}}\! dm(\alpha) \, S^{\lambda \, \ast}_{ab}(\alpha)  \, \, E[|g'| 0,|g|\, \alpha]  \,\, N[|g'| \, 0,|g|\, \alpha]  \,\,\,. 
\end{eqnarray}
\end{subequations}
The integration over $D_{{\cal G}}$ in Eq.~\ref{projected_kernels2} amounts to performing a mixing along the phase of the order parameter in order to lift the degeneracy associated with the fictitious Goldstone mode. As stated earlier, Eqs.~\ref{decompo_kernel}-\ref{projected_kernels2} prove that the extraction of the symmetry-restored energy kernel ${\cal E}^{\lambda}_{ab}[|g'|,|g|]$ can be rigorously formulated~\cite{Duguet:2010cv} on the basis of a general EDF kernel $E[g',g]$ that satisfies the minimal set of properties introduced in Sec.~\ref{constraints}, i.e. it is not necessary for such a kernel to derive from a pseudo Hamilton operator (see Sec.~\ref{pseudopotentialMR} for further discussions). In such a general situation, one cannot and should not invoke a projected state as is (incorrectly) done in standard presentations of the MR-EDF formalism. The above derivation does demonstrate that the projected state can indeed be bypassed without any difficulty.

As $S^{\lambda}_{ab}(0) = \delta_{ab}$ for any $\lambda$, setting $\alpha =0$ into Eq.~\ref{decompo_kernel} provides a sum rule relating symmetry-restored energy and norm kernels to un-rotated symmetry-breaking kernels, i.e.
\begin{subequations}
\label{sumrule}
\begin{eqnarray}
N[|g'| \, 0,|g| \, 0] &=& \sum_{\lambda a} {\cal N}^{\lambda}_{aa}[|g'|,|g|] \,\, , \label{norm_sumrule} \\
E[|g'| \, 0,|g| \, 0]  \,\, N[|g'| \, 0,|g| \, 0] &=& \sum_{\lambda a} {\cal E}^{\lambda}_{aa}[|g'|,|g|] \,\, {\cal N}^{\lambda}_{aa}[|g'|,|g|] \,\, , \label{energy_sumrule} 
\end{eqnarray}
\end{subequations}
where the independence of ${\cal E}^{\lambda}_{aa}[|g'|,|g|]$ and ${\cal N}^{\lambda}_{aa}[|g'|,|g|]$ on $a$ has not been explicitly utilized yet. Exploiting it and particularizing Eq.~\ref{sumrule} to $|g'|=|g|$ provides two sum rules
\begin{subequations}
\label{sumrule2}
\begin{eqnarray}
1 &=& \sum_{\lambda}  d_{\lambda} \,  {\cal N}^{\lambda}_{|g|} \,\, , \label{norm_sumrule2} \\
E^{SR}_{|g|}  &=& \sum_{\lambda} d_{\lambda} \, {\cal N}^{\lambda}_{|g|} \,\, {\cal E}^{\lambda}_{|g|}  \,\, , \label{energy_sumrule2} 
\end{eqnarray}
\end{subequations}
the second of which relates, for a given value of $|g|$, the SR energy to the complete set of symmetry-restored energies ${\cal E}^{\lambda}_{|g|}$. In Eq.~\ref{sumrule2} simplified notations ${\cal E}^{\lambda}_{|g|}\equiv {\cal E}^{\lambda}_{aa}[|g|,|g|]$ and ${\cal N}^{\lambda}_{|g|}\equiv {\cal N}^{\lambda}_{aa}[|g|,|g|]$ have been used.

First and foremost, sum rule~(\ref{energy_sumrule2}) provides a consistency checks in numerical codes used to extract MR energies. However, such a decomposition of the SR energy has shown to be very helpful in pinning down profound issues with the formalism when specifying to $U(1)$ symmetry. Refer to Sec.~\ref{difficulties} for the corresponding discussion.

\subsubsection{Specification to $U(1)$}
\label{sumrulessection}

Of particular interest is the specification of Eqs.~\ref{decompo_kernel}-\ref{sumrule2} to the $U(1)$ group, i.e. to particle-number restoration (PNR). Singling out the order parameter $g\equiv ||\kappa|| \, e^{i\varphi}$ associated with the breaking of nucleon number and omitting the other collective variables at play, one obtains the Fourier decomposition of the kernels
\begin{subequations}
\label{decompo_kernelN}
\begin{eqnarray}
N[||\kappa'|| \, 0 ,  ||\kappa|| \, \varphi] &\equiv& \sum_{N \in  \mathbb{Z}} \, {\cal N}^{N}[||\kappa'|| , ||\kappa||] \, \, e^{iN\varphi}  \, , \label{decompo_kernelN2} \\
E[||\kappa'|| \, 0 ,  ||\kappa|| \, \varphi]  \, \, N[||\kappa'|| \, 0 ,  ||\kappa|| \, \varphi] &\equiv& \sum_{N \in  \mathbb{Z}} \, {\cal E}^{N}[||\kappa'|| , ||\kappa||] \, {\cal N}^{N}[||\kappa'|| , ||\kappa||] \, \, e^{iN\varphi}  \, . \label{decompo_kernelN1} 
\end{eqnarray}
\end{subequations}
From a mathematical viewpoint, the sum in Eq.~\ref{decompo_kernelN} runs a priori over all Irreps of $U(1)$, i.e. over both positive {\it and} negative integers. Following Eq.~\ref{projected_kernels2}, one extracts particle-number restored kernels through
\begin{subequations}
\label{projected_kernelsN2}
\begin{eqnarray}
\label{projected_norm_kernelN2}   {\cal N}^{N}[||\kappa'|| , ||\kappa||] &=& \frac{1}{2\pi} \int_{0}^{2\pi} \! d\varphi \,
       e^{-i N \varphi} \,\,  N[||\kappa'|| \, 0 ,  ||\kappa|| \, \varphi] \,\,\, , \\
\label{projected_energy_kernelN2} {\cal E}^{N}[||\kappa'|| , ||\kappa||] \, {\cal N}^{N}[||\kappa'|| , ||\kappa||] &=& \frac{1}{2\pi} \int_{0}^{2\pi} \! d\varphi \,
       e^{-i N \varphi} \,\, E[||\kappa'|| \, 0 ,  ||\kappa|| \, \varphi]  \, \, N[||\kappa'|| \, 0 ,  ||\kappa|| \, \varphi] \,\,\, .
\end{eqnarray}
\end{subequations}
Setting $\varphi=0$ into Eq.~\ref{decompo_kernelN} provides a sum rule relating particle-number-restored energy and norm kernels to un-rotated particle-number-breaking kernels, i.e.
\begin{subequations}
\label{sumruleN}
\begin{eqnarray}
N[||\kappa'|| \, 0 ,  ||\kappa|| \, 0] &\equiv& \sum_{N \in  \mathbb{Z}} \, {\cal N}^{N}[||\kappa'|| , ||\kappa||] \, , \label{decompo_kernelN3} \\
E[||\kappa'|| \, 0 ,  ||\kappa|| \, 0]  \, \, N[||\kappa'|| \, 0 ,  ||\kappa|| \, 0] &\equiv& \sum_{N \in  \mathbb{Z}} \, {\cal E}^{N}[||\kappa'|| , ||\kappa||] \, {\cal N}^{N}[||\kappa'|| , ||\kappa||]  \, . \label{decompo_kernelN4} 
\end{eqnarray}
\end{subequations}
Further setting $||\kappa'||=||\kappa||$ provides two sum rules
\begin{subequations}
\label{sumruleN2}
\begin{eqnarray}
1 &=& \sum_{N \in  \mathbb{Z}}   {\cal N}^{N}_{||\kappa||} \,\, , \label{norm_sumruleN2} \\
E^{SR}_{||\kappa||}  &=& \sum_{N \in  \mathbb{Z}}  {\cal N}^{N}_{||\kappa||} \, {\cal E}^{N}_{||\kappa||}  \,\, , \label{energy_sumruleN2} 
\end{eqnarray}
\end{subequations}
the second of which relates, for a given value of $||\kappa||$, the SR energy to the whole set of particle-number restored energies ${\cal E}^{N}_{||\kappa||}$.

\subsubsection{Specification to $SO(3)$}
\label{sumrules}

Of particular interest is the specification of Eqs.~\ref{decompo_kernel}-\ref{sumrule2} to the $SO(3)$ group, i.e. to angular-momentum restoration (AMR). Singling out the order parameter associated with the breaking of angular momentum and omitting the other collective variables at play, one obtains the expansion of the kernels
\begin{subequations}
\label{decompo_kernelJ}
\begin{eqnarray}
N[\rho'_{\lambda\mu} \, 0 , \rho_{\lambda\mu} \, \Omega] &\equiv& \sum_{JMK} \, {\cal N}^{J}_{MK}[\rho'_{\lambda\mu} , \rho_{\lambda\mu}]\, \, {\cal D}^{J}_{MK}(\Omega) \, , \label{decompo_kernelJ2} \\
E[\rho'_{\lambda} \, 0 , \rho_{\lambda\mu} \, \Omega]  \,\, N[\rho'_{\lambda\mu} \, 0 , \rho_{\lambda\mu} \, \Omega] &\equiv& \sum_{JMK} \, {\cal E}^{J}_{MK}[\rho'_{\lambda\mu} , \rho_{\lambda\mu}]  \,\, {\cal N}^{J}_{MK}[\rho'_{\lambda\mu} , \rho_{\lambda\mu}] \, \, {\cal D}^{J}_{MK}(\Omega) \, . \label{decompo_kernelJ1} 
\end{eqnarray}
\end{subequations}
Following Sec.~\ref{section_symmetry_restoration}, one extracts angular-momentum restored kernels through
\begin{subequations}
\label{projected_kernelsJ2}
\begin{eqnarray}
\label{projected_norm_kernelJ2}  {\cal N}^{J}_{MK}[\rho'_{\lambda\mu} , \rho_{\lambda\mu}] &=& \frac{2J\!+\!1}{16\pi^2} \int_{D_{SO(3)}}\! d\Omega \, {\cal D}^{J \, \ast}_{MK}(\Omega)  \,\, N[\rho'_{\lambda\mu} \, 0 , \rho_{\lambda\mu} \, \Omega]  \,\,\, , \\
\label{projected_energy_kernelJ2} {\cal E}^{J}_{MK}[\rho'_{\lambda\mu} , \rho_{\lambda\mu}]  \,\, {\cal N}^{J}_{MK}[\rho'_{\lambda\mu} , \rho_{\lambda\mu}] &=& \frac{2J\!+\!1}{16\pi^2} \int_{D_{SO(3)}}\! d\Omega \, {\cal D}^{J \, \ast}_{MK}(\Omega)  \, \, E[\rho'_{\lambda\mu} \, 0 , \rho_{\lambda\mu} \, \Omega]  \,\, N[\rho'_{\lambda\mu} \, 0 , \rho_{\lambda\mu} \, \Omega] \,\,\, .
\end{eqnarray}
\end{subequations}
Setting $\Omega=0$ into Eq.~\ref{decompo_kernelJ} provides a sum rule relating angular-momentum restored energy and norm kernels to un-rotated angular-momentum breaking kernels, i.e.
\begin{subequations}
\label{sumruleJ}
\begin{eqnarray}
N[\rho'_{\lambda\mu} \, 0 , \rho_{\lambda\mu} \, 0] &\equiv& \sum_{JM} \, {\cal N}^{J}_{MM}[\rho'_{\lambda\mu} , \rho_{\lambda\mu}] \, , \label{norm_sumruleJ} \\
E[\rho'_{\lambda\mu} \, 0 , \rho_{\lambda\mu} \, 0]  \,\, N[\rho'_{\lambda\mu} \, 0 , \rho_{\lambda\mu} \, 0] &\equiv& \sum_{JM} \, {\cal E}^{J}_{MM}[\rho'_{\lambda\mu} , \rho_{\lambda\mu}]  \,\, {\cal N}^{J}_{MM}[\rho'_{\lambda\mu} , \rho_{\lambda\mu}] \, . \label{energy_sumruleJ} 
\end{eqnarray}
\end{subequations}
Further setting $\rho'_{\lambda\mu} = \rho_{\lambda\mu}$ provides two sum rules
\begin{subequations}
\label{sumruleJ2}
\begin{eqnarray}
1 &=& \sum_{J}  (2J\!+\!1) \, {\cal N}^{J}_{\rho_{\lambda\mu}}  \,\, , \label{norm_sumruleJ2} \\
E^{SR}_{\rho_{\lambda\mu}}  &=& \sum_{J}  (2J\!+\!1) \, {\cal N}^{J}_{\rho_{\lambda\mu}} \,  {\cal E}^{J}_{\rho_{\lambda\mu}}  \,\, , \label{energy_sumruleJ2}
\end{eqnarray}
\end{subequations}
the second of which relates, for a given value of $\rho_{\lambda\mu}$, the SR energy to the whole set of angular-momentum restored energies ${\cal E}^{J}_{\rho_{\lambda\mu}}$.

\subsection{Full fledged MR mixing}

In practice, PNR and AMR are often combined. To make formula bearable, we come back to a generic symmetry group. Starting from the symmetry-restored kernels extracted through Eq.~\ref{projected_kernels2}, one mixes the components\footnote{Such a mixing does not appear in the case of the $U(1)$ group given that its Irreps are of dimension 1.} of the targeted Irrep and further performs the mixing over the norm of the order parameter to define the MR energy through
\begin{equation}
\label{Intro_met:MR-Min} 
E_{\lambda k}^{MR} \equiv \text{Min}_{ f^{\lambda k\ast}_{|g'|a} }  \left\{ \frac{\sum_{|g|,|g'|}\sum_{a,b}  \, f_{|g'|a}^{\lambda k\ast}  \,\, f_{|g|b}^{\lambda k} \, \,\, {\cal E}^{\lambda}_{ab}[|g'|,|g|]\,\,  {\cal N}^{\lambda}_{ab}[|g'|,|g|]}{\sum_{|g|,|g'|}\sum_{a,b}  \, f_{|g'|a}^{\lambda k\ast} \,\,  f_{|g|b}^{\lambda k} \, \, \, {\cal N}^{\lambda}_{ab}[|g'|,|g|]} \right\} \,\,\,  .
\end{equation}
Mixing coefficients $f_{|g|b}^{\lambda k}$ are determined by solving the Hill-Wheeler equation of motion~\cite{Hill53} obtained as a result of minimization~\ref{Intro_met:MR-Min}
\begin{equation}
\label{hill-wheeler} 
\sum_{|g| \, b} {\cal E}^{\lambda}_{ab}[|g'|,|g|] \,\, {\cal N}^{\lambda}_{ab}[|g'|,|g|] \, f_{|g|b}^{\lambda k} = E_{\lambda k}^{MR} \sum_{|g| \, b}  {\cal N}^{\lambda}_{ab}[|g'|,|g|]  \, f_{|g|b}^{\lambda k}  \,\,\, .
\end{equation}
Equation~\ref{hill-wheeler} denotes an eigenvalue problem, expressed in a non-orthogonal basis, whose eigen-solution is nothing but the MR energy $E_{\lambda k}^{MR}$. As a matter of fact, Eq.~\ref{hill-wheeler} provides a complete set of excitation energies $\{E_{\lambda k}^{MR} ; k=0,1,2\ldots\}$ for each value of the symmetry quantum number $\lambda$. As such, one accesses the low-lying collective spectroscopy along with associated correlations in the ground state. 

\subsection{Pseudo-potential-based energy kernel}
\label{pseudopotentialMR}

In the particular case of a pseudo-potential-based EDF kernel, the MR energy (Eq.~\ref{Intro_met:MR-Min}) can be factorized into a more conventional form invoking a MR {\it wave function}. The derivation provided below does {\it not} hold when employing an EDF kernel that does not strictly derive from a pseudo Hamiltonian, e.g. for any of the modern Skyrme, Gogny and relativistic parametrizations. As such, the MR energy $E_{\lambda k}^{MR}$ cannot be expressed in terms of a MR wave-function in the most general EDF context, e.g. when using a {\it density-dependent} "Hamiltonian". Such a fact is systematically overlooked in standard presentations of the EDF theory, which constitutes a problem given the intimate connection between such a feature and the pathologies alluded to  in Sec.~\ref{difficulties}.

In virtue of Eq.~\ref{defpseudoEDF}, one can first re-express the symmetry-restored energy and norm kernels (Eq.~\ref{projected_kernels2}) according to
\begin{subequations}
\label{projected_kernels_pseudoH}
\begin{eqnarray}
\label{projected_norm_kernel2}  {\cal N}^{\lambda}_{ab}[|g'|,|g|] &=&  \langle \Phi^{(|g'|0)} | P^{\lambda}_{ab} | \Phi^{(|g|0)} \rangle  \,\,\, ,\\
\label{projected_energy_kernel2} {\cal E}^{\lambda}_{ab}[|g'|,|g|]\,\, {\cal N}^{\lambda}_{ab}[|g'|,|g|] &=& \langle \Phi^{(|g'|0)} | H_{\text{pseudo}} \, P^{\lambda}_{ab} | \Phi^{(|g|0)} \rangle  \,\,\,,
\end{eqnarray}
\end{subequations}
where the transfer operator is introduced as
\begin{equation}
P^{\lambda}_{ab} \equiv \frac{d_\lambda}{v_{{\cal G}}} \int_{D_{{\cal G}}}\! dm(\alpha) \, S^{\lambda \, \ast}_{ab}(\alpha)  \, R(\alpha) \, . \label{transferoperator}
\end{equation}
Further considering that $P^{\lambda}_{ac} P^{\zeta}_{db}= \delta_{\lambda\zeta} \delta_{cd} P^{\lambda}_{ab}$ and that $[H_{\text{pseudo}}, P^{\lambda}_{ac}]=0$, as well as that $P^{\lambda}_{ac}=(P^{\lambda}_{ca})^{\dagger}$, one can finally factorize the full fledged MR energy according to
\begin{equation}
\label{Intro_met:MR-Min_pseudo} 
E_{\lambda k}^{MR} \equiv \text{Min}_{| \Psi^{\lambda c}_{k} \rangle}  \, \left\{\frac{\langle \Psi^{\lambda c}_{k} | H_{\text{pseudo}}  | \Psi^{\lambda c}_{k} \rangle}{\langle \Psi^{\lambda c}_{k} | \Psi^{\lambda c}_{k} \rangle} \right\} \,\,\,  ,
\end{equation}
where the MR {\it wave-function} is defined by
\begin{equation}
\label{MR-WF_pseudo} 
| \Psi^{\lambda c}_{k} \rangle \equiv \sum_{|g|}\sum_{b}   f_{|g|b}^{\lambda k} \, P^{\lambda}_{cb} \, | \Phi^{(|g|0)} \rangle  \,\,\,  ,
\end{equation}
and where the mixing coefficients are obtained through Eq.~\ref{hill-wheeler}. In such a context, one recovers the textbook Hamiltonian-based GCM~\cite{ring80a} performed along the variable $|g|$ on the basis of symmetry-projected HFB wave-functions.

\subsection{Other observables}
\label{observables}

Other observables besides binding energies and low-lying excitation spectra can be extracted from MR-EDF calculations, once Eq.~\ref{hill-wheeler} has been solved. Typical quantities of interest are expectation values and transition matrix elements of electromagnetic and electroweak operators. Recently, ground-state density distributions have also been extracted~\cite{Yao:2012cx,Yao:2013tca} whereas transition densities or pair transfer form factors could be calculated in the future. 

The archetypal quantity one wishes to compute is the $B(E2)$~\cite{Bender:2008zv}
\begin{eqnarray}
B (E2; J_{k'}' \to J_k) & = & \frac{e^2}{2J'+1}
      \sum_{M =-J }^{+J }
      \sum_{M'=-J'}^{+J'}
      \sum_{\mu=-2}^{+2}
      | \langle \Psi^{JM}_{k} | Q_{2 \mu} | \Psi^{J'M'}_{k'}  \rangle \big|^2 \, , \label{BE2}
\end{eqnarray}
where the electric quadrupole moment operator $Q_{2\mu} = e \,\sum_p r_p^2 \, Y_{2\mu}(\Omega_p)$ is written for point protons with
their bare electric charge $e$. Independent of whether one uses a pseudo-potential EDF kernel or not, auxiliary observables are computed as  matrix elements of bare operators in between MR wave-functions. The latter can always been built according to Eq.~\ref{MR-WF_pseudo} as soon as Eq.~\ref{hill-wheeler} is solved to extract $f_{|g|b}^{\lambda k}$.  In view of the overall accuracy of the method, the current agreement of computed, e.g., $B(E2)$ or $B(E3)$ values with experimental data is considered to be reasonably good and justifies this common practice. Would the accuracy of the method improve significantly, one could consider going beyond such a paradigm by, e.g., designing density functional kernels for auxiliary observables as well. 

In the present context, computing Eq.~\ref{BE2} eventually boils down to evaluating the matrix element of a tensor operator, e.g. $Q_{2 \mu}$, in between two reference states on which different transition operators are applied. Coming back to our general notations, this corresponds to computing
\begin{eqnarray}
\langle \Phi^{(|g'|0)} | P^{\lambda'}_{a'c'} T^{\lambda''}_{\mu} P^{\lambda}_{ca}
| \Phi^{(|g|0)} \rangle & = & \frac{2\lambda+1}{2\lambda'+1} \; (\lambda \lambda'' \lambda' | c \mu c' )
      \sum_{\nu =-\lambda}^{+\lambda} (\lambda \lambda'' \lambda' | a, a - \nu, \nu) \; \nonumber \\
&   & \quad \times
       \langle \Phi^{(|g'|0)} | P^{\lambda'}_{a' \nu} \; T^{\lambda''}_{a - \nu} | \Phi^{(|g|0)} \rangle
     \, ,
\end{eqnarray}
where the matrix element appearing on the right-hand side can eventually be evaluated, after expanding $P^{\lambda'}_{a' \nu}$ according to Eq.~\ref{transferoperator}, on the basis of the generalized Wick theorem~\cite{balian69a}.

\subsection{Dynamical correlations}
\label{GS_correlations}

Let us now summarize the way correlations are incorporated in the nuclear EDF approach. The power of the method relies on (i) the parametrization of the "bulk" of correlations, i.e. the part of the binding energy that varies smoothly with neutron and/or proton numbers, under the form of a functional of the one-body density matrices and on (ii) the grasping of correlations that vary quickly with the filling of nuclear shells through the breaking of symmetries along with the subsequent treatment of the fluctuations of the associated order parameters. Incorporating the second type of correlations within symmetry-conserving approaches, e.g. the CI method, would necessitate tremendous computational efforts in heavy open-shell nuclei. 

\begin{table}[hb]
\caption{\label{scalescorrelations} Schematic classification of correlation energies as they naturally appear in the nuclear EDF method. The quantity $A_{{\rm val}}$ denotes the number of valence nucleons while $G_{{\rm deg}}$ characterizes the degeneracy of the valence major shell.}
\small\rm
\begin{tabular}{|l|l|l|l|}
\hline
Correlation energy & Treatment & Scales as &  Varies with \\
\hline
Bulk & Summed into EDF kernel & $\sim 8\, A$ MeV  & $A$ \\
Static collective & Finite order parameter $|g|$  & $\lesssim 25$ MeV  & $A_{{\rm val}}, G_{{\rm deg}}$  \\
Dynamical collective & Fluctuations of $g$ & $\lesssim 5$ MeV &   $A_{{\rm val}}, G_{{\rm deg}}$ \\
\hline
\end{tabular}
\end{table}

\begin{figure}[htbp]
\begin{center}
\includegraphics[width = 0.45\textwidth, keepaspectratio]{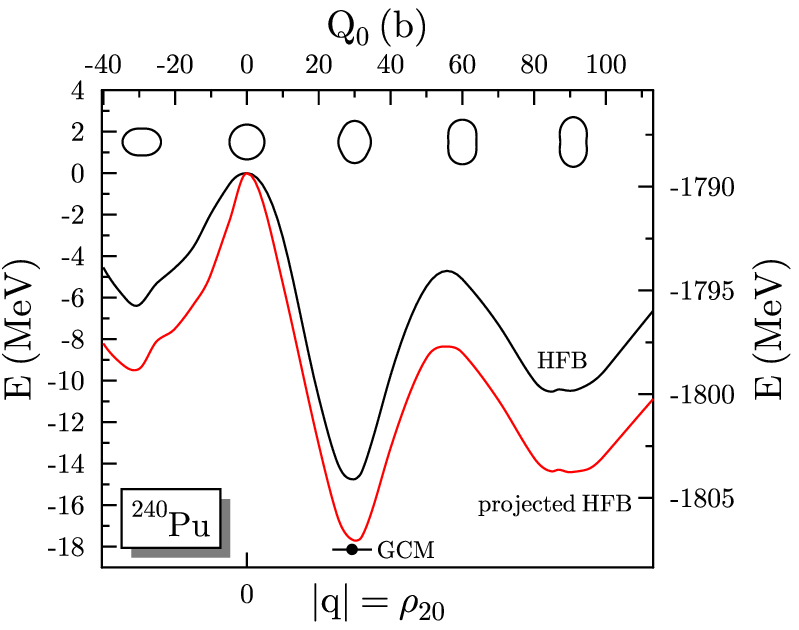} \hspace{0.7cm} \includegraphics[width = 0.45\textwidth, keepaspectratio]{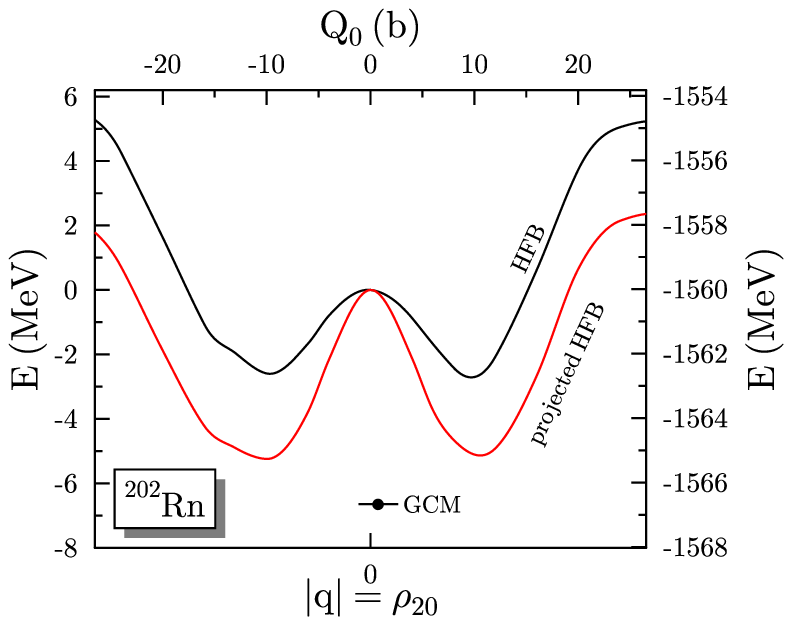} \includegraphics[width = 0.45\textwidth, keepaspectratio]{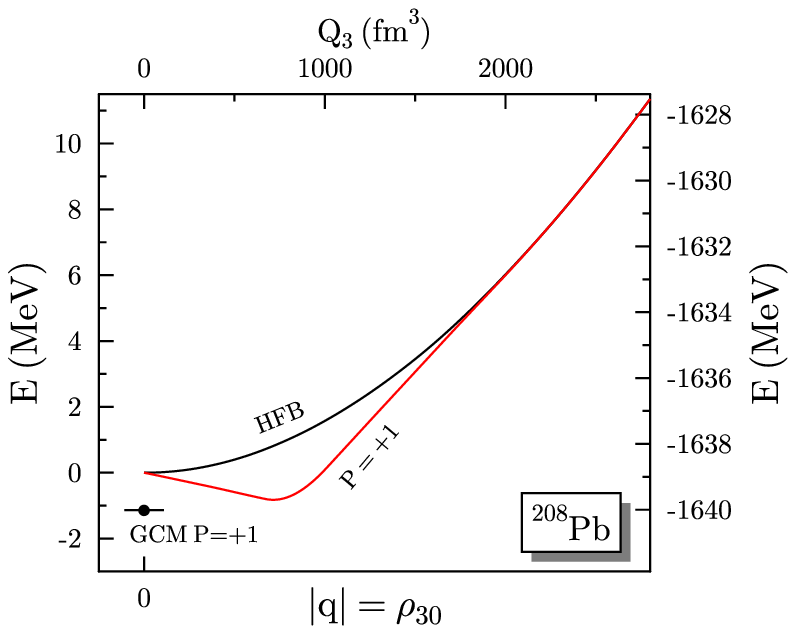}  \hspace{0.7cm} \includegraphics[width = 0.45\textwidth, keepaspectratio]{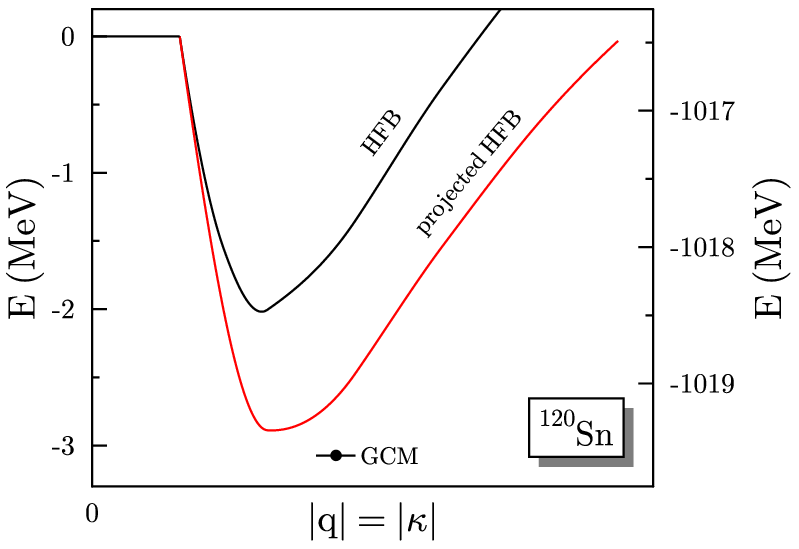}
\caption{Upper panels: energy of $^{240}$Pu and $^{202}$Rn as a function of the axial quadrupole degree of freedom ($|g|\equiv \rho_{20}$): single-reference calculation (full black line), with the added effect of particle number and ($J=0$) angular momentum restorations (full red line) as well as of the shape mixing along $|g|\equiv \rho_{20}$ (black circle labelled as "GCM"). Lower left panel: energy of $^{208}$Pb as a function of the axial octupole degree of freedom ($|g|\equiv \rho_{30}$): single-reference calculation (full black line), with the added effect of (positive) parity restoration (full red line) and  mixing of shapes along $|g|\equiv \rho_{30}$ (black circle labelled as "GCM"). Lower right panel: energy of $^{120}$Sn as a function of the pairing degree of freedom ($|g|\equiv ||\kappa||$): single-reference calculation (full black line), with the added effect of neutron number restoration (full red line) and mixing along $|g|\equiv ||\kappa||$ (black circle labelled as "GCM"). Left vertical axes are rescaled with respect to the symmetry conserving, i.e. non-deformed, reference point. Please note that $|q|$ stands for $|g|$ in the present figure. Taken from Ref.~\cite{bender13a}.}
\label{FIGcorrelations}
\end{center}
\end{figure}

Of course, the success of the approach eventually relies on the validity of the empirical decoupling between the bulk of correlations and those that are more explicitly accounted for. To some extent, the different scales that characterize these two categories of correlations play in favour of such an empirical decoupling. Let us come back to the four nuclei considered in Fig.~\ref{PES} to illustrate this point. Figure~\ref{FIGcorrelations} separates the binding energy of $^{240}$Pu, $^{202}$Rn, $^{208}$Pb and $^{120}$Sn into 
\begin{enumerate}
\item the symmetry conserving SR energy, 
\item the symmetry-unrestricted SR energy, 
\item the symmetry-restored MR energy,
\item the full fledged MR-EDF energy. 
\end{enumerate}
The symmetry conserving SR-EDF result (full black line at $|g|=0$) provides the "bulk" part of the energy and accounts for, at least, $98\%$ of the binding energy. Authorizing the breaking of symmetries (absolute minimum of the full black line) does not bring anything to stable double closed-shell nuclei such as $^{208}$Pb. However, the spontaneous breaking of rotational symmetry brings up to $20$~MeV correlation energy in heavy double open-shell nuclei such as $^{240}$Pu, which accounts for about $2\%$ of the binding. In a transitional nucleus such as $^{202}$Rn, the symmetry breaking only accounts for $2$~MeV but it signals that such a nucleus should not even be considered at the SR level because of the anticipated large amplitude fluctuations.  Superfluidity associated with the breaking of neutron and/or proton numbers typically accounts for $2$\,MeV in singly-open shell nuclei such as $^{120}$Sn. Most important, including pairing is mandatory to describe other observables, e.g. the odd-even mass staggering, individual excitations of even-even nuclei or the moment of inertia of rotating systems. Restoring symmetries (absolute minimum of the full red line) brings in additional correlations, even in nuclei whose SR minimum is symmetry conserving.  Typically, restoring angular momentum ($^{240}$Pu and $^{202}$Rn), parity ($^{208}$Pb) or neutron number ($^{120}$Sn) add between $1$\,MeV and $3$\,MeV correlation energy, depending on how much the symmetry is broken in the first place. Last but not least, the fluctuations of $|g|$ ("GCM" circle) differentiate nuclei that are stiff (i.e. $^{240}$Pu, $^{208}$Pb, $^{120}$Sn) from those that are soft (e.g. $^{202}$Rn) with respect to the collective degree of freedom under study. While the correlation energy is of the order of one or two hundreds keV in the former, it can be as large as $1$\,MeV in the latter. Although the examples discussed here are only illustrative, they are quite representative of the various behaviours one may encounter. Eventually, Tab.~\ref{scalescorrelations} recall the various categories of correlations at play and summarizes schematically the scale and the scaling that characterize them. For systematic studies on how correlations impact binding energies and other observables in the context of MR-EDF calculations, see Refs.~\cite{Bender:2005ri,Bender:2008gi,Robledo:2011nf}.

\subsection{State-of-the art calculations}

As of today, full fledged MR-EDF calculations are limited to even-even nuclei. In their most advanced form, they simultaneously restore neutron number, proton number and angular momentum from triaxially deformed Bogoliubov states and further perform the mixing of quadrupole shapes ($|g|=\rho_{2\mu}$ with $\mu=-2,0,2$). Such calculations are available for non-relativistic Skyrme~\cite{Bender:2008zv} and Gogny~\cite{Rodriguez:2010by} functionals as well as for relativistic Lagrangians~\cite{Yao:2009zv}. Still, those cutting-edge calculations are currently limited to light nuclei such that approximations are needed (e.g. limiting oneself to axially deformed shapes) to tackle heavy nuclei. An important effort is also being pursued to restore both good angular momentum and isospin from triaxially deformed Slater determinants~\cite{Satula:2009sx}. This is relevant to the evaluation of isospin mixing and isospin-breaking corrections to super-allowed $\beta$-decay in view of testing the unitarity of the CKM matrix~\cite{Satula:2012zz}. The versatility of the method also permits to address delicate questions such as the quest of neutrino-less double $\beta$-decay to pin down the Dirac or Majorana character of neutrinos~\cite{Rodriguez:2012mx}.

The current forefront corresponds to extending MR-EDF schemes in several (complementary) directions. First and foremost, it is crucial to have the ability to perform MR-EDF calculations of odd-even and odd-odd nuclei. This poses a great technical challenge~\cite{Bally:2011iz} but will extend the reach of the method tremendously and greatly enhance the synergy with upcoming experimental studies. Along the same line, MR-EDF schemes must be extended such as to include {\it diabatic} effects~\cite{duguet02thesis}, i.e. configurations generated through an even number of quasi-particle excitations. This is expected to improve significantly the description of, e.g., the first $2^{+}$ excited state in near-spherical nuclei and to allow a clean description of K isomers. Also of importance is the implementation of the MR method on the basis of references states generated through {\it cranked} SR calculations, i.e. calculations employing a constraints on $\langle \Phi^{(g)} |J_{x,y,z}| \Phi^{(g)}\rangle\neq 0$~\cite{baye84a,Zdunczuk:2006qh,avez13a}. By accounting for Coriolis effects, this is expected to improve moments of inertia that are systematically too low in MR calculations based on uncranked states. Eventually, state-of-the-art calculations should combine quadrupole and octupole degrees of freedom~\cite{meyer95} as well as the mixing over $||\kappa||$~\cite{Bender:2006tb,Vaquero:2011hq}. The latter also impacts moment of inertia significantly and authorizes the description of pairing fluctuations and pairing vibrations near closed shell, as well as the computation of pair transfer overlap functions.

All such extensions are particularly timely given that upcoming RIB facilities are accessing an increasingly larger number of short-lived atomic nuclei. Among the latter, exotic nuclei with a large neutron excess are likely to require more systematically the inclusion of MR correlations from the outset, i.e. to be less-good "mean-field" nuclei than those located near the valley of $\beta$ stability.

\subsection{Approximations to full fledged MR-EDF}
\label{approxMR}

Several approximations to or variants of the full fledged MR-EDF approach are being pursued with great success. It is beyond the scope of the present lecture notes to review them. Let us however mention the most important ones and refer the reader to recent associated works. 

The quasi-particle random phase approximation (QRPA) that can be motivated in many different ways, one of which is the approximation of the MR kernels in the limit where  $| \Phi^{(g')} \rangle$ and $| \Phi^{(g)} \rangle$ differ harmonically from a common reference state~\cite{jancovici64,brink68}. Quasi-particle random phase approximation, along with its extensions, provides vibrational excitations of various multipolarities and associated ground-state correlations. This includes low-lying states as well as giant resonances. A limitation of such an approximation is its inability to describe violently anharmonic systems undergoing large amplitude motion. There is a significant on-going effort to develop the method in deformed nuclei~\cite{Peru:2008gd,Yoshida:2009jn,pena09a,Losa:2010bm,Terasaki:2010zg} on the basis of complete EDF parametrizations and efficient algorithms~\cite{Nakatsukasa:2007qj,Toivanen:2009mr,Avogadro:2011gd}. This will permit to address many upcoming challenges including the quest of potentially new exotic vibrational modes~\cite{Paar:2010ww}.

Second is the collective (e.g. Bohr) Hamiltonian that can be motivated in two different ways,  one of which is the (topological) Gaussian overlap approximation~\cite{reinhard87,Hagino:2002ya,Rohozinski:2012mu} of the transition EDF kernels. In practice, however, inertia parameters are not computed from available full fledged MR-EDF calculations. Indeed, the latter are not complete enough at this point in time to compute inertia parameters reliably. Five-dimensional collective Hamiltonians built from non-relativistic Skyrme~\cite{prochniak04a,Prochniak:2009fs} and Gogny~\cite{Libert:1999sw,Delaroche:2009fa} functionals as well as from relativistic Lagrangians~\cite{Niksic:2008cn} are available. Work is currently being pursued to improve on the Inglis-Belyaev moments of inertia and cranking mass parameters by means of Thouless Valentin~\cite{Li:2012mn,Hinohara:2012zz}. Within such a scheme, low-lying collective spectra of heavy even-even nuclei can be computed while including the full quadrupole dynamics.

Last but not least, it is worth mentioning the recent revival of the interacting boson model (IBM) within a microscopic setting, i.e. based on the mapping of triaxial HFB energy landscapes generated from a Gogny functional~\cite{Nomura:2010gg} or a relativistic Lagrangian~\cite{Nomura:2011xk}. Such a method allows the efficient description of low-lying collective spectra of complex heavy nuclei.

As for full fledged MR-EDF calculations, modern accounts of the three above methods are only available for even-even nuclei. Extensions to odd-even and odd-odd nuclei must be envisioned in the future.

\subsection{Pathologies of MR-EDF calculations}
\label{difficulties}

\begin{figure}
\includegraphics*[width=2.95in,clip=]{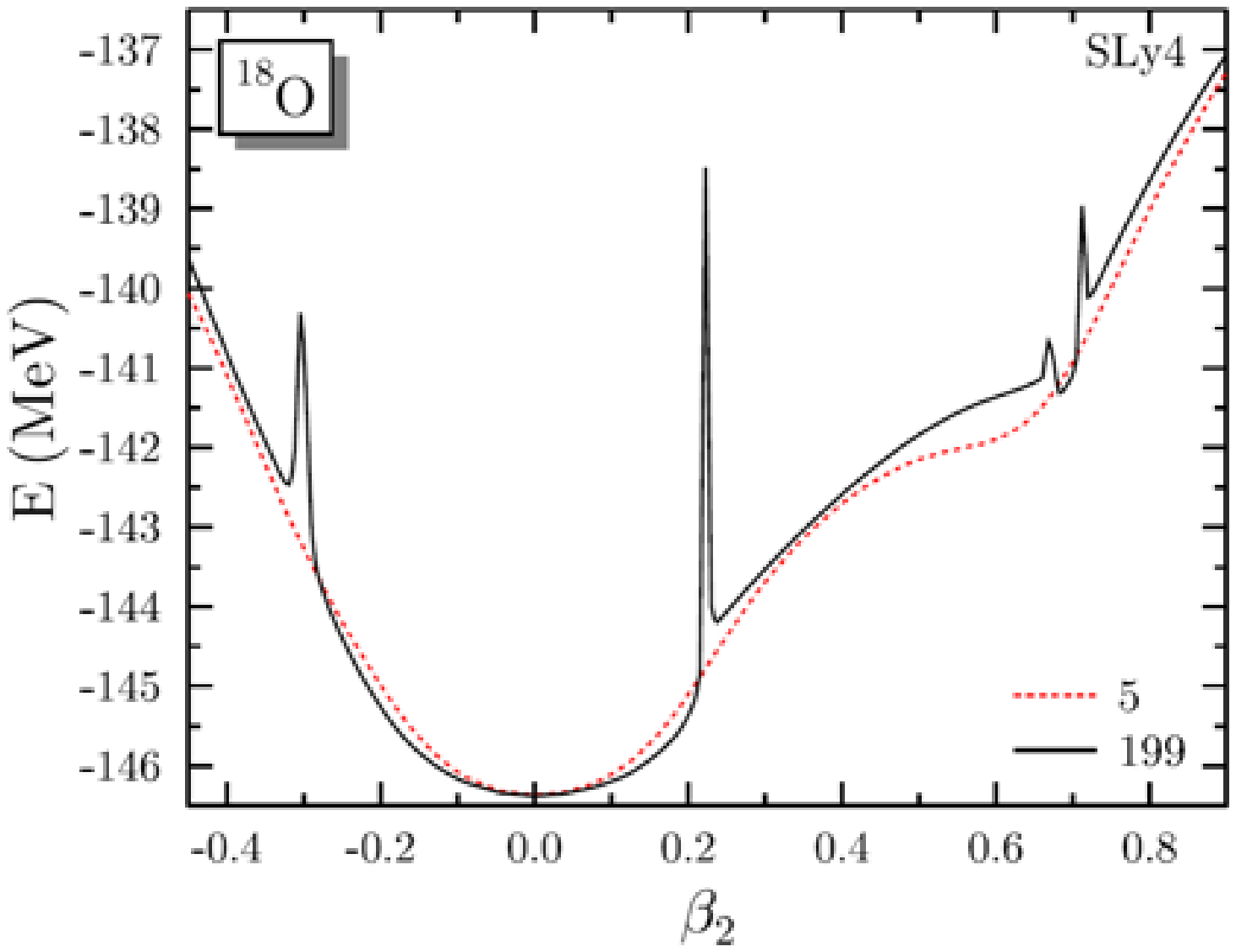}
\includegraphics*[width=2.95in,clip=]{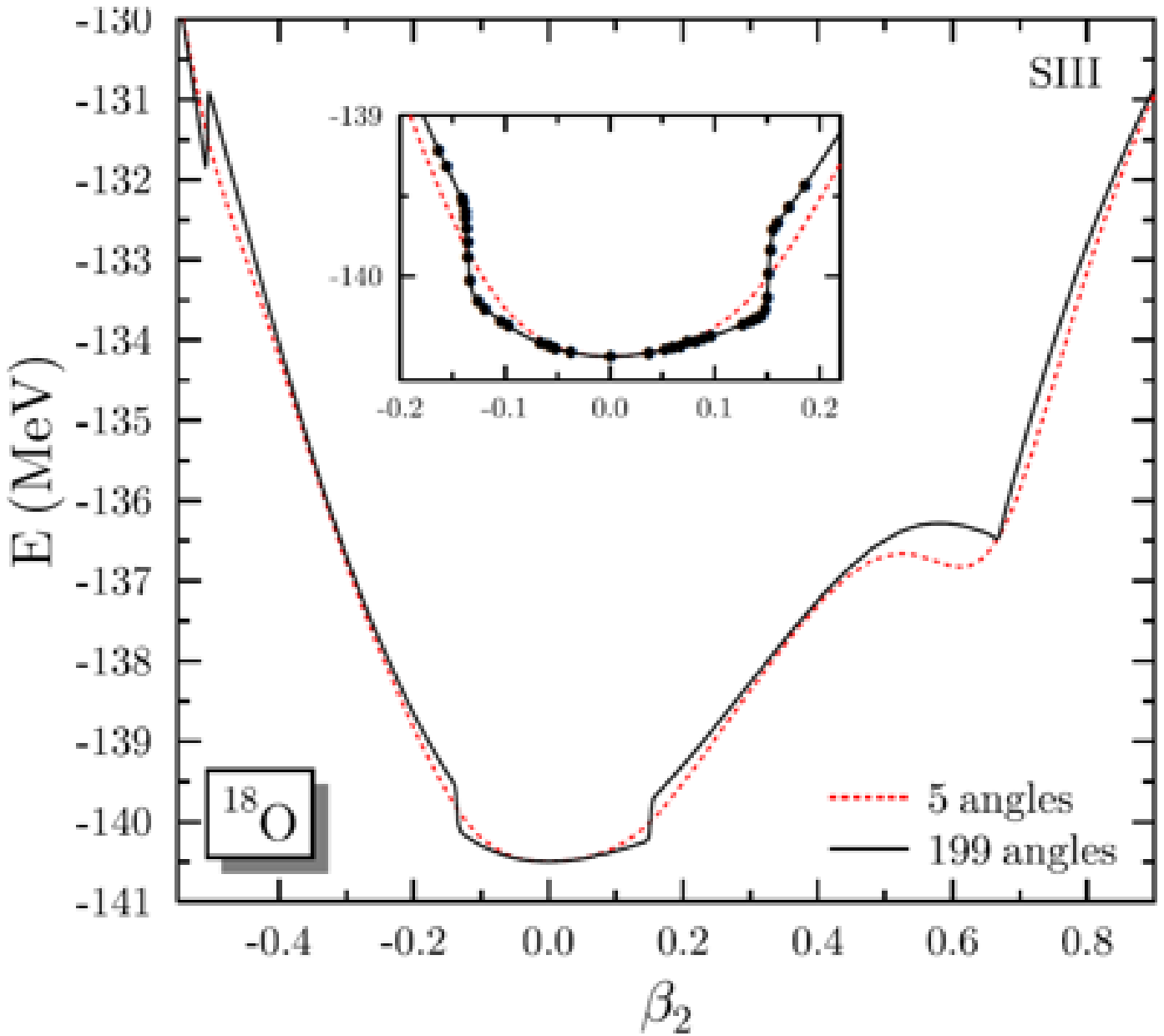}
\caption{\label{divergence}
(Color online)
Proton-number restored energy ${\cal E}^{Z}_{\rho_{20}}$ of $^{18}$O as a function of the axial quadrupole deformation ($\beta_2$ is a dimensionless measure of $\rho_{20}$) using 5 and 199 discretization points in the integral over the gauge angle (Eq.~\ref{projected_kernelsN2}). Left panel: calculations performed with the
SLy4 Skyrme parametrization and a density-independent pairing interaction. Right panel: calculations performed with the
SIII Skyrme parametrization and a density-independent pairing interaction. Taken from Ref.~\cite{Bender:2008rn}.
}
\end{figure}

In spite of the mathematically sound formulation of the MR-EDF method provided above, pathologies were identified under the form of spurious divergences~\cite{almehed01a,anguiano01a} and steps~\cite{dobaczewski07} in potential energy curves obtained from PNR calculations. Examples are given in Fig.~\ref{divergence} for two different Skyrme parametrizations of the EDF kernel. The occurrence of such anomalies were analysed in details in Refs.~\cite{dobaczewski07,Bender:2008rn,Duguet:2008rr} and put in connection with non-analyticities of the energy kernel over the complex plane, after performing the continuation $z=e^{i\varphi}$, where $\varphi$ denotes the gauge angle characterizing the off-diagonal energy kernel at play (see Sec.~\ref{energy_kernel}). In particular, the problem manifests differently depending on the analytical structure of the EDF kernel~\cite{Bender:2008rn}. The left panel of Fig.~\ref{divergence} is characteristic of the general case where divergences occur whenever a proton and/or neutron single-particle level crosses the Fermi energy~\cite{dobaczewski07}. Additionally, the potential energy surface displays finite steps across any such divergence. The right panel of Fig.~\ref{divergence} illustrates the particular case of a functional that is strictly bilinear in the density matrices of a given isospin species. In such a situation, no divergence occurs and one is only left with finite discontinuities. 

A step towards the formulation of a remedy to the problem was made in Refs.~\cite{Lacroix:2008rj,Bender:2008rn,Duguet:2008rr}. Firstly, the problem was shown to relate to the breaking of Pauli's principle discussed in Sec.~\ref{pseudopotential}. Specifically, spurious contributions associated with self-interaction and self-pairing processes are multiplied with dangerous weights in the {\it off-diagonal} energy kernel $E[g',g]$, which results in the anomalies illustrated in Fig.~\ref{divergence}. Secondly, divergences and steps were shown to constitute the visible part of the problem only, i.e. PNR energies are not only contaminated where divergences and steps occur but also  {\it away} from them.

\begin{figure}
\includegraphics*[width=2.95in,clip=]{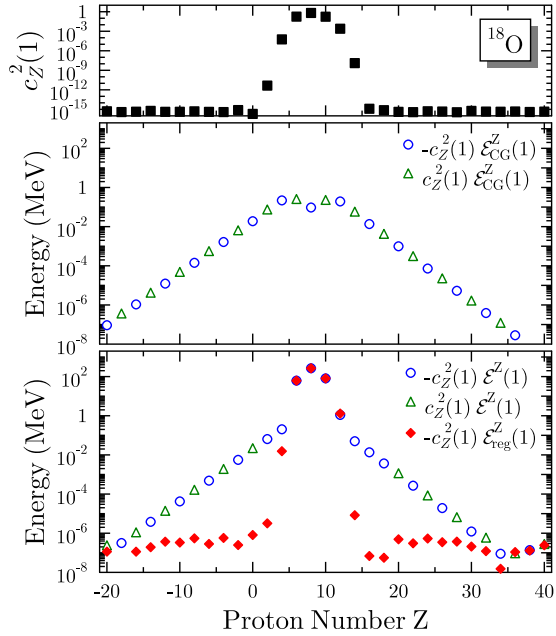}
\caption{\label{fig:decomposition:r1.0}
(Color online) Proton-number-restored kernels as a function of the $Z$ one restores. Upper panel: norm kernel ${\cal N}^{Z}$. Middle panel: spurious contribution to the (weighted) energy kernels. Lower panel: uncorrected  ${\cal N}^{Z} \, {\cal E}^{Z}$ and corrected  ${\cal N}^{Z} \, {\cal E}^{Z}_{\text{REG}}$ proton-number-restored energy kernels. All results are
obtained using the same SR state calculated for $^{18}$O at a deformation of $\beta_2 = 0.371$. The neutron number is not restored. Taken from Ref.~\cite{Bender:2008rn}.
}
\end{figure}

Another striking manifestation of spurious self-interaction and self-pairing processes in PNR calculations was identified in Ref.~\cite{Bender:2008rn}. Whereas contributions to sum rule~\ref{energy_sumruleN2} corresponding to $N\leq 0$ are zero in the absence of self-interaction and self-pairing, i.e. when working within the pseudo-potential-based approach, non-analyticities of the energy kernel over the complex plane translate into\footnote{The overlap kernel being analytical over the complex plane, it is straightforward to prove that ${\cal N}^{N}=0$ for $N\leq 0$.} having ${\cal N}^{N} \, {\cal E}^{N}\neq 0$ for $N\leq 0$. Such a feature is illustrated in Fig.~\ref{fig:decomposition:r1.0} for the interaction energy part (i.e. the kinetic energy contribution is omitted) obtained from PNR calculation of $^{18}$O. The distribution of absolute values of ${\cal N}^{Z} \, {\cal E}^{Z}$ as a function of $Z$ does not follow the distribution of the weights ${\cal N}^{Z}$ displayed in the upper panel. Instead, it has a long tail that spreads visibly to $Z = -20$ and $Z = 34$, before it cannot be distinguished from numerical noise anymore. In these tails, ${\cal N}^{Z} \, {\cal E}^{Z}$ displays alternating signs, which is clearly unphysical.

The fact that PNR calculations do provide non-zero (weighted) energies for negative or null particle numbers is certainly the most illuminating proof that having a mathematically well-founded formalism is necessary but not sufficient to make it physically meaningful, i.e. while mathematics makes sum rule~\ref{energy_sumruleN2} run over all Irreps a priori, physics requires that the expansion coefficients associated with negative integers are zero, which is not guaranteed in general and is not the case for {\it any} existing modern parametrization of the EDF kernel. 

\begin{figure}
\includegraphics*[width=0.75\textwidth,clip=]{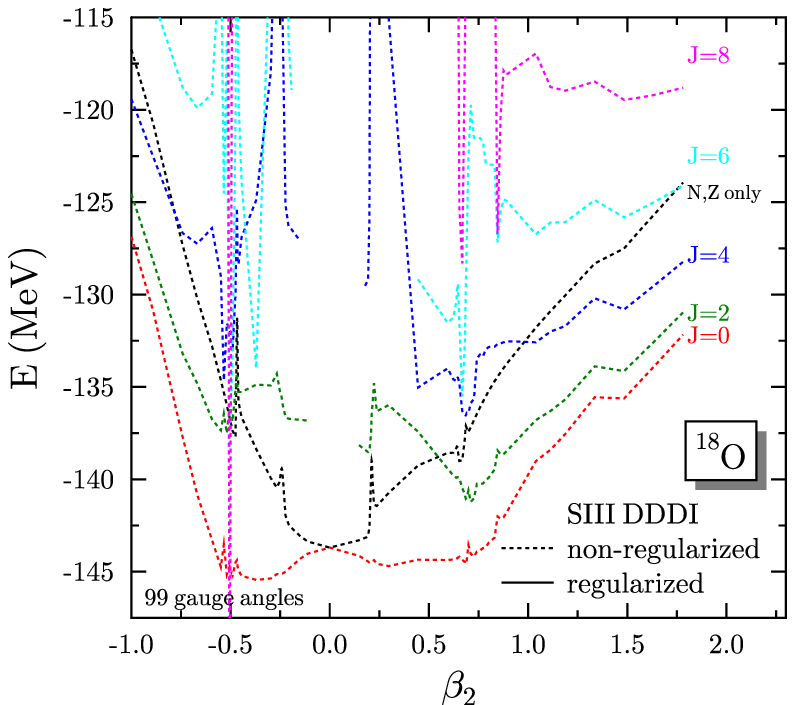}
\caption{(Color online) Proton-number- and angular-momentum-restored energies of $^{18}$O for various values of $J$ as a function of the axial quadrupole deformation. The integral over the gauge angle (Eq.~\ref{projected_kernelsN2}) uses 99 discretization points. Calculations are performed with the SIII Skyrme parametrization and a density-dependent pairing interaction. Solid lines defined in the legend are not shown in the present figure but are (will be) visible in the original reference~\cite{bender12a}. The curve labelled with "N,Z only" only performs the restoration of particle number.}
\label{018nonregularized1}
\end{figure}

Although most clearly highlighted through PNR calculations, i.e. in calculations realizing the mixing over the gauge angle, pathologies due to the violation of Pauli's principle contaminate {\it any} type of MR mixing. Figure~\ref{018nonregularized1} displays the result of a MR-EDF calculation of $^{18}$O including both PNR and AMR, and compares it to the result obtained via PNR only.

It is interesting to note at this point that certain approximations to full fledged MR-EDF calculations~\cite{bender03b}, i.e. calculations based on a collective Hamiltonian or on QRPA, avoid the dramatic pathologies discussed above by bypassing the problem from the outset, i.e. thanks to the approximation to the off-diagonal kernels that define them. However, such methods are not free from less dramatic, i.e. smooth and finite, contaminations associated with the presence of spurious self-interaction and self-pairing in the energy kernel. This question deserves attention in the future.

\subsection{Towards pseudo-potential-based energy kernels}

In order to resolve the difficulties illustrated above, a regularization of the off-diagonal energy kernel was designed for parametrizations that are strictly polynomial in the density matrices~\cite{Lacroix:2008rj}. The method was meant to eliminate a posteriori the pathologies contaminating MR-EDF calculations without fully enforcing the Pauli principle from the outset. Exposing the regularization method is beyond the scope of the present document and we refer the reader to Ref.~\cite{Lacroix:2008rj} for details. As of today, the regularization method has been implemented not only in pure PNR calculations~\cite{Bender:2008rn} but also for the most general MR-EDF calculations available~\cite{bender12a}. This includes the most advanced ones aiming at the description of odd nuclei~\cite{Bally:2011iz}. In spite of solving the problem for pure PNR calculations,  the regularization method leaves implementations that go beyond it, e.g. calculations mixing PNR and AMR, with unwanted pathologies~\cite{bender12a}. 

As of today, the only viable route to a sound MR-EDF formalism relies on energy kernels that strictly derive from a pseudo-potential~\cite{sadoudi11thesis}, i.e. kernels that enforce the Pauli principle from the outset to bypass spurious self-interaction and self-pairing processes. Several efforts~\cite{sadoudi11thesis,Sadoudi:2012jg,Dobaczewski:2012cv,Bennaceur:2013fua} in this direction are currently being pursued as alluded to in Sec.~\ref{modernparam}. This constitutes a turning point in the construction of nuclear EDF parametrizations. It is beyond the scope of the present lecture notes to expose such developments. Let us however briefly explain why such a route is not straightforward to follow. As a matter of fact, {\it none} of the modern, i.e. Skyrme, Gogny or relativistic, parametrizations belong to the category of {\it strict} pseudo-potential-based EDF kernels. The reason for such a situation is precisely that practitioners have moved away from the strict pseudo-potential-based philosophy throughout the last four decades because of its apparent lack of flexibility and its inability to produce high-quality EDF parametrizations. The challenge is thus to develop pseudo-potentials that are more general than those considered in the past such that they can provide a high-quality phenomenology. The pseudo potentials must however be simple enough for the fit of its free parameters to be meaningfully handled. Several new families of EDF parametrizations strictly deriving from pseudo potentials and allowing for safe MR-EDF calculations can be expected to be published in the coming years.

\subsection{Towards non-empirical energy kernels}
\label{nonempiricalEDF}

On the longer term, it is mandatory to go beyond the empirical formulation of the nuclear EDF method in order to augment its predictive power. This requires the design of ab-initio many-body methods from which both SR- and MR implementations of the EDF method, i.e. both diagonal and off-diagonal energy functional kernels, can be derived through a set of controlled approximations. This is meant to lead to so-called \emph{non-empirical} energy functionals possessing a link to the underlying nuclear Hamiltonian describing few-body scattering and bound-state observables.  The objective is not to replace but rather complement the development of empirical EDFs based on trial and error by combining the predictive character of an ab-initio method with the gentle numerical scaling of EDF calculations. Indeed, while empirical EDFs already achieve an accuracy for known observable that will be difficult, if not impossible, to reach with purely non-empirical functionals, they lack predictive power away from the experimentally known region of the nuclear chart.

The first way to improve on such limitations consists of using "pseudo-data" generated from ab-initio calculations for nuclei located in the experimentally unknown region (i) for the fitting procedure of EDF parametrizations and (ii) to benchmark extrapolations from such EDF parametrizations. In this way, unknown couplings of the empirical EDF parametrization can be "microscopically" constrained. Eventually, the goal is to discriminate between different functional forms. The benefit of such an indirect approach is that any ab-initio method that can provide precise enough benchmarks for the systems and observables of interest can be employed. However, no direct/explicit connection with vacuum interactions is realized such that no specific insight about the {\it form} of new functional terms that could capture the missing physics is easily gained in this way, i.e. the predictive power of EDF calculations away from the benchmarks remains bound to the quality of the postulated functional form such that improvements still rely on trial and error.

A greater challenge is to connect explicitly the {\it form} of the energy functional kernel, in addition to the {\it value} of its couplings, to vacuum nuclear interactions. One is essentially looking for {\it microscopically-educated guesses}. Ground-breaking, though very incomplete, works in this direction have been undertaken recently~\cite{Gebremariam:2009ff,Gebremariam:2010ni,Kaiser:2010pp,Holt:2011nj,Holt:2011jd,Kaiser:2012mn}. Eventually, a fine-tuning of the couplings, within the intrinsic error bars with which they will have been produced, can be envisioned~\cite{Stoitsov:2010ha}. In this context, microscopically-educated functionals are to be derived through analytical approximations of the ground-state energy computed via a given ab-initio method of reference (preferably the same as the one providing benchmarks for observable quantities). It is a challenging task whose complexity depends on the nuclear Hamiltonian and the many-body method one starts from. In particular, ab-initio methods that are amenable to such a mapping must share certain key features of the nuclear EDF method, the most important of which being the notion of spontaneous symmetry breaking. Let us take the part of the EDF that drives superfluidity as an example, i.e. the part that depends on the anomalous pairing tensor $\kappa^{g'\!g}_{ij}$ (see Sec.~\ref{sectionsymbreaking}). Such a functional dependence of the EDF kernel exists only because pairing correlations are grasped through the breaking of good particle-number associated with $U(1)$ gauge symmetry. Deriving microscopically-educated EDF kernels can thus only be achieved starting from an ab-initio method that also incorporates pairing correlations through the breaking of U(1) gauge symmetry.

\section{Conclusions}

Very significant advances have been made in the last 15 years within the frame of the nuclear energy density functional method. In doing so, the focus of the field has shifted in several respects, with the consequences that
\begin{enumerate}
\item routine applications have moved from SR to MR calculations,
\item one can address, e.g. neutron-rich, nuclei that do not fit the mean-field paradigm,
\item applications are now equally dedicated to ground and excited states,
\item one can provide both
\begin{enumerate}
\item the detailed quantitative picture of a given system of interest,
\item study trends through large-scale MR calculations,
\end{enumerate}
\item advances in the field are bound to making consistent progress regarding
\begin{enumerate}
\item the foundations of the approach and its formal consistency,
\item the rooting of EDFs into basic many-body methods and interactions,
\item the building of EDFs from improved fitting protocols,
\item the building of EDF parametrizations from enlarged data sets,
\item the  further development of powerful numerical tools,
\end{enumerate}
while points (a), (b) and (c) were essentially discarded 15 years ago,
\item applications more strongly impact astrophysics and particle physics.
\end{enumerate}
The field is expected to move forward in these directions in the next 10 years. Most probably, this will be the era of the strong overlapping with emerging ab-initio methods for mid-mass nuclei and of the materialization of powerful numerical tools dedicated to the description of odd-even and odd-odd nuclei. In addition to these already on-going trends, one can expect surprises to emerge that will guide the development of the EDF methods in new directions.

\section{Acknowledgments}

It is a great pleasure to thank deeply all those I have had the chance to collaborate with on topics related to the matter of the present lecture notes, i.e. B. Avez, M. Bender, K. Bennaceur, P. Bonche, B. A. Brown, P.-H. Heenen, D. Lacroix, T. Lesinski, J. Meyer, V. Rotival, J. Sadoudi, N. Schunck and C. Simenel. I also wish to thank M. Bender for providing me with several of the figures that are used in the present lecture notes.

\begin{appendix}

\section{$F$-functions}
\label{sec:F-functions}

Kinetic densities are expressed in INM in terms of functions $F^{(0)}_m(I_\tau,I_\sigma,I_{\sigma\tau})$, $F^{(\tau)}_m(I_\tau,I_\sigma,I_{\sigma\tau})$, $F^{(\sigma)}_m(I_\tau,I_\sigma,I_{\sigma\tau}) $ and $F^{(\sigma\tau)}_m(I_\tau,I_\sigma,I_{\sigma\tau}) $ defined through~\cite{bender02a}
\begin{subequations}
\begin{align}
F^{(0)}_{m}%(I_\tau,I_\sigma,I_{\sigma\tau}) 
\equiv& \frac{1}{4} 
\Big[ (1+I_\tau+I_\sigma+I_{\sigma\tau})^m 
+ (1+I_\tau-I_\sigma-I_{\sigma\tau})^m
\nonumber \\ &  
+(1-I_\tau+I_\sigma-I_{\sigma\tau})^m
+(1-I_\tau-I_\sigma+I_{\sigma\tau})^m \Big]
, \\
F^{(\tau)}_{m}%(I_\tau,I_\sigma,I_{\sigma\tau}) 
\equiv& \frac{1}{4} 
\Big[ (1+I_\tau+I_\sigma+I_{\sigma\tau})^m 
+ (1+I_\tau-I_\sigma-I_{\sigma\tau})^m
\nonumber \\ & 
-(1-I_\tau+I_\sigma-I_{\sigma\tau})^m
-(1-I_\tau-I_\sigma+I_{\sigma\tau})^m \Big]
, \\
F^{(\sigma)}_{m}%(I_\tau,I_\sigma,I_{\sigma\tau}) 
\equiv& \frac{1}{4} 
\Big[ (1+I_\tau+I_\sigma+I_{\sigma\tau})^m 
- (1+I_\tau-I_\sigma-I_{\sigma\tau})^m
\nonumber \\ & 
+(1-I_\tau+I_\sigma-I_{\sigma\tau})^m
-(1-I_\tau-I_\sigma+I_{\sigma\tau})^m \Big]
, \\ 
F^{(\sigma\tau)}_{m}%(I_\tau,I_\sigma,I_{\sigma\tau}) 
\equiv& \frac{1}{4} 
\Big[ (1+I_\tau+I_\sigma+I_{\sigma\tau})^m 
- (1+I_\tau-I_\sigma-I_{\sigma\tau})^m
\nonumber \\ & 
-(1-I_\tau+I_\sigma-I_{\sigma\tau})^m
+(1-I_\tau-I_\sigma+I_{\sigma\tau})^m \Big]
.
\end{align}
\end{subequations}
Their first derivatives with respect to spin, isospin and spin-isospin excesses are
\begin{subequations}
\begin{align}
\frac{\partial F^{(\tau)}_{m}}{\partial I_\tau} =& \frac{\partial F^{(\sigma)}_{m}}{\partial I_\sigma} = \frac{\partial F^{(\sigma\tau)}_{m}}{\partial I_{\sigma\tau}} = m F^{(0)}_{m-1}
\,\,\,, \\
\frac{\partial F^{(0)}_{m}}{\partial I_\tau} =& \frac{\partial F^{(\sigma)}_{m}}{\partial I_{\sigma\tau}} = \frac{\partial F^{(\sigma\tau)}_{m}}{\partial I_\sigma} = m F^{(\tau)}_{m-1}
\,\,\,, \\
\frac{\partial F^{(0)}_{m}}{\partial I_\sigma} =& \frac{\partial F^{(\tau)}_{m}}{\partial I_{\sigma\tau}} = \frac{\partial F^{(\sigma\tau)}_{m}}{\partial I_\tau} = m F^{(\sigma)}_{m-1}
\,\,\,, \\
\frac{\partial F^{(0)}_{m}}{\partial I_{\sigma\tau}} =& \frac{\partial F^{(\tau)}_{m}}{\partial I_\sigma} = \frac{\partial F^{(\sigma)}_{m}}{\partial I_\tau} = m F^{(\sigma\tau)}_{m-1} \,\,\,,
\end{align}
\end{subequations}
while their second derivatives are
%\begin{subequations}
\begin{align}
%\frac{\partial^2 F^{(0)}_{m}}{\partial I_i^2} =& m (m-1) F^{(0)}_{m-2} \,\,\,, \\
\frac{\partial^2 F^{(j)}_{m}}{\partial I_i^2} =& m (m-1) F^{(j)}_{m-2} \,\,\,,
\end{align}
%\end{subequations}
for any $i,j \in \{0,\tau,\sigma,\sigma\tau\}$. Remarkable values are 
\begin{subequations}
\begin{align}
& F^{(0)}_{0}(I_\tau,I_\sigma,I_{\sigma\tau}) =1 \,\,\, , \,\,\, F^{(i)}_{0}(I_\tau,I_\sigma,I_{\sigma\tau}) =0 \,\,\,,  \\
& F^{(0)}_{1}(I_\tau,I_\sigma,I_{\sigma\tau}) =1 \,\,\, , \,\,\, F^{(i)}_{1}(I_\tau,I_\sigma,I_{\sigma\tau}) =I_i \,\,\,,
\end{align}
\end{subequations}
and
\begin{subequations}
\begin{align}
& F^{(0)}_{m} (0,0,0) = 1 \,\,\,,   \\ 
& F^{(i)}_{m} (0,0,0) = 0 \,\,\,,   \\
& F^{(\tau)}_{m} (0,1,0) = F^{(\tau)}_{m} (0,0,1) = 0 \,\,\,,  \\
& F^{(\sigma)}_{m} (1,0,0) = F^{(\sigma)}_{m} (0,0,1) = 0 \,\,\,,   \\
& F^{(\sigma\tau)}_{m} (1,0,0) = F^{(\sigma\tau)}_{m} (0,1,0) = 0 \,\,\,,  \\
& F^{(0)}_{m} (1,0,0) = F^{(0)}_{m} (0,1,0) = F^{(0)}_{m} (0,0,1) = 2^{m-1} \,\,\,,  \\
& F^{(\tau)}_{m} (1,0,0) = F^{(\sigma)}_{m} (0,1,0) = F^{(\sigma\tau)}_{m} (0,0,1) = 2^{m-1} \,\,\,,  \\
& F^{(0)}_{m} (1,1,1) = F^{(i)}_{m} (1,1,1) = 4^{m-1} \,\,\,,
\end{align}
\end{subequations}
where $i \in \{\tau,\sigma,\sigma\tau\}$.

\end{appendix}

%\bibliography{LNP_duguet}

\begin{thebibliography}{99}

\bibitem{KalantarNayestanaki:2011wz}
N. Kalantar-Nayestanaki {\it et al.}, 
Rept. Prog. Phys. \textbf{75}, 016301 (2012)

\bibitem{nogga00}
A. Nogga, H. Kamada, W. Gl{\"o}ckle,
Phys. Rev. Lett. \textbf{85}, 944 (2000)

\bibitem{nogga04b}
A. Nogga, S. K. Bogner, A. Schwenk,
Phys. Rev. C \textbf{70}, 061002 (2004)

\bibitem{faessler75}
A. Faessler, S. Krewald, G. J. Wagner,
Phys. Rev. C \textbf{11}, 2069 (1975)

\bibitem{fujita57}
J. Fujita, H. Miyazawa,
Prog. Theor. Phys. \textbf{17}, 360 (1957)

\bibitem{zuo02a}
W. Zuo {\it et al.}, 
Nucl. Phys. A \textbf{706}, 418 (2002)

\bibitem{coester70}
F. Coester {\it et al.}, 
Phys. Rev. C \textbf{1}, 769 (1970)

\bibitem{brockmann90a}
R. Brockmann, R. Machleidt,
Phys. Rev. C \textbf{42}, 1965 (1990)

\bibitem{sonzogni07}
A. Sonzogni, \textit{NNDC Chart of Nuclides}, 2007, http://www.nndc.bnl.gov/chart/

\bibitem{sorlin08}
O. Sorlin, M.-G.  Porquet,
Prog. Part. Nucl. Phys. \textbf{61}, 602 (2008)

\bibitem{tanihata85a}
I. Tanihata {\it et al.},
Phys. Rev. Lett. \textbf{55}, 2676 (1985)

\bibitem{fukuda91}
M. Fukuda {\it et al.},
Phys. Lett. B \textbf{268}, 339 (1991)

\bibitem{hansen87}
P.G. Hansen, B. Jonson,
Europhys. Lett. \textbf{4}, 409 (1987)

\bibitem{jensen04}
A. S. Jensen {\it et al.},
Rev. Mod. Phys. \textbf{76}, 215 (2004)

\bibitem{blank07a}
B. Blank, M. Ploszajczak,
Rept. Prog. Phys. \textbf{71}, 046301 (2008)

\bibitem{pfutzner12a}
M. Pf\"{u}tzner {\it et al.},
Rev. Mod. Phys. \textbf{84}, 567 (2012)

\bibitem{blaum06a}
K. Blaum,
Phys. Rep. \textbf{425}, 1 (2006)

\bibitem{schlitt96}
B. Schlitt {\it et al.},
Hyp. Int. \textbf{99}, 117 (1996)

\bibitem{wang11a}
M. Wang {\it et al.},
Journal of Physics: Conference Series \textbf{312}, 092064 (2011)

\bibitem{Towner:2010zz}
I. S. Towner, J. C.  Hardy,
Rept. Prog. Phys. \textbf{73}, 046301 (2010)

\bibitem{Zagrebaev:2012hy}
V. Zagrebaev, A. Karpov, W. Greiner,
J. Phys. Conf. Ser. \textbf{420}, 012001 (2013)

\bibitem{moller02}
P. M{\"o}ller {\it et al.},
At. Data Nucl. Data. Tables \textbf{59}, 185 (2002)

\bibitem{royer06}
G. Royer, C. Gautier,
Phys. Rev. C \textbf{73}, 067302 (2006)

\bibitem{friar88}
J. L. Friar {\it et al.},
Phys. Lett. B \textbf{311}, 4 (1988)

\bibitem{nogga97}
A. Nogga {\it et al.},
Phys. Lett. B \textbf{409}, 19 (1997)

\bibitem{Pieper:2004qw}
S. C. Pieper, R. B. Wiringa, J. Carlson,
Phys. Rev. C \textbf{70}, 054325 (2004)

\bibitem{Pastore:2013ria}
S. Pastore {\it et al.}, 
arXiv:1302.5091 (2013)

\bibitem{Navratil:2009ut}
P. Navratil {\it et al.}, 
J. Phys. G \textbf{36}, 083101 (2009)

\bibitem{Epelbaum:2012qn}
E. Epelbaum {\it et al.}, 
Phys. Rev. Lett. \textbf{109}, 252501 (2012)

\bibitem{Hagen:2010gd}
G. Hagen {\it et al.}, 
Phys. Rev. C \textbf{82}, 034330  (2010)

\bibitem{Binder:2012mk}
S. Binder {\it et al.}, 
arXiv:1211.4748 (2012)

\bibitem{Dickhoff:2004xx}
W. H. Dickhoff, C. Barbieri,
Prog. Part. Nucl. Phys. \textbf{52} 377 (2004)

\bibitem{Cipollone:2013zma}
A. Cipollone, C. Barbieri, P. Navratil,
arXiv1303.4900 (2013)

\bibitem{Tsukiyama:2010rj}
K. Tsukiyama, S. K. Bogner, A. Schwenk,
Phys. Rev. Lett. \textbf{106}, 222502 (2011)

\bibitem{Hergert:2012nb}
H. Hergert {\it et al.}, 
Phys. Rev. C \textbf{87}, 034307 (2013)

\bibitem{Soma:2011aj}
V. Som\`a, T. Duguet, C. Barbieri,
Phys. Rev. C \textbf{84}, 064317 (2011)

\bibitem{Soma:2012zd}
V. Som\`a, C. Barbieri, T. Duguet,
Phys. Rev. C \textbf{87}, 011303 (2013)

\bibitem{Hergert:2013uja}
H. Hergert {\it et al.},
Phys. Rev. Lett. \textbf{110}, 242501 (2013)

\bibitem{signoracci13a}
A. Signoracci, T. Duguet, G. Hagen, unpublished (2013)

\bibitem{caurier04}
E. Caurier {\it et al.},
Rev. Mod. Phys. \textbf{77}, 427 (2005)

\bibitem{Dean:2004ck}
D. J. Dean {\it et al.},
Prog. Part. Nucl. Phys. \textbf{53}, 419 (2004)

\bibitem{brown06a}
B. A. Brown, W. A. Richter,
Phys. Rev. C \textbf{74}, 034315 (2006)

\bibitem{zuker03}
A. P. Zuker,
Phys. Rev. Lett. \textbf{90}, 042502 (2003)

\bibitem{Otsuka:2009cs}
T. Otsuka {\it et al.},
Phys. Rev. Lett. \textbf{105}, 032501 (2010)

\bibitem{Holt:2010yb}
J. D. Holt {\it et al.},
J. Phys. G \textbf{39}, 085111 (2012)

\bibitem{Holt:2011fj}
J. D. Holt, A. Schwenk,
Eur. Phys. J. A \textbf{49}, 39 (2013)

\bibitem{Holt:2013tda}
J. D. Holt, J. Engel,
Phys. Rev. C \textbf{87}, 064315 (2013)

\bibitem{bender03b}
  M. Bender, P.-H. Heenen, P.-G. Reinhard,
  Rev. Mod. Phys. \textbf{75}, 121 (2003)

\bibitem{Niksic:2011sg}
T. Niksic, D. Vretenar, P. Ring,
Prog. Part. Nucl. Phys. \textbf{66}, 519 (2011)

\bibitem{negele72a}
J. W. Negele, D. Vautherin,
Phys. Rev. C \textbf{5}, 1472 (1972)

\bibitem{ring80a}
P. Ring and P. Schuck, {\it The Nuclear Many-Body Problem}, 1980, Springer-Verlag, New-York

\bibitem{Robledo07a}
L. M. Robledo,
Int. J. Mod. Phys. E \textbf{16}, 337 (2007)

\bibitem{dobaczewski07}
J. Dobaczewski {\it et al.},
Phys. Rev. C \textbf{76}, 054315 (2007)

\bibitem{Lacroix:2008rj}
D. Lacroix, T. Duguet, M. Bender,
Phys. Rev. C \textbf{79}, 044318 (2009)

\bibitem{Bender:2008rn}
M. Bender, T. Duguet, D. Lacroix,
Phys. Rev. C \textbf{79}, 044319 (2009)

\bibitem{Duguet:2008rr}
T. Duguet {\it et al.},
Phys. Rev. C \textbf{79}, 044320 (2009)

\bibitem{rotival07a}
V. Rotival, T. Duguet,
Phys. Rev. C \textbf{79}, 054308 (2009)

\bibitem{chabanat98}
E. Chabanat {\it et al.},
Nucl. Phys. A \textbf{635}, 231 (1998)

\bibitem{rotival07b}
V. Rotival, K. Bennaceur, T. Duguet, 
Phys. Rev. C \textbf{79}, 054309 (2009)

\bibitem{lesinski06a}
T. Lesinski {\it et al.},
Phys. Rev. C \textbf{74}, 044315 (2006)

\bibitem{Lesinski:2007zz}
T. Lesinski {\it et al.},
Phys. Rev. C \textbf{76}, 014312 (2007)

\bibitem{Kortelainen:2008rp}
M. Kortelainen {\it et al.},
Phys. Rev. C \textbf{77}, 064307 (2008)

\bibitem{Bender:2009ty}
M. Bender {\it et al.},
Phys. Rev. C \textbf{80}, 064302 (2009)

\bibitem{Margueron:2007uf}
J. Margueron, H. Sagawa, K. Hagino,
Phys. Rev. C \textbf{77}, 054309 (2008)

\bibitem{Niksic:2008vp}
T. Niksic, D. Vretenar, P. Ring,
Phys. Rev. C \textbf{78}, 034318 (2008)

\bibitem{Carlsson:2008gm}
B. G. Carlsson, J. Dobaczewski, M. Kortelainen,
Phys. Rev. C \textbf{78}, 044326 (2008)

\bibitem{Goriely:2009zz}
S. Goriely {\it et al.},
Phys. Rev. Lett. \textbf{102}, 242501 (2009)

\bibitem{Kortelainen:2010hv}
M. Kortelainen{\it et al.},
Phys. Rev. C \textbf{82}, 024313 (2010)

\bibitem{Kortelainen:2011ft}
M. Kortelainen{\it et al.},
Phys. Rev. C \textbf{85}, 024304 (2012)

\bibitem{Duguet:2010cv}
T. Duguet, J. Sadoudi,
J. Phys. G: Nucl. Part. Phys. \textbf{37}, 064009 (2010)

\bibitem{duguet02a}
T. Duguet {\it et al.},
Phys. Rev. C \textbf{65}, 014310 (2002)

\bibitem{Bally:2011iz}
B. Bally {\it et al.}, 
Int. J. Mod. Phys. E \textbf{21}, 1250026 (2012)

\bibitem{rodriguezguzman04a}
R. R. Rodriguez-Guzman, K. W. Schmid,
Eur. Phys. J. A \textbf{19}, 45 (2004)

\bibitem{varshalovich88a}
D. A. Varshalovich, A. N. Moskalev, V. K. Khersonskii, {\it Quantum Theory of Angular Momentum}, 1988, World Scientific, Singapor

\bibitem{sadoudi11thesis}
J. Sadoudi,  {\it Constraints on the nuclear energy density functional and new possible analytical forms}, 2011, Universit\'e Paris XI, France, http://tel.archives-ouvertes.fr/docs/00/04/49/86/PDF/tel-00001784.pdf

\bibitem{Robledo:2009yd}
L. M. Robledo, 
Phys. Rev. C \textbf{79}, 021302 (2009)

\bibitem{Robledo:2011ce}
L. M. Robledo,
Phys. Rev. C \textbf{84}, 014307 (2011)

\bibitem{Avez:2011wr}
B. Avez, M. Bender,
Phys. Rev. C \textbf{85}, 034325 (2012)

\bibitem{Oi:2011qp}
M. Oi, M. Takahiro,
Phys. Lett. B \textbf{707}, 305 (2012)

\bibitem{Gao:2013vaa}
Z.-C. Gao, Q.-L. Hu, Y. S. Chen,
arXiv:1306.3051 (2013)

\bibitem{robledo10a}
L. M. Robledo,
J. Phys. G \textbf{37}, 064020 (2010)

\bibitem{kamlah68a}
A. Kamlah,
Z. Phys. \textbf{216}, 52 (1968)

\bibitem{jancovici64}
B. Jancovici, D. H. Schiff,
Nucl Phys. \textbf{58}, 678 (1964)

\bibitem{brink68}
D. M. Brink, A. Weiguny,
Nucl. Phys. A \textbf{120}, 59 (1968)

\bibitem{balian69a}
R. Balian, E. Br{\'e}zin,
Nuovo Cimento \textbf{64}, 37 (1969)

\bibitem{perlinska04a}
E. Perlinska {\it et al.},
Phys. Rev. C \textbf{69}, 014316 (2004)

\bibitem{doba95a}
J. Dobaczewski, J. Dudek,
Phys. Rev. C \textbf{52}, 1827 (1995)

\bibitem{perdew81a}
J. P. Perdew, A. Zunger,
Phys. Rev. B \textbf{23}, 5048 (1981)


\bibitem{Chamel:2010ac}
N. Chamel,
Phys. Rev. C \textbf{82}, 061307 (2010)

\bibitem{Ruz07aDFT}
A. Ruzsinsky {\it et al.},
J. Phys. Chem. \textbf{126}, 104102 (2007)

\bibitem{Duguet:2003yi}
T. Duguet,
Phys. Rev. C \textbf{69}, 054317 (2004)

\bibitem{Yamagami:2008ks}
M. Yamagami, Y. R. Shimizu, T. Nakatsukasa,
Phys. Rev. C \textbf{80}, 064301 (2009)

\bibitem{Chamel:2010rw}
N. Chamel,
Phys. Rev. C \textbf{82}, 014313 (2010)

\bibitem{Yamagami:2012ga}
M. Yamagami {\it et al.},
Phys. Rev. C \textbf{86}, 034333 (2012)

\bibitem{Margueron:2009rn}
J. Margueron, H. Sagawa,
J. Phys. G \textbf{36}, 125102 (2009)

\bibitem{Cochet:2003ex}
B. Cochet {\it et al.},
Nucl. Phys. A \textbf{731}, 34 (2004)

\bibitem{Cochet:2003sy}
B. Cochet {\it et al.},
Int. J. Mod. Phys. E \textbf{13}, 187 (2004)

\bibitem{Lesinski:2008cd}
T. Lesinski {\it et al.},
Eur. Phys. J. A \textbf{40}, 121 (2009)

\bibitem{Lesinski:2011rn}
T. Lesinski {\it et al.},
J. Phys. G \textbf{39}, 015108 (2012)

\bibitem{Zalewski:2010ni}
M. Zalewski, P. Olbratowski, W. Satula,
Phys. Rev. C \textbf{81}, 044314 (2010)

\bibitem{Fantina:2010iq}
A. F. Fantina {\it et al.},
J. Phys. G \textbf{38}, 025101 (2011)

\bibitem{MoyadeGuerra:2011zz}
E. Moya de Guerra, O. Moreno, P. Sarriguren,
J. Phys. Conf. Ser. \textbf{312}, 092045(2011)

\bibitem{Davesne:2013aja}
D. Davesne, A. Pastore, J. Navarro,
arXiv:1307.2349 (2013)

\bibitem{Sadoudi:2012jg}
J. Sadoudi {\it et al.},
Phys. Scripta T \textbf{154}, 014013 (2013)

\bibitem{Dobaczewski:2012cv}
J. Dobaczewski, K. Bennaceur, F. Raimondi,
J. Phys. G \textbf{39}, 125103 (2012)

\bibitem{Bennaceur:2013fua}
K. Bennaceur, J. Dobaczewski, F. Raimondi,
arXiv:1305.7210 (2013)

\bibitem{staszczak10a}
A. Staszczak {\it et al.},
Eur. Phys. J. A \textbf{46}, 85 (2010)

\bibitem{bender13a}
M. Bender, private communication (2013)

\bibitem{baranger70a}
M. Baranger,
Nucl. Phys. A \textbf{149}, 225 (1970)

\bibitem{Duguet:2011sq}
T. Duguet, G. Hagen,
Phys. Rev. C \textbf{85}, 034330 (2012)

\bibitem{sadoudi12b}
J. Sadoudi, T. Duguet, unpublished (2013)

\bibitem{duguet12a}
T. Duguet, unpublished (2013)

\bibitem{Chamel:2012br}
N. Chamel, S. Goriely, J. M. Pearson, {\it Pairing: from atomic nuclei to neutron-star crusts}, in {\it Fifty Years of Nuclear BCS: Pairing in Finite Systems}, p. 284, 2013, Ed. R. Broglia and W. Zelevinsky, World Scientific Publishing Co. Pte. Ltd.

\bibitem{bender02a}
M. Bender, J. Dobaczewski, J. Engel, and W. Nazarewicz,
Phys. Rev. C \textbf{65}, 054322 (2002).

\bibitem{hohenberg64}
P. Hohenberg, W. Kohn,
Phys. Rev. \textbf{136},  B864 (1964)

\bibitem{fertig00a}
H. A. Fertig, W. Kohn,
Phys. Rev. A \textbf{62}, 052511 (2000)

\bibitem{gross88a}
E. K. U. Gross, L. N. Oliveira, W. Kohn,
Phys. Rev. A \textbf{37}, 2809 (1988)

\bibitem{gorling93a}
A. Gorling, 
Phys. Rev. A \textbf{47}, 2783 (1993)

\bibitem{Eng07a}
J. Engel,
Phys. Rev. C \textbf{75}, 014306 (2007)

\bibitem{Messud:2009jh}
J. Messud, M. Bender, E. Suraud, 
Phys. Rev. C \textbf{80}, 054314 (2009)

\bibitem{lesinski13a}
T. Lesinski,
arXiv:1301.0807 (2013)

\bibitem{Hill53}
D. L. Hill, J. A. Wheeler,
Phys. Rev. \textbf{89}, 1106 (1953)

\bibitem{Yao:2012cx}
J.-M. Yao {\it et al.},
Phys. Rev. C \textbf{86}, 014310 (2012)

\bibitem{Yao:2013tca}
J.-M. Yao, H. Mei, Z. P. Li,
Phys. Lett. B \textbf{723}, 459 (2013)

\bibitem{Bender:2008zv}
M. Bender, P.-H. Heenen,
Phys. Rev. C \textbf{78}, 024309 (2008)

\bibitem{Bender:2005ri}
M. Bender, G. F. Bertsch, P.-H. Heenen,
Phys. Rev. C \textbf{73}, 034322 (2006)

\bibitem{Bender:2008gi}
M. Bender, G. F. Bertsch, P.-H. Heenen,
Phys. Rev. C \textbf{78},  054312 (2008)

\bibitem{Robledo:2011nf}
L. M. Robledo, G. F. Bertsch,
Phys. Rev. C \textbf{84}, 054302 (2011)

\bibitem{Rodriguez:2010by}
T. R. Rodriguez, J. L. Egido,
Phys. Rev. C \textbf{81}, 064323 (2010)

\bibitem{Yao:2009zv}
J.-M. Yao {\it et al.},
Phys. Rev. C \textbf{81}, 044311 (2010)

\bibitem{Satula:2009sx}
W. Satula {\it et al.},
Phys. Rev. C \textbf{81}, 054310 (2010)

\bibitem{Satula:2012zz}
W. Satula {\it et al.},
Phys. Rev. C \textbf{86}, 054316 (2012)

\bibitem{Rodriguez:2012mx}
T. R. Rodriguez, G. Martinez-Pinedo,
Phys. Rev. C \textbf{85}, 044310 (2012)

\bibitem{duguet02thesis}
T. Duguet, {\it Probl\`eme \`a N corps nucl\'eaire et force effective dans les m\'ethodes de champ moyen auto-coh\'erent}, 2002, http://tel.archives-ouvertes.fr/docs/00/04/49/86/PDF/tel-00001784.pdf

\bibitem{baye84a}
D. Baye, P.-H. Heenen,
Phys. Rev. C \textbf{29}, 1056 (1984)

\bibitem{Zdunczuk:2006qh}
H. Zdunczuk, J. Dobaczewski, W. Satula,
Int. J. Mod. Phys. E \textbf{16}, 377 (2007)

\bibitem{avez13a}
B. Avez {\it et al.}, unpublished (2013)

\bibitem{meyer95}
J. Meyer {\it et al.},
Nucl. Phys. A \textbf{588}, 597 (1995)

\bibitem{Bender:2006tb}
M. Bender, T. Duguet,
Int. J. Mod. Phys. E \textbf{16}, 222 (2007)

\bibitem{Vaquero:2011hq}
N. L. Vaquero, T. R. Rodriguez, J. L. Egido,
Phys. Lett. B \textbf{704}, 520 (2011)

\bibitem{Peru:2008gd}
S. Peru, H. Goutte, 
Phys. Rev. C \textbf{77}, 044313 (2008)

\bibitem{Yoshida:2009jn}
K. Yoshida,
Eur. Phys. J. A \textbf{42},  583 (2009)

\bibitem{pena09a}
D. Pena Arteaga, P. Ring,
arXiv:0912.0908 (2009)

\bibitem{Losa:2010bm}
C. Losa {\it et al.},
Phys. Rev. C \textbf{81}, 064307 (2010)

\bibitem{Terasaki:2010zg}
J. Terasaki, J. Engel,
Phys. Rev. C \textbf{82}, 034326 (2010)

\bibitem{Nakatsukasa:2007qj}
T. Nakatsukasa, T. Inakura, K. Yabana
Phys. Rev. C \textbf{76}, 024318 (2007)

\bibitem{Toivanen:2009mr}
J. Toivanen {\it et al.},
Phys. Rev. C \textbf{81}, 034312 (2010)

\bibitem{Avogadro:2011gd}
P. Avogadro, T. Nakatsukasa,
Phys. Rev. C \textbf{84}, 014314 (2011)

\bibitem{Paar:2010ww}
N. Paar,
J. Phys. G \textbf{37}, 064014 (2010)

\bibitem{reinhard87}
P.-G. Reinhard, K. Goeke,
Rep. Prog. Phys. \textbf{50}, 1 (1987)

\bibitem{Hagino:2002ya}
K. Hagino, P.-G. Reinhard, G. F. Bertsch, 
Phys. Rev. C \textbf{65}, 064320 (2002)

\bibitem{Rohozinski:2012mu}
S. G. Rohozinski,
J. Phys. G \textbf{39}, 095104 (2012)

\bibitem{prochniak04a}
L. Prochniak {\it et al.},
Nucl. Phys. A \textbf{730}, 59 (2004)

\bibitem{Prochniak:2009fs}
L. Prochniak, S. G. Rohozinski,
J. Phys. G \textbf{36}, 123101 (2009)

\bibitem{Libert:1999sw}
J. Libert, M. Girod, J.-P. Delaroche,
Phys. Rev. C \textbf{60}, 054301 (1999)

\bibitem{Delaroche:2009fa}
J.-P. Delaroche {\it et al.},
Phys. Rev. C \textbf{81}, 014303 (2010)

\bibitem{Niksic:2008cn}
T. Niksic {\it et al.},
Phys. Rev. C \textbf{79}, 034303 (2009)

\bibitem{Li:2012mn}
Z. P. Li {\it et al.},
Phys. Rev. C \textbf{86}, 034334 (2012)

\bibitem{Hinohara:2012zz}
N. Hinohara {\it et al.},
Phys. Rev. C \textbf{85}, 024323 (2012)

\bibitem{Nomura:2010gg}
K. Nomura {\it et al.},
Phys. Rev. C \textbf{83}, 014309 (2011)

\bibitem{Nomura:2011xk}
K. Nomura {\it et al.},
Phys. Rev. C \textbf{84}, 014302 (2011)

\bibitem{almehed01a}
D. Almehed, S. Frauendorf, F. D{\"o}nau,
Phys. Rev. C \textbf{63}, 044311 (2001)

\bibitem{anguiano01a}
M. Anguiano, J. L. Egido, L. M. Robledo,
Nucl. Phys. A \textbf{683}, 227 (2001)

\bibitem{bender12a}
M. Bender {\it et al.},
unpublished (2013)

\bibitem{Gebremariam:2009ff}
B. Gebremariam, T. Duguet, S. K. Bogner,
Phys. Rev. C \textbf{82}, 014305 (2010)

\bibitem{Gebremariam:2010ni}
B. Gebremariam, S. K. Bogner, T. Duguet,
Nucl. Phys. A \textbf{851}, 17 (2011)

\bibitem{Kaiser:2010pp}
N. Kaiser,
Eur. Phys. J. A \textbf{45}, 61 (2010)

\bibitem{Holt:2011nj}
J. W. Holt, N. Kaiser, W. Weise,
Eur. Phys. J. A \textbf{47}, 128 (2011)

\bibitem{Holt:2011jd}
J. W. Holt, N. Kaiser, W. Weise,
Prog. Part. Nucl. Phys. \textbf{67}, 353 (2012)

\bibitem{Kaiser:2012mn}
N. Kaiser,
Eur. Phys. J. A \textbf{48}, 36 (2012)

\bibitem{Stoitsov:2010ha}
M. Stoitsov {\it et al.},
Phys. Rev. C \textbf{82}, 054307 (2010)


\end{thebibliography}

\end{document}